\documentclass[fleqn,usenatbib]{mnras}

\usepackage{newtxtext,newtxmath}

\usepackage[T1]{fontenc}

\DeclareRobustCommand{\VAN}[3]{#2}
\let\VANthebibliography\thebibliography
\def\thebibliography{\DeclareRobustCommand{\VAN}[3]{##3}\VANthebibliography}

\usepackage{graphicx}	
\usepackage{amsmath}	
\usepackage{subfig}
\usepackage{orcidlink}
\usepackage{breakurl}

\title[B supergiant stellar wind structure at low Z]{Optically-thick structure in early B type supergiant stellar winds at low metallicities}

\author[T. N. Parsons et al.]{Timothy N. Parsons$^{1}$\thanks{E-mail: timothy.parsons.15@ucl.ac.uk (TNP)}\orcidlink{0000-0002-7714-7823},
Raman K. Prinja$^{1}$\orcidlink{0000-0002-7714-7823},
Matheus Bernini-Peron$^{2}$\orcidlink{0000-0003-1113-0727},
Alex W. Fullerton$^{3}$,
\newauthor
Derck L. Massa$^{4}$\orcidlink{0000-0002-9139-2964},
Lidia M. Oskinova$^{5}$\orcidlink{0000-0003-0708-4414},
Daniel Pauli$^{5}$\orcidlink{0000-0002-5453-2788},
Matthew J. Rickard$^{1}$\orcidlink{0000-0002-5255-8116} and
Andreas A.C. Sander$^{2}$\orcidlink{0000-0002-2090-9751}\\
$^{1}$Department of Physics and Astronomy, University College London, Gower Street, London WC1E 6BT, UK \\
$^{2}$ Zentrum f\"{u}r Astronomie der Universit\"{a}t Heidelberg, Astronomiches Rechen-Institut, M\"{o}nchhofstr. 12-14, 69120 Heidelberg, Germany \\
$^{3}$Space Telescope Science Institute, 3700 San Martin Drive, Baltimore, MD 21218, USA \\
$^{4}$Space Science Institute (SSI), 4750 Walnut Street, Suite 205, Boulder, CO 80301, USA \\
$^{5}$Institut f\"{u}r Physik und Astronomie, Universit\"{a}t Potsdam, Karl-Liebknecht-Str. 24/25, 14476 Potsdam, Germany 
}

\date{Accepted 2023 December 21. Received 2023 December 20; in original form 2023 October 28}

\pubyear{2023}

\begin{document}
\label{firstpage}
\pagerange{\pageref{firstpage}--\pageref{lastpage}}
\maketitle

\begin{abstract}
Accurate determination of mass-loss rates from massive stars is important to understanding stellar and galactic
evolution and enrichment of the interstellar medium. Large-scale structure and variability in stellar winds have significant effects on mass-loss rates. Time-series observations provide direct quantification of such variability. Observations of this nature are available for some Galactic early supergiant stars but not yet for stars in lower metallicity environments such as the Magellanic Clouds. We utilise ultraviolet spectra from the Hubble Space Telescope ULLYSES program to demonstrate that the presence of structure in stellar winds of supergiant stars at low metallicities may be discerned from single-epoch spectra. We find evidence that, for given stellar luminosities and mean stellar wind optical depths, structure is more prevalent at higher metallicities. We confirm, at Large Magellanic Cloud (0.5~$Z_\odot$), Small Magellanic Cloud (0.2~$Z_\odot$) and lower (0.14 -- 0.1~$Z_\odot$) metallicities, earlier Galactic results that there does not appear to be correlation between the degree of structure in stellar winds of massive stars and stellar effective temperature. Similar lack of correlation is found with regard to terminal velocity of stellar winds. Additional and revised values for radial velocities of stars and terminal velocities of stellar winds are presented. Direct evidence of temporal variability, on timescales of several days, in stellar wind at low metallicity is found. We illustrate that narrow absorption components in wind-formed profiles of Galactic OB stellar spectra remain common in early B supergiant spectra at low metallicities, providing means for better constraining hot, massive star mass-loss rates.
\end{abstract}




\section{Introduction}

Their relative rarity notwithstanding, massive stars are the key drivers of galactic evolution and of enrichment of the interstellar medium (ISM). Stars with zero-age main sequence masses greater than approximately $8 \,M_\odot$ evolve rapidly and undergo significant mass-loss, during both main sequence and post-main sequence phases. These stars, with some exceptions in the very high mass regime, end their lives as core-collapse supernovae, thereby returning significant amounts of highly-processed material to the surrounding ISM on relatively short timescales.
   
Unlike lower mass stars, massive stars exhibit strong stellar winds. These winds are accelerated, not simply by continuum-derived radiation pressure, but very significantly by ``line-driving'', whereby resonance line absorption of ultraviolet (UV) photons can impart significant momentum, and substantial velocities, to stellar wind material \citep[see, for example, discussion of this process in][]{Mihalas1978, Puls2000, Hubeny2015}.

In this work, we aim primarily to identify the presence of large-scale structure in the stellar winds of massive stars at low metallicities, specifically early B-type supergiants in the Large and Small Magellanic Clouds (LMC, SMC) and some very low-metallicity Local Group dwarf galaxies (IC 1613 and NGC 3109), from single-epoch UV spectra. We seek to demonstrate that the presence of optically-thick structure in these winds may be reliably identified even in the absence of extensive time-series observations. The presence of such structure is likely to have significant consequence for accurate estimation of mass-loss rates from these stars.

As demonstrated by \citet{Oskinova2007}, among others, neglecting the effects of macro-scale optically thick clumping in stellar winds can lead to a significant underestimation of the rates of mass-loss from hot, massive stars: a key factor driving stellar and galactic evolution. \citet{Fullerton2006} and \citet{Prinja2013} show that this discrepancy may be at an order-of-magnitude scale. The presence and nature of stellar wind structure has been identified and analysed in considerable detail, both temporally and spatially, in respect of some Galactic B supergiant stars using time series data obtained with the \textit{International Ultraviolet Explorer} (\textit{IUE}) \citep[see, for example,][]{Massa1995, Fullerton1997, Prinja2002}. To date however, no similar data are available for similar stars at lower metallicities. In addition, we identify examples where evidence of temporal variability may be identified in our data, indicating the value that future time-series observations would be able to provide.

In order to undertake the wind structure analysis described, we make use of stellar physical parameters for our target sample of stars largely from existing literature. These are referenced in detail in Table \ref{table:Alltargets} below. Some of these values are revised or newly-derived, as described below. The derivation and more precise quantification of many such physical parameters represents a substantial, and ongoing, programme and reference is made in particular to the \textit{XShootU} collaboration in this regard \citep[see][]{Vink2023}.

We further attempt to identify whether any correlations are observed between effective temperatures, stellar luminosity, metallicity and rotational velocity on the one hand and the extent of wind structure which may be discerned from the method here discussed and currently available data.

Use of the ``Sobolev with exact integration'' (SEI) method described in Section \ref{SEImethod} and the conclusions drawn from its application to existing single-epoch data are motivated by the need to provide a more reliable starting point for the determination of mass-loss over the full lifetime of massive stars in low-metallicity conditions. In this way, we aim to provide better constraints to models of galactic evolution in the early Universe.

Finally, we use the results presented in Section \ref{results} to identify some suitable targets for obtaining time-series UV observations of massive stars in low-metallicity environments for the purpose of obtaining direct empirical evidence to test the conclusions reached in this paper. Such targeted observations, together with the results presented here, would provide powerful constraints upon the true extent of variability in mass-loss rates from massive stars at low metallicities and, by extension, in early phases of galactic evolution.

\section{Observational data, target selection and data reduction}

\subsection{Target stars and ULLYSES data}

We use UV spectra obtained with the \textit{Cosmic Origins Spectrograph} (COS) and \textit{Space Telescope Imaging Spectrograph} (STIS) instruments on the \textit{Hubble Space Telescope} (HST). The primary data come from the HST Ultraviolet Legacy Library of Young Stars as Essential Standards (ULLYSES) observing program \citep{Roman2020}, up to and including Data Release 5 (June 2022). We concentrate on early B supergiant stars in the LMC and SMC. The target stars examined are all early B-type supergiant stars (spectral types B0 - B5). B supergiants are the most numerous hot, luminous stars \citep{Prinja2005}. Stars of these spectral types generally provide well-developed wind profiles in their UV spectra which are, in many cases, not saturated, particularly in the case of stars in the SMC where the lower metallicity environment can generally be expected to produce weaker stellar winds \citep{Kudritzki2000, Trundle2004, Bouret2015, Rickard2022}.

In addition to the foregoing, we have included a small number of stars of similar spectral types from ULLYSES Data Release 6 (March 2023), being early B supergiant stars observed as part of the ``low Z'' programme of observing such stars in certain very low metallicity Local Group dwarf galaxies (IC 1613, NGC 3109 and Sextans A). These are among suitable targets identified in previous studies of these galaxies, such as \citet{Bresolin2007} and \citet{Lorenzo2022}.

\begin{table*} 
\centering
\caption{\label{table:Alltargets}Tabulation of LMC (first block) and SMC (second block) B0 - B5 supergiants in the ULLYSES DR5 sample and Low Z early B supergiants (third block) in the ULLYSES DR6 sample which are suitable for analysis, spectral types and values for principal stellar physical parameters. Targets marked (*) are not susceptible to separate line profile fitting for each doublet element and a combined fit with floating inferred oscillator strength ratio is used, as described in the text. New and varied values of stellar radial velocity and terminal wind velocity derived for this work are indicated in the table.}
\begin{tabular}{llrrrrrrrrrrr} \hline \hline \\
Star & Spectral type & \textit{Ref} & $T_{\rm eff}$ & log $g$ & log $L$ & \textit{Ref} & $\varv\textsubscript{rad}$ & \textit{Ref} & $\varv\textsubscript{$\infty$}$ & \textit{Ref} & $\varv$ sin $i$ & \textit{Ref} \\
 & & & (K) & (cgs) & $(L/L_\odot)$ & & (km s$^{-1}$) & & (km s$^{-1}$) & & (km s$^{-1}$) & \\[0.5ex]
\hline
Sk -69 43 * & B0.5 Ia & \textit{4} & 22,845 & 2.62 & 5.50 & \textit{24} & 242 & \boldmath$x$ & 900 & \boldmath$xx$ & 102 & \textit{19} \\
Sk -68 140 * & B0.7 Ib-Iab Nwk & \textit{22} & 23,500 & 2.75 & 5.64 & \textit{22} & 266 & \textit{21} & 1,150 & \textit{30} & 53 & \textit{22} \\
Sk -66 35 & BC1 Ia & \textit{5} & 22,000 & 2.6\phantom{0} & 5.73 & \textit{24} & 255 & \boldmath$x$ & 810 & \textit{30} & 80 & - \\
Sk -68 129 * & B1 I & \textit{30} & 22,200 & 2.5\phantom{0} & 5.25 & \textit{30} & 275 & \boldmath$x$ & 1,380 & \textit{30} & 96 & \textit{19} \\
Sk -67 14 * & B1.5 Ia & \textit{8} & 22,890 & 2.68 & 5.74 & \textit{24} & 280 & \boldmath$x$ & 900 & \textit{30} & 88 & \textit{19} \\
Sk -68 26 & BC2 Ia & \textit{5} & 18,160 & 2.19 & 5.71 & \textit{24} & 290 & \boldmath$x$ & 390 & \textit{30} & 80 & - \\
Sk -67 78 & B3 Ia & \textit{5} & 16,150 & 2.03 & 5.2\phantom{0} & \textit{24} & 315 & \boldmath$x$ & 470 & \boldmath$xx$ & 50 & - \\
Sk -70 50 & B3 Ia & \textit{5} & 15,000 & 2.0\phantom{0} & 5.2\phantom{0} & - & 240 & \boldmath$x$ & 360 & \boldmath$xx$ & 50 & \textit{29} \\
Sk -69 140 * & B4 I & \textit{1} & 15,000 & 2.25 & 4.73 & \textit{30} & 275 & \boldmath$xx$ & 1,100 & \boldmath$x$ & 50 & - \\
Sk -70 16 & B4 I & \textit{1} & 15,000 & 2.25 & 4.62 & \textit{30} & 258 & \boldmath$xx$ & 910 & \textit{30} & 50 & - \\
Sk -68 8 & B5 Ia+ & \textit{4} & 14,200 & 2.2\phantom{0} & 5.54 & \textit{30} & 290 & \textit{28} & 300 & \textit{30} & 50 & \textit{30} \\
NGC 2004 ELS 3 & B5 Ia & \textit{14} & 14,450 & 2.1\phantom{0} & 5.1\phantom{0} & \textit{17} & 309 & \textit{14} & 440 & \textit{30} & 42 & \textit{17} \\[0.5ex]
\hline
AzV 215 * & B0 (Ib)/BN0 Ia & \textit{9,6} & 27,000 & 2.9\phantom{0} & 5.63 & \textit{12} & 145 & \boldmath$x$ & 1,540 & \textit{30} & 92 & \textit{19} \\
AzV 235 * & B0 Iaw & \textit{2} & 27,500 & 2.9\phantom{0} & 5.72 & \textit{11} & 170 & \textit{18} & 1,480 & \boldmath$x$ & 90 & \textit{19} \\
AzV 317 * & B0 Iw & \textit{3} & 27,000 & 3.00 & 5.40 & \textit{25} & 118 & \textit{23} & 1,240 & \boldmath$x$ & 80 & - \\
AzV 104 & B0.5 Ia & \textit{6} & 27,500 & 3.1\phantom{0} & 5.31 & \textit{12} & 190 & \boldmath$x$ & 600 & \textit{30} & 56 & \textit{19} \\
AzV 488 * & B0.5 Iaw & \textit{2} & 27,500 & 2.9\phantom{0} & 5.74 & \textit{11} & 186 & \boldmath$x$ & 1,200 & \textit{30} & 87 & \textit{19} \\
AzV 242 & B0.7 Iaw & \textit{2} & 25,000 & 2.85 & 5.67 & \textit{13} & 187 & \boldmath$x$ & 950 & \textit{11} & 83 & \textit{11,19} \\
AzV 264 & B1 Ia & \textit{6} & 22,500 & 2.55 & 5.44 & \textit{13} & 131 & \textit{23} & 720 & \boldmath$x$ & 116 & \textit{19} \\
AzV 266 & B1 I & \textit{23} & 22,000 & 2.50 & 5.09 & \textit{25} & 149 & \textit{23} & 990 & \boldmath$x$ & 61 & \textit{19,23} \\
AzV 96 & B1 I/B1.5 Ia & \textit{23,6}& 20,000 & 2.55 & 5.39 & \textit{13} & 166 & \textit{23} & 800 & \boldmath$x$ & 70 & \textit{19} \\
AzV 78 & B1.5 Ia+ & \textit{6} & 21,500 & 2.40 & 5.92 & \textit{12} & 170 & \boldmath$x$ & 590 & \textit{30} & 46 & \textit{13} \\
Sk 191 & B1.5 Ia & \textit{6} & 22,500 & 2.55 & 5.77 & \textit{12} & 120 & \boldmath$x$ & 460 & \boldmath$x$ & 118 & \textit{19} \\
AzV 210 & B1.5 Ia & \textit{6} & 20,500 & 2.4\phantom{0} & 5.41 & \textit{12} & 178 & \boldmath$x$ & 750 & \textit{11} & 51 & \textit{27} \\
AzV 18 & B2 Ia & \textit{6} & 19,000 & 2.3\phantom{0} & 5.44 & \textit{12} & 150 & \boldmath$x$ & 400 & \boldmath$x$ & 49 & \textit{12} \\
NGC 330 ELS 4 & B2.5 Ib & \textit{14} & 17,000 & 2.3\phantom{0} & 4.77 & \textit{17} & 158 & \textit{18} & 460 & \boldmath$x$ & 36 & \textit{17} \\
AzV 187 & B3 Ia & \textit{6} & 14,600 & 2.0\phantom{0} & 4.91 & - & 135 & \boldmath$x$ & 470 & \boldmath$xx$ & 45 & - \\
AzV 234 & B3 Iab & \textit{27} & 15,700 & 2.15 & 4.91 & \textit{27} & 165 & \boldmath$xx$ & 280 & \textit{30} & 43 & \textit{27} \\
NGC 330 ELS 2 & B3 Ib & \textit{14} & 14,590 & 2.15 & 4.73 & \textit{17} & 154 & \textit{14} & 460 & \boldmath$x$ & 14 & \textit{17} \\
AzV 324 & B4 Iab & \textit{7} & 14,600 & 2.00 & 4.89 & \textit{30} & 153 & \boldmath$xx$ & 270 & \boldmath$x$ & 50 & - \\
AzV 22 & B5 Ia & \textit{6} & 14,500 & 1.9\phantom{0} & 5.04 & \textit{12} & 144 & \textit{26} & 300 & \textit{30} & 46 & \textit{12} \\
AzV 445 & B5 (Iab) & \textit{6} & 14,500 & 2.00 & 5.0\phantom{0} & - & 175 & \boldmath$x$ & 350 & \boldmath$xx$ & 50 & - \\[0.5ex]
\hline
NGC 3109 EBU07 * & B0-1 Ia & \textit{16} & 27,000 & 2.9\phantom{0} & 5.82 & \textit{16} & 382 & \textit{16} & 1,350 & \boldmath$xx$ & 80 & - \\
IC 1613 BUG2007 B3 & B0 Ia & \textit{15} & 24,500 & 2.65 & 5.53 & \textit{15} & -250 & \textit{15} & 1,100 & \boldmath$x$ & 100 & \textit{20} \\
IC 1613 BUG2007 A10 & B0.5 Ia/B1 Ia & \textit{20,15} & 25,000 & 2.7\phantom{0} & 5.71 & \textit{15} & -255 & \textit{15} & 1,075 & \textit{20} & 50 & \textit{20} \\
IC 1613 BUG2007 B4 & B1.5 Ia & \textit{15} & 22,500 & 2.6\phantom{0} & 5.20 & \textit{15} & -243 & \textit{15} & 875 & \textit{20} & 50 & \textit{20} \\[0.5ex]
\hline
\end{tabular} 
\\[1.5ex]
References: \textbf{1} \citet{Rousseau1978}, \textbf{2} \citet{Walborn1983}, \textbf{3} \citet{Garmany1987}, \textbf{4} \citet{Fitzpatrick1988}, \textbf{5} \citet{Fitzpatrick1991}, \textbf{6} \citet{Lennon1997}, \textbf{7} \citet{Smith1997}, \textbf{8} \citet{Walborn2002}, \textbf{9} \citet{Evans2004b}, \textbf{10} \citet{Evans2004c}, \textbf{11} \citet{Evans2004d}, \textbf{12} \citet{Trundle2004}, \textbf{13} \citet{Trundle2005}, \textbf{14} \citet{Evans2006}, \textbf{15} \citet{Bresolin2007}, \textbf{16} \citet{Evans2007}, \textbf{17} \citet{Trundle2007}, \textbf{18} \citet{Evans2008}, \textbf{19} \citet{Penny2009}, \textbf{20} \citet{Garcia2014}, \textbf{21} \citet{Evans2015}, \textbf{22} \citet{McEvoy2015}, \textbf{23} \citet{Lamb2016}, \textbf{24} \citet{Urbaneja2017}, \textbf{25} \citet{Castro2018}, \textbf{26} \citet{Neugent2018}, \textbf{27} \citet{Dufton2019}, \textbf{28} \citet{Jonsson2020} \textbf{29}  derived from the method in \citet{Bestenlehner2023}, \textbf{30} \citet{Hawcroft2023}.

(\boldmath$x$) = Derived for this work (and varying earlier published results), (\boldmath$xx$) = New result derived for this work, (-) = No data available and result estimated from other parameters or comparable stars.

Note: source references listed after luminosity data apply also to temperature and surface gravity data, except: Sk -69 43 log $L$ figure is estimated; Sk -69 140, Sk -70 16, AzV 266 and AzV 324log $g$ figures are estimated, AzV 96 $T_{\rm eff}$ figure is derived from UV spectral fitting; AzV 266 $T_{\rm eff}$ figure is from Ref \textit{13}.
\end{table*}

Table \ref{table:Alltargets} sets out the list of early B supergiant LMC and SMC stars from ULLYSES Data Release 5 which are suitable for analysis using the technique presented here, as well as some very low metallicity stars from other Local Group dwarf galaxies from ULLYSES Data Release 6. Stellar identifications are from \citet{Azzopardi1975} and \citet{Azzopardi1982} (AzV), \citet{Sanduleak1968, Sanduleak1970} (Sk), \citet{Evans2006} (ELS), \citet{Evans2007} (EBU) and \citet{Bresolin2007} (BUG). A number of early B supergiants in the ULLYSES data are excluded from these lists for one or more of the following reasons:
\begin{enumerate}
    \item saturation of absorption profiles;
    \item terminal wind velocity being too high for separate line-fitting to be reliable over a reasonable velocity range;
    \item the presence of prominent emission line characteristics or other peculiar spectral features within the resonance line profile;
    \item prominent interstellar absorption lines affecting critical parts of the resonance line absorption profile (particularly prevalent in the case of LMC stars with low terminal wind velocities, comparable in magnitude to the radial velocity of the star);
    \item lack of well-developed P Cygni-type resonance line profiles.
\end{enumerate}

\subsection{Data reduction method} \label{datared}
The high level science products from ULLYSES Data Releases 5 and 6 comprise \texttt{.fits} files containing wavelength and flux data for a given star from either, or both, of the COS and STIS instruments described above (as well as shorter wavelength \textit{FUSE} data in some cases). These may be made from more than one observation in each case. The ULLYSES data include a metatable of principal stellar parameters for target stars extracted from earlier literature (although the metatable is not complete for all targets). Primary sources for stellar data are referenced in Table \ref{table:Alltargets}.

Radial velocities for each target star were required as part of the data reduction process undertaken for this work. In preference to using ``bulk'' radial velocities for each galaxy, radial velocity data for individual target stars were each checked (without initial reference to existing literature values) using the \texttt{DIPSO} spectral analysis code, part of the \texttt{STARLINK} software package \citep{Currie2014}, by overplotting of the limited number of available photospheric absorption lines in the ULLYSES UV data: in each case, using at least two of Si \textsc{iii} 1312.59\AA, C \textsc{ii} 1323.95\AA{} and Si \textsc{iii} 1417.24\AA. These values were then compared to existing literature and, in some cases where there are significant differences identified and the results obtained via this method are sufficiently unambiguous, have been used in preference to literature figures. Where values were not given by existing literature, this method has been used to discern a suitable value. This process has provided radial velocity measurements to a precision of order 5 km s$^{-1}$, which is sufficient for the fitting techniques applied here.

Reliable terminal wind velocities are an essential aspect of line profile fitting using the SEI method. For those target stars contained within the ULLYSES Data Release 3 sample (August 2021), results from the comprehensive analysis contained in \citet{Hawcroft2023} have generally been used. In a small number of those cases, a different value has been used, generally arising from the use in this work of individual stellar radial velocity values, derived as described in the preceding paragraph, rather than the applicable bulk galactic values adopted in that work. For target stars contained within more recent ULLYSES data releases, we have applied similar techniques to those described in \citet{Hawcroft2023} to measure terminal velocities or to revisit and/or confirm other existing literature results. As noted in that work, applying the fitting process used there and in this work to unsaturated profiles will provide only a lower bound on terminal wind velocity, rather than the more precise determination possible from using saturated profiles. Where possible, we have therefore further constrained terminal velocity results by applying the technique of measuring the velocity displacement of narrow absorption component (NAC) features in unsaturated Si \textsc{iv} 1393.76, 1402.77\AA{} profiles, as described by \citet{Prinja1990} or by direct measurement from the width of saturated C \textsc{iv} 1548.20, 1550.77\AA{} profiles, as also described by \citet{Prinja1990} and by \citet{Hawcroft2023}.

Checks have also been made against other results in the literature where the given value of a parameter appears uncertain, or is absent. Use has also been made of results produced by application of the spectroscopic analysis pipeline recently introduced by \citet{Bestenlehner2023}. Further investigations have been undertaken where those results suggest stellar parameters significantly differing from the existing literature.

For each target star, a suitable photospheric spectrum is selected from the \texttt{TLUSTY} model grids \citep[see][and later guides]{Hubeny1988, Lanz2007}, based upon the effective temperature and surface gravity for the star, as derived from the process discussed above. Metallicity is taken to be 0.5 \(Z_\odot\) for LMC stars and 0.2 \(Z_\odot\) for SMC stars. In each case, a further visual check of the selected \texttt{TLUSTY} spectrum is made against the observed spectrum to identify any clearly discrepant spectra. In a small number of cases, the definite presence, or absence, of photospheric lines from particular ionized states in one spectrum compared to the other has required the selection of an alternative \texttt{TLUSTY} model spectrum to that suggested by the effective temperature provided in the literature. For the ``low Z'' target stars, metallicity is taken to be 0.1 \(Z_\odot\) when selecting \texttt{TLUSTY} grid models. \citet{Garcia2014} utilised SMC metallicity (0.2 \(Z_\odot\)) to select photospheric models for stars in IC 1613, however this difference does not significantly affect the results derived here.

Once a suitable photospheric spectrum is selected, that spectrum is ``spun up'' using a short \texttt{FORTRAN} procedure, \texttt{ROTIN3}, which forms part of the wider \texttt{TLUSTY} software package. This provides appropriate rotational broadening of the spectral lines, based upon the $\varv \sin i$ rotational velocity parameter for the star. The resulting spectrum is also normalised through division by a suitable continuum spectrum from the \texttt{TLUSTY} model grids.

The observed spectrum for each star is normalised manually using the \texttt{DIPSO} spectral analysis code previously referred to. A suitable continuum is adopted and the observed spectrum divided by this.  The widely-acknowledged difficulty in normalising UV spectra (due to extensive line-blanketing making the true continuum difficult to locate) means that this process does involve some trial-and-error in order to produce a reasonably well-normalised spectrum. The resulting spectrum is then smoothed for clarity using the \texttt{qsm} Gaussian smoothing function available within \texttt{DIPSO}.

The two normalised spectra are then combined, with the observed spectrum being suitably blue-shifted (or in the case of stars in IC 1613, red-shifted) to account for the stellar radial velocity previously derived. This is carried out using an \texttt{IDL} procedure, \texttt{auto\_windlines2}, which uses the Sobolev with exact integration method described in the following section together with weighted non-linear least-squares minimisation, using the Marquardt-Levenburg method, to produce a suitable model fit to the observed wind profile for the ion or ions selected \citep[see][]{Lamers1987, Massa2003, Prinja2010}.

This procedure provides identification of the photospheric contribution to the spectrum in the region(s) examined and seeks to provide a best fit model spectrum for a specified terminal wind velocity ($\varv_\infty$) with principal outputs being the radial optical depth of the wind across the normalised velocity range of the wind ($\tau_{\rm rad}(w)$, where $w=\varv/\varv_\infty$), the wind acceleration parameter ($\beta$) and a wind turbulence factor ($w_{\rm turb}$) expressed as a fraction of normalised terminal wind velocity ($\varv/\varv_\infty$ = 1). A single $\beta$ law to describe the wind velocity at any radius is assumed \citep{Castor1975}, such that the wind velocity at a given radius is expressed as:

\begin{equation}
    \varv(r) = \varv_\infty \left(1-\frac{b}{r} \right)^\beta .
\end{equation}

The value of the constant, \textit{b}, sets the wind velocity at or near the stellar photosphere. A fixed value of \textit{b} = 1 was used for each fit, representing a wind launched from zero velocity at the photosphere. Adopting this value permits the wind velocity equation to be written as:

\begin{equation}
    \varv(r) = \varv_\infty \left(1-\frac{R_*}{r} \right)^\beta ,
\end{equation}
where $R_*$ is the photospheric radius of the star.

It can readily be seen that higher values of $\beta$ describe winds with a lower rate of radial acceleration, and vice-versa for lower values of $\beta$. Empirically-derived typical $\beta$ values for OB stars range between 0.7-1.5, with supergiants tending to the upper end of this range. Note that $\beta$ values may fall outside this range and can be significantly larger for later type stars \citep{Kudritzki2000}. Higher values of $\beta$ for blue supergiants have also been derived in earlier literature, although in some cases these are derived assuming a smooth, non-clumped wind \citep[see, for example,][]{Trundle2005}. \citet{Bernini2023} adopted $\beta=2$ when including both clumping in stellar winds and X-ray effects in their description of winds of cool B supergiants (spectral types B2 - B5), based on values previously obtained by \citet{Crowther2006} and the fact that \citet{Haucke2018} identified a trend of $\beta\ge2$ for cooler B supergiants (spectral types B3 and later).

The weighted least-squares line fitting process used as part of the SEI method permits the values for the acceleration parameter and the wind turbulence factor to float as part of the process of finding the best fit model profile. The terminal wind velocity needs however to be specified to be an appropriate value in order for the fitting process to operate effectively, as described above.

   \begin{figure*}
\begin{center}
 \subfloat[ ]{\includegraphics[width=3.34in]{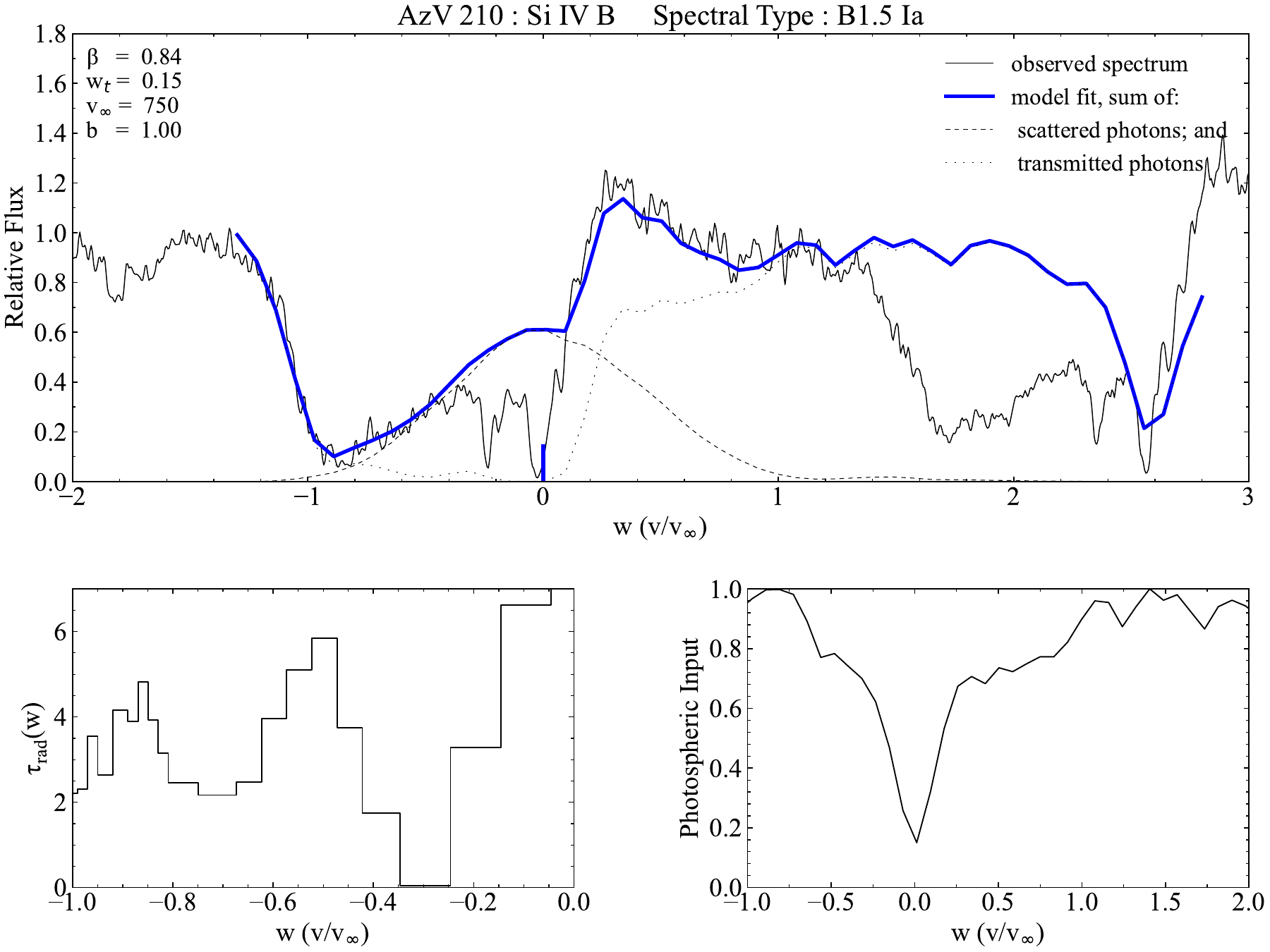} }
 \qquad
 \subfloat[ ]{\includegraphics[width=3.34in]{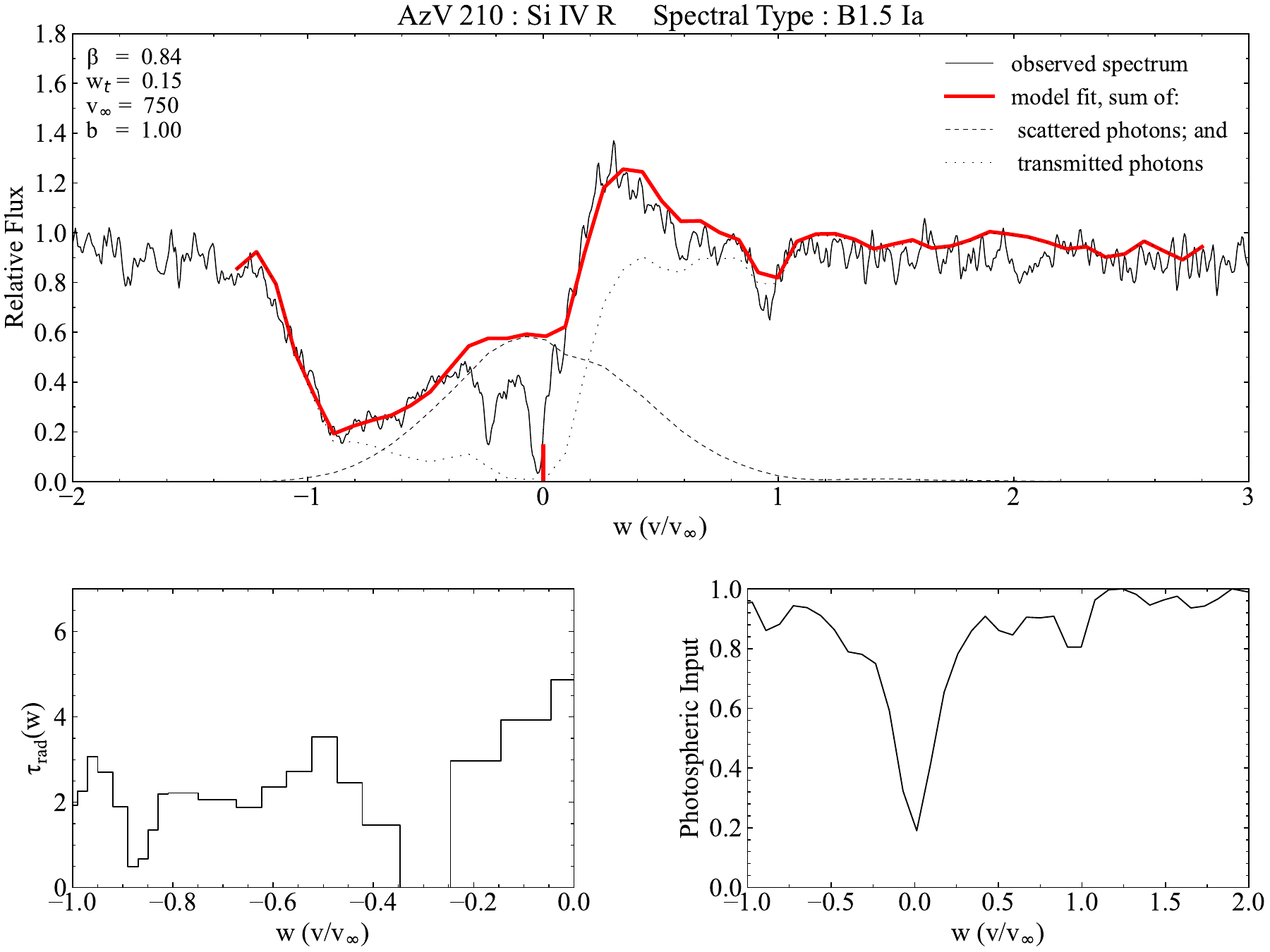} }
 \caption{SEI-derived model fits for (a) blue and (b) red components of the Si \textsc{iv} 1393.76, 1402.77 \AA{} resonance line doublet feature in the UV spectrum of SMC star AzV 210 (spectral type B1.5 Ia). As described in the introduction to Section \ref{SEImethod}, the lower panels of each plot respectively show the radial optical depths of the stellar wind at differing normalised velocities which produce the best-fit model line (lower left panels) and the photospheric contribution to the spectrum as derived from the model photospheric spectrum selected based upon the physical parameters of the star (lower right panels). A detailed explanation of each of the parameters derived and displayed with the plots is set out in Subsection \ref{datared}. Rest wavelengths of the blue and red elements of the doublet are highlighted in those colours. Note also that the decoupled treatment of each element means that the model fit for each element ``ignores'' the presence of the other element, as can be seen in the upper panel of plot (a). This does not affect the model fit within the relevant doublet element.}
 \label{fig210a}
\end{center}
    \end{figure*}

\section{The SEI method for line profile fitting} \label{SEImethod}

Use is made of a modified form of the SEI method introduced by \citet{Hamann1981} and developed, particularly, by \citet{Lamers1987}. This employs a single application of the Sobolev approximation to determine the source function and combines this with an exact solution to the formal integral for the radiative transfer equation. \citet{Massa2003} utilised this method to derive wind profiles and parameters for a large sample of O type stars in the LMC, which those authors then used to derive mass-loss rates for those stars using the formulae and methods introduced by \citet{Vink2000} and \citet{Vink2001}. The technique has recently also been used to estimate the terminal wind velocities of a large sample of the HST ULLYSES \citep{Roman2020} observing program's LMC and SMC target stars, through line fitting to the C \textsc{iv} 1548.20, 1550.77\AA{} feature \citep{Hawcroft2023}.

Wind-driven features visible in the UV spectra of early supergiant stars frequently exhibit characteristic P Cygni-type line profiles indicative of significant mass-loss from these stars and of the presence of a dense, radially accelerating envelope of surrounding material. We observe that certain resonance line (i.e. ground state transition) doublet features present well-developed and separated, but unsaturated, profiles in the case of many early B supergiants, particularly at lower than solar metallicities. The Si \textsc{iv} 1393.76, 1402.77\AA{} feature in particular often provides considerable diagnostic data capable of being derived from the shape, depth and width of the absorption parts of that feature. We concentrate therefore on this feature in the following analysis. In some cases, use is made of the Al \textsc{iii} 1854.72, 1862.79\AA{} feature.

The data reduction process required to use the SEI technique to produce a best-fit model of the observed P Cygni profiles is described in detail in Subsection \ref{datared} above. To illustrate our approach, we show in Figure \ref{fig210a} an example fitted spectrum for the SMC star AzV 210 (spectral type B1.5 Ia) produced using this technique. The plots show the observed spectrum in velocity space centred on the rest wavelength of the blue element of the Si \textsc{iv} resonance line doublet at 1393.76\AA{} in the left upper plot and the same for the red element at 1402.77\AA{} in the right upper plot. The best-fit model line for each feature is shown in the appropriate colour overplotted onto the applicable observed spectrum. The contributions to the model line from transmitted and scattered photons are also plotted. These plots further show the fundamental parameters of the stellar wind which are either derived from the fitting process or are adopted as fixed inputs. Each of these parameters is described in Subsection \ref{datared} above. The lower panels show, on the left in each case, the radial optical depths of the stellar wind at differing normalised velocities which produce the best-fit model line. This process is described in Subsection \ref{optdepths} below. The lower panel on the right in each group of plots shows the photospheric contribution to the spectrum as derived from the model photospheric spectrum selected based upon the physical parameters of the star. 

\subsection{Radial optical depths} \label{optdepths}

When carrying out the line fitting of the above-referenced Si \textsc{iv} or Al \textsc{iii} doublets, we also obtain and plot the radial optical depths ($\tau_{\rm rad}$) of the stellar wind which give rise to the best-fit line profile. These optical depths are recorded in 21 ``bins'' at normalised velocities ($w = \varv(r)/\varv_\infty$) ranging from zero at the stellar photosphere up to and including the terminal wind velocity ($\varv_\infty$, i.e. $w=1$). Toward the higher velocity region of the stellar wind, these bins are narrower in width, and more concentrated in number.

Examining the measured ratios between the optical depths of the components of a doublet provides a strong indicator of clumping in the stellar wind. If a source is obscured by a uniform distribution of gas which absorbs only in the lines of the doublet being examined, the ratio of the optical depths inferred from the observed absorption lines is simply the ratio of the oscillator strengths (\textit{f}) of the doublet elements. Fundamental atomic theory provides a fixed ratio between the relative strengths of each of these transitions: for this Si \textsc{iv} doublet, 2.013:1 \citep{Morton1991} \citep[or 2.015:1 from][]{Mas2019} and, for this Al \textsc{iii} doublet, 2.010:1 \citep{Morton1991} \cite[or 2.012 from][]{Mas2019}.

If however the source is partially obscured due to the presence of highly optically thick clumps or other features, the ratio of the optical depths will depend only upon the fraction of the source which remains uncovered. Thus, we can expect the \textit{apparent} ratio of the radial optical depths required to produce a fit to the observed spectrum to diverge from the canonical figure for the \textit{f}-ratio and to approach unity as those opaque clumps cover an increasing proportion of the stellar disk (\citealt{Ganguly1999} (in the context of active galactic nuclei); \citealt{Massa2008}; \citealt{Prinja2010}). The numerical relationship between covering factor and observed optical depth ratios is described in detail in \citet{Massa2008}.

\subsection{Coupled and decoupled doublet elements} \label{coupled}

A development of the SEI technique, using a weighted least-squares approach to fit model line profiles separately to individual elements of the Si \textsc{iv} 1393.76, 1402.77\AA{} resonance line doublet was introduced by \citet{Prinja2010}. Provided the terminal velocity ($\varv_\infty$) of the stellar wind being considered does not exceed approximately half of the separation of the doublet's elements, those authors demonstrated that it is possible to treat each element as being radiatively decoupled and, therefore, to derive separate model line fits for each. In the case of Si \textsc{iv}, this means we can reliably produce separate fits for stars with $\varv_\infty \lesssim$ 970 km s$^{-1}$. This technique is employed here, where possible, to derive radial optical depths for each element of the Si \textsc{iv} doublet. In a small number of cases where that doublet feature is saturated, we have been able to make use of the unsaturated Al \textsc{iii} 1854.72, 1862.79\AA{} resonance line doublet, which can provide similar diagnostic capabilities for stars with $\varv_\infty \lesssim$ 650 km s$^{-1}$.

In order to increase the size of the data set, in cases where the terminal wind velocity significantly exceeds the figure permitting reliable separate fitting of the elements of the doublet, a single fit is performed across the whole doublet. In these cases however, the ratio of the oscillator strengths is allowed to be a floating parameter in the fitting process in order to mimic the separate fitting process described above, adopting the approach of \citet{Prinja2010}. Targets where this technique is adopted are identified in the tables of stellar data and in subsequent plots. Although this single-fitting method does not permit individual fitting of each element of the doublet, it nevertheless provides evidence of wind structure surrounding additional stars once we have established the appropriateness of the separate fitting process for establishing the presence of such structure, as described in this work.

\section{Results} \label{results}
We set out below a discussion of the principal conclusions derived from the foregoing analysis and, in some cases, limitations on the usable data. Individual stars are ordered, as in Table \ref{table:Alltargets}, by host galaxy (LMC, SMC and others) and by spectral type, earliest first, within each group.

\begin{figure*}
\begin{center}
 \includegraphics[scale=0.45]{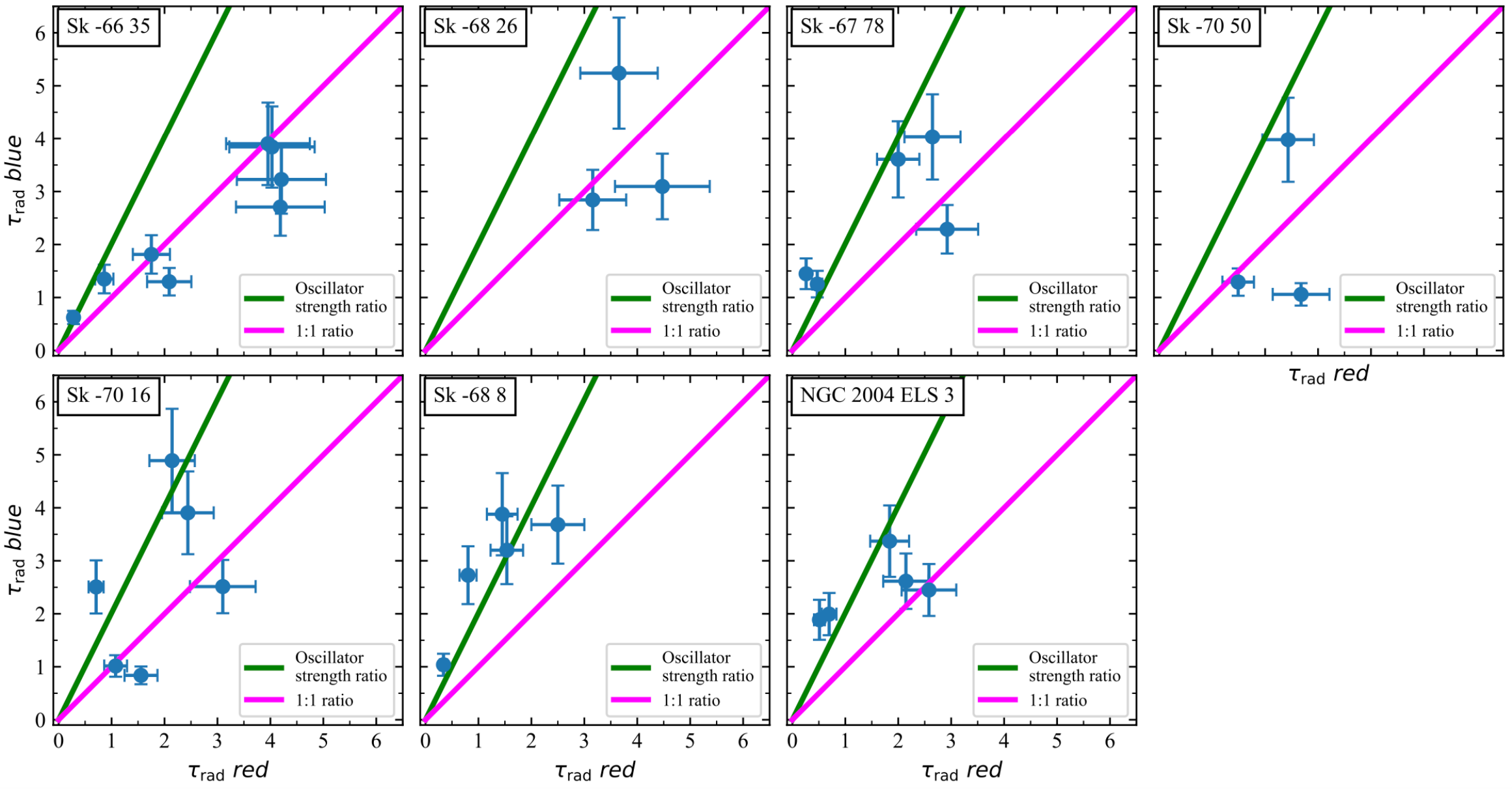}
 \caption{Plots of Blue and red radial optical depths for LMC stars in the sample for which separate SEI fits for doublet elements are possible. In this and the equivalent plots for SMC (Fig. \ref{fig:taurads_smc}) and Low Z (Fig. \ref{fig:taurads_lowz}) stars, the optical depth ratios at different radial velocities derived from SEI model fitting are shown. The diagonal lines show the extremes of optical depth ratios to be expected from minimal optically-thick coverage of the stellar disk (green) to extensive coverage (magenta).}
 \label{fig:taurads_lmc}
\end{center}
\end{figure*}

Figure \ref{fig:taurads_lmc} contains plots of the fitted model optical depths of the red element of the relevant doublet against the fitted model optical depth of the blue element of that doublet for each normalised radial velocity ``bin'' for those LMC target stars where separate fitting of each element has been possible. The optical depths plotted are the optical depths of the stellar wind at varying radial velocities which best reproduce the observed absorption profile in the blue or red, as the case may be, element of the observed resonance-line doublet feature in the UV stellar spectrum. The aim here is to demonstrate, in a more quantitative fashion, that the wind structure which may be qualitatively observed by comparing the similarities between the shapes of the blue and red element SEI-derived plots of optical depths for a given star's stellar wind is indeed present and not an artifact of the SEI-fitting process. ``Structure'' is taken here to be evident from the different optical depths able to be inferred at different radial velocities within the stellar wind (and therefore distances from the stellar photosphere). The individual plots are ordered by spectral type, from earliest to latest.

\begin{figure*}
\begin{center}
 \includegraphics[scale=0.45]{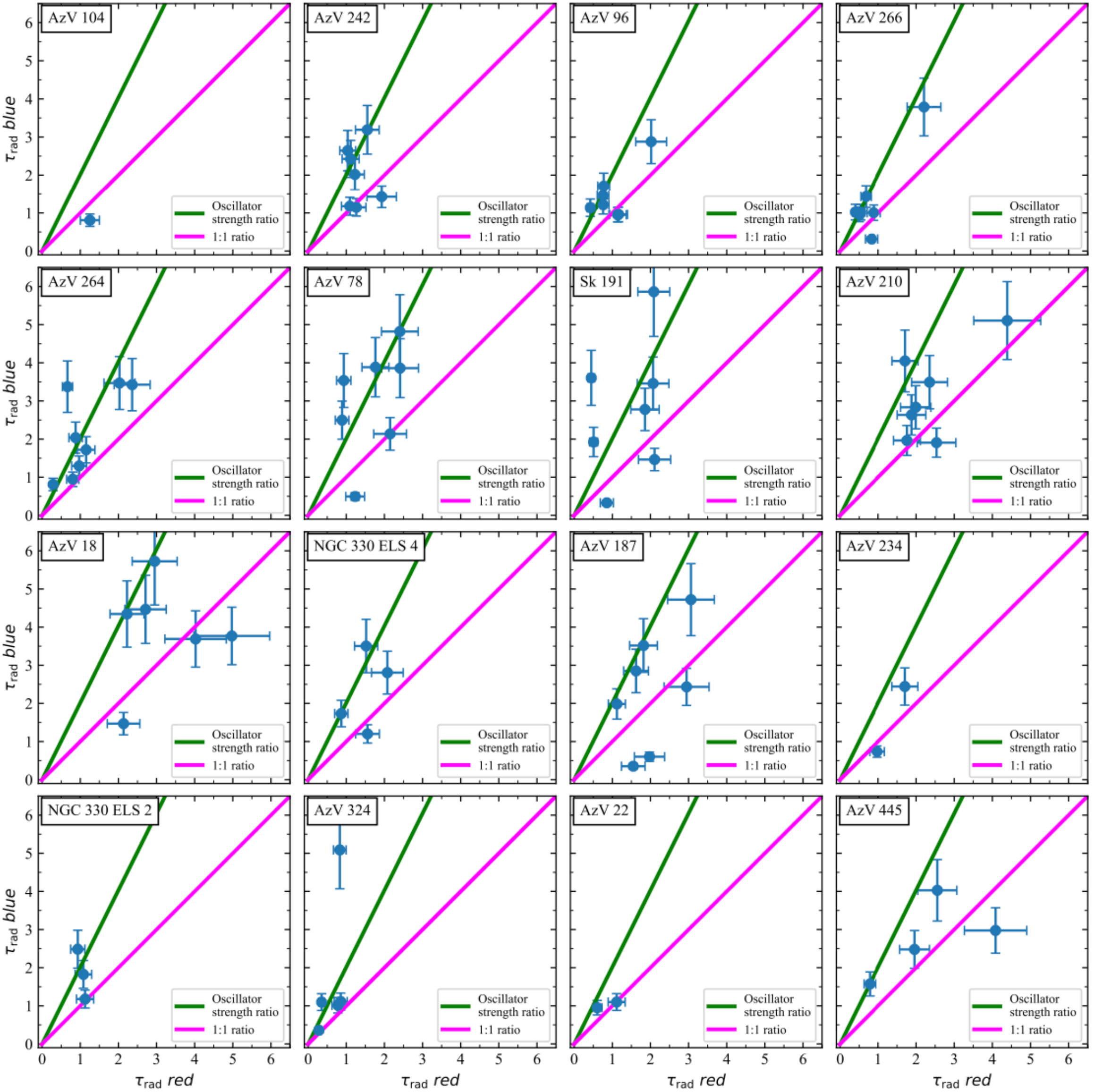}
 \caption{Plots of Blue and red radial optical depths for SMC stars in the sample for which separate SEI fits for doublet elements are possible. The optical depth ratios at different radial velocities derived from SEI model fitting are shown. The diagonal lines show the extremes of optical depth ratios to be expected from minimal optically-thick coverage of the stellar disk (green) to extensive coverage (magenta).}
 \label{fig:taurads_smc}
\end{center}
\end{figure*}

\begin{figure*}
\begin{center}
 \includegraphics[scale=0.45]{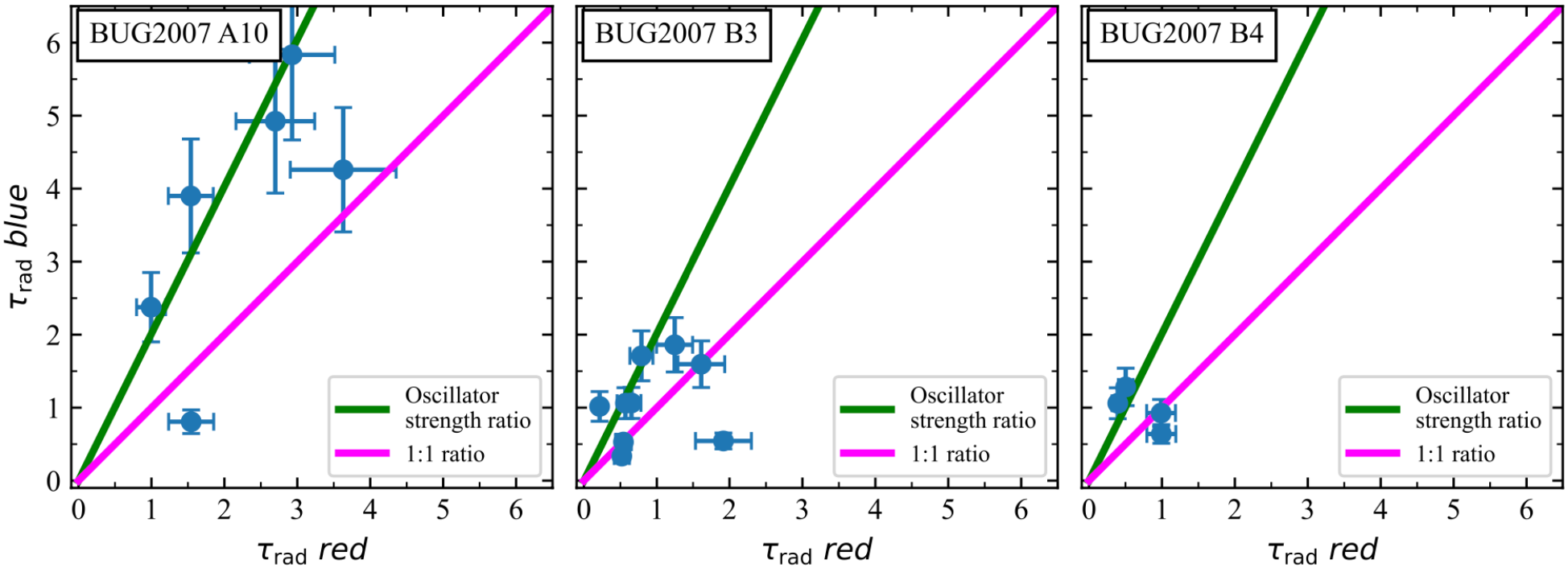}
 \caption{Plots of Blue and red radial optical depths for Low Z stars in the sample for which separate SEI fits for doublet elements are possible. The optical depth ratios at different radial velocities derived from SEI model fitting are shown. The diagonal lines show the extremes of optical depth ratios to be expected from minimal optically-thick coverage of the stellar disk (green) to extensive coverage (magenta).}
 \label{fig:taurads_lowz}
\end{center}
\end{figure*}

Figure \ref{fig:taurads_smc} shows the same plots for all of the SMC stars examined for which separate fitting of doublet elements has been possible, once again ordered by spectral type, from earliest to latest. Figure \ref{fig:taurads_lowz} sets out the same comparison for the ``low Z'' sample.

In each of the plots of optical depths of red against blue doublet elements, there are varying numbers of data points between stars. This results from the necessary exclusion of data points with very small or very large optical depths. Consistently with the approach taken by \citet{Prinja2010}, optical depths in the range 0.3(blue)/0.2(red) - 6.0(blue)/5.0(red) are considered, in order to avoid poor quality and overly-saturated situations. The plotted uncertainties for individual optical depth data points have been broadly estimated by manually adjusted optical depths in selected radial velocity bins for some stars to identify the extent of variation required to produce  a clearly discernible change in the model fitted line profile. This generally varies between 10 and 20 per cent of the radial optical depth value. In order to avoid using unjustified levels of precision in the plotted results, the larger uncertainty figure has been adopted throughout.

\subsection{Numerical results}
We set out in Table \ref{table:Allratios} a summary of key stellar parameters, repeated from earlier tables for ease of reference, together with our results for the mean radial optical depth ratios derived from the stellar data and procedures described above and the mean optical depths of the blue element of the applicable resonance line doublet used for each star. In each case, the mean ratio has been calculated using the optical depth ratios for normalised radial velocities between 0.35 and 0.75, both inclusive, of the respective terminal wind velocity for each target star. From these results, we consider whether there is a correlation of stellar effective temperature, luminosity and other stellar parameters with the prevalence of structure in the stellar wind, as signified by the mean optical depth ratio. The uncertainties for mean optical depth ratios are obtained in one of two ways. For the stars where a single fit of the entire doublet has been carried out, the (asymmetric) uncertainty is taken directly from the uncertainty in the oscillator strength ratio obtained by allowing this parameter to float in the fitting process, as described in Subsection \ref{coupled} above. For stars where each doublet element has been separately fitted, the uncertainty is the standard error of the mean optical depth ratio for the star.

In most cases, utilising only this central ``continuum'' range of wind velocities excludes those parts of the wind absorption profile for which either the SEI method breaks down (lowest velocities) or the profile is significantly affected by ISM absorption lines. It should be noted that in some instances, particularly of LMC stars with low terminal wind velocities, there is nevertheless some contamination of this continuum region by Milky Way interstellar lines when the stellar spectrum is shifted into the rest frame of the star in question.

\begin{table*}
\centering
\caption{\label{table:Allratios}Tabulation of LMC (first block), SMC (second block) and Low Z (third block) target star parameters: spectral type, effective temperature and luminosity; with results for mean optical depth ratios and mean blue doublet component optical depths  (for the range $w$ = 0.35-0.75) derived from the SEI fitting techniques described in this work for the relevant line profiles. ``O.D.R'' refers to the ratio of the optical depths of blue and red components of the relevant line profile, being that for Si \textsc{iv}, unless otherwise noted.}
\begin{tabular}{llrcrrrr} \hline \hline \\
Star  & Spectral type & $T_{\rm eff}$ & log $L$ & Mean O.D.R. & Unc. & Mean Opt. Depth & \\
 & & (K) & $(L/L_\odot)$ & & & (Blue) \\ [0.5ex]
\hline
Sk -69 43 & B0.5 Ia & 22,845 & 5.50 & 1.61 & {\raisebox{0.5ex}{\tiny$^{+\,2.13}_{-0.58}$}} & (Al \textsc{iii}) 1.59 \\
Sk -68 140 & B0.7 Ib-Iab Nwk & 25,100 & 5.64 & 1.38 & {\raisebox{0.5ex}{\tiny$^{+\,0.22}_{-0.17}$}} & 2.79 \\
Sk -66 35 & BC1 Ia & 22,000 & 5.73 & 1.10 & $\pm$0.18 & 2.34 \\
Sk -68 129 & B1 I & 22,000 & 5.25 & 1.58 & {\raisebox{0.5ex}{\tiny$^{+\,0.66}_{-0.36}$}} & 2.08 \\
Sk -67 14 & B1.5 Ia & 22,890 & 5.74 & 1.65 & {\raisebox{0.5ex}{\tiny$^{+\,1.20}_{-0.49}$}} & (Al \textsc{iii})1.74 \\
Sk -68 26 & BC2 Ia & 18,160 & 5.71 & 1.01 & $\pm$0.18 & 3.73 \\
Sk -67 78 & B3 Ia & 16,150 & 5.50 & 2.47 & $\pm$0.49 & 2.53 \\
Sk -70 50 & B3 Ia & 15,000 & 5.54 & 0.96 & $\pm$0.27 & 2.11 \\
Sk -69 140 & B4 I & 15,000 & 4.73 & 1.56 & {\raisebox{0.5ex}{\tiny$^{+\,0.37}_{-0.25}$}} & 2.63 \\
Sk -70 16 & B4 I & 15,000 & 4.62 & 1.62 & $\pm$0.42 & 2.61 \\
Sk -68 8 & B5 Ia+ & 14,200 & 5.54 & 2.54 & $\pm$0.31 & 2.91 \\
NGC 2004 ELS 3 & B5 Ia & 14,450 & 5.1\phantom{0} & 2.11 & $\pm$0.43 & 2.46 \\[0.5ex]
\hline
AzV 215 & B0 (Ib)/BN0 Ia & 27,000 & 5.63 & 1.75 & {\raisebox{0.5ex}{\tiny$^{+\,0.37}_{-0.26}$}} & 3.99 \\
AzV 235 & B0 Iaw & 27,500 & 5.72 & 1.65 & {\raisebox{0.5ex}{\tiny$^{+\,0.53}_{-0.32}$}} & 1.42 \\
AzV 317 & B0 Iw & 27,000 & 5.40 & 1.79 & {\raisebox{0.5ex}{\tiny$^{+\,0.67}_{-0.38}$}} & 3.86 \\
AzV 104 & B0.5 Ia & 27,500 & 5.31 & 0.64 & - & 0.81 \\
AzV 488 & B0.5 Iaw & 27,500 & 5.74 & 1.24 & {\raisebox{0.5ex}{\tiny$^{+\,0.39}_{-0.24}$}} & 3.41 \\
AzV 242 & B0.7 Iaw & 25,000 & 5.67 & 1.59 & $\pm$0.24 & 2.00 \\
AzV 264 & B1 Ia & 22,500 & 5.44 & 2.04 & $\pm$0.32 & 2.14 \\
AzV 266 & B1 I & 22,000 & 5.09 & 1.69 & $\pm$0.24 & 1.36 \\
AzV 96 & B1 I/B1.5 Ia & 20,000 & 5.39 & 1.64 & $\pm$0.24 & 1.47 \\
AzV 78 & B1.5 Ia+ & 21,500 & 5.92 & 1.98 & $\pm$0.39 & (Al \textsc{iii}) 3.03 \\
Sk 191 & B1.5 Ia & 22,500 & 5.77 & 2.14 & $\pm$0.49 & (Al \textsc{iii}) 2.77 \\
AzV 210 & B1.5 Ia & 20,500 & 5.41 & 1.37 & $\pm$0.17 & 3.14 \\
AzV 18 & B2 Ia & 19,000 & 5.44 & 1.32 & $\pm$0.22 & 3.91 \\
NGC 330 ELS 4 & B2.5 Ib & 17,000 & 4.77 & 1.60 & $\pm$0.30 & 2.31 \\
AzV 187 & B3 Ia & 14,600 & 4.91 & 1.26 & $\pm$0.22 & 2.35 \\
AzV 234 & B3 Iab & 15,700 & 4.91 & 1.09 & $\pm$0.24 & 1.59 \\
NGC 330 ELS 2 & B3 Ib & 14,590 & 4.73 & 1.80 & $\pm$0.38 & 1.83 \\
AzV 324 & B4 Iab & 14,600 & 4.89 & 2.19 & $\pm$0.51 & 1.73 \\
AzV 22 & B5 Ia & 14,500 & 5.04 & 1.28 & $\pm$0.21 & 1.02 \\
AzV 445 & B5 (Iab) & 14,500 & 5.0\phantom{0} & 1.39 & $\pm$0.23 & 2.76 \\[0.5ex]
\hline
IC 1613 BUG2007 B3 & B0 Ia & 24,500 & 5.53 & 1.59 & $\pm$0.33 & 1.08 \\
IC 1613 BUG2007 A10 & B1 Ia & 25,000 & 5.71 & 1.74 & $\pm$0.28 & 3.68 \\
NGC 3109 EBU07 & B0-1 Ia & 25,000 & 5.6\phantom{0}& 1.36 & {\raisebox{0.5ex}{\tiny$^{+\,0.96}_{-0.40}$}} & 1.39 \\
IC 1613 BUG2007 B4 & B1.5 Ia & 22,500 & 5.20 & 1.71 & $\pm$0.46 & 0.98 \\[0.5ex]
\hline
\end{tabular}
\end{table*}

\subsection{General observations}
For those target stars where separate fitting of individual doublet elements is possible, we observe that in almost all cases there is a broad visual correlation between the shapes of the radial optical depth plots for the best fit model profile for each element. This provides a strong indication that the fitting process effectively identifies physical characteristics of the stellar winds of these stars.

The following sections discuss features of interest discernible for individual stars as well as highlighting some parts of individual stellar spectra which may not be usable for the techniques presented here due to the presence of other features. The SEI-derived model fits, observed spectra and plots of radial optical depths of the model fits are set out in the Appendices, as referenced in the individual descriptions. The plots of radial optical depths for stars requiring a single SEI model fit across both doublet elements are plots for the blue element of the relevant doublet in each case. For stars where separate fitting of each doublet element is possible, separate spectra and plots for each doublet element are shown.

References to the Si \textsc{iv} doublet refer to the 1393.76, 1402.77\AA{} feature, references to the Al \textsc{iii} doublet refer to the 1854.72, 1862.79\AA{} feature and references to the C \textsc{iv} doublet refer to the 1548.20, 1550.77\AA{} feature.

\subsection{Narrow absorption components}
Narrow absorption components (NACs) at or near the blue edge of wind-formed absorption profiles in Galactic OB stellar spectra have been shown by \citet{Prinja1986} and \citet{Howarth1989} to be ubiquitous features of those spectra. To date, no comprehensive investigation of the presence or absence of these features in low-metallicity OB stellar spectra has been undertaken. The spectra presented in the Appendices to this work do however provide evidence \citep[as has been noted by][]{Hawcroft2023} that NACs are indeed present, and able to be detected with current instruments and techniques, in the spectra of many early B supergiant stars at LMC, SMC and lower metallicities.

The spectra presented here indicate that a wider-ranging and systematic investigation of the prevalence of NACs at lower metallicities is, therefore, a key next step in the investigation of the formation mechanisms of NACs and of stellar wind characteristics close to the terminal velocity of the wind. As has already been illustrated by \citet{Prinja2013}, the stellar wind within NACs becomes ``smooth'' such that porosity effects are minimised. As a consequence, measurements of mass-loss rates by reference to optical depths and optical depth ratios in NACs are likely to be key methods to constrain to accurate quantification of mass-loss from hot, massive stars. The persistence of these features at lower metallicity points the way to these methods also being applicable to stars in such environments. The discussions of individual stellar spectra in Subsections \ref{specLMC}, \ref{specSMC} and \ref{speclowZ} highlight those spectra which demonstrate the presence of NACs. Likely and possible NAC features are identified in each relevant plot. In each case, the NAC location marked in each portion of a doublet feature in a particular spectrum is at the same distance (velocity) from the rest wavelength of the relevant doublet element in that spectrum.

\subsection{Temporal variation in a low-metallicity B supergiant}
We highlight in the following one star in our sample (SMC star AzV 96, spectral type B1 I/B1.5 Ia) which demonstrates stellar wind variability observable within individual spectra which make up the ULLYSES high level science product spectrum. SEI model fits to the ULLYSES spectrum for AzV 96 produced a considerable lack of consistency between the optical depth profiles for each of the elements of the Si \textsc{iv} doublet. The ULLYSES spectrum is composed of an average of individual observations, one of which was obtained several days following the two other, contiguous, observations. Inspection of the individual spectra discloses variation in the shape of the observed P Cygni profile of both elements of the Si \textsc{iv} resonance line doublet over that time scale. More particularly, a broad discrete absorption component (DAC) feature can be seen to migrate bluewards in the absorption trough of both elements of the doublet. This is consistent with other observations of Galactic stars \citep{Kaper1996, Massa2015} and of SMC O type stars \citep{Rickard2022} and represents the first observed evidence for outward migration of a DAC in the atmosphere of a low-metallicity B supergiant. The spectra may be compared by considering Figs. \ref{fig96} and \ref{fig96a} in Appendix \ref{AppB}.

\subsection{Notes on individual stellar spectra: LMC} \label{specLMC}

\subsubsection{Sk -69 43}
(Fig. \ref{fig6943}).
The Al \textsc{iii} profile is used due to saturation of the Si \textsc{iv} profile. As expected for a B0.5 star, the Al \textsc{iii} profile is relatively weak. There is some evidence for wind structure from the derived optical depth ratio, however the uncertainties on this result are large, as a consequence of the very low optical depth values, particularly for the red element of the doublet.
\subsubsection{Sk -68 140}
(Fig. \ref{fig68140}).
The Si \textsc{iv} doublet profile displays very prominent deep absorption features close to the terminal wind velocity in each element, as well as strong emission peaks. The absorption features are wider than a NAC, but may show a broader DAC which has migrated close to the terminal wind velocity. A similar morphology is seen in the Si \textsc{iv} profile of AzV 215 and was noted by \citet{Evans2004a} in relation to that star, as discussed in sub-section \ref{specav215} below. Time-series analysis would be helpful in determining whether this is the case. The single profile fitting technique used due to the high terminal wind velocity does suggest the presence of significant structure in the stellar wind. The SEI fitting method does however have difficulty reproducing the very strong emission profile observed in the spectrum.
\subsubsection{Sk -66 35}
(Fig. \ref{fig6635}).
The relatively low terminal wind velocity permits separate fitting of each element of the Si \textsc{iv} doublet. There is clear evidence for extensive structure throughout almost the entire velocity range of the stellar wind and the mean optical depth ratio producing the best fit model profiles approaches unity (1.10). The absorption profiles are very deep at mid-range normalised wind velocities but rapidly become optically thin at higher velocities. The fitting technique applied in this paper had difficulty in reproducing this very deep central absorption feature, even when utilising unusually high values for the acceleration parameter ($\beta$ = 2.56). Again, time-series analysis of this star would be helpful in determining whether the deep absorption feature is seen to migrate blueward in a similar fashion to DACs in the winds of other stars.
\subsubsection{Sk -68 129}
(Fig. \ref{fig68129}).
A prominent NAC feature is visible in each element of the Si \textsc{iv} profile, giving rise to a characteristic increased optical depth around a wind velocity of $w$ = 0.9, with significant reduction in optical depth at slightly lower velocities.
\subsubsection{Sk -67 14}
(Fig. \ref{fig6714}).
The Al \textsc{iii} profile shows deep absorption in the higher velocity portion of the stellar wind, although this is partly obscured in the red element of the doublet by overlap with the blue element emission peak.
\subsubsection{Sk -68 26}
(Fig. \ref{fig6826}).
The results for wind velocities greater than $w$ = 0.6 are affected by the blue element of the Si \textsc{iv} profile approaching saturation. The depth of the profile suggests a particularly optically-thick wind for a star of this spectral type (BC2 Ia) as the terminal velocity is approached. It is to be noted that the acceleration parameter providing the best fit is relatively high for a supergiant ($\beta$ = 1.81), suggesting a low rate of acceleration which potentially contributes to a concentration of material toward the terminal velocity. The mean optical depth ratio of 1.01 suggests a high degree of structure present in the wind or extensive coverage of the stellar disk by optically-thick material, or both in combination. This result may be exaggerated by the near-saturation of the blue element of the profile, however there remains clear evidence for the presence of structure at lower velocities where the profile is not approaching saturation. Although not presented here, the Al \textsc{iii} profile is unhelpful for this purpose as it is highly saturated, as may be expected for a star of spectral type BC2.
\subsubsection{Sk -67 78}
(Fig. \ref{fig6778}).
The profile indicates significant absorption at higher radial wind velocities, more evidently in the blue Si \textsc{iv} doublet element. The region around $w$ = 0.7 is affected by the blue-shifted Milky Way ISM absorption feature, however optical depth ratios mostly do not show a significant departure from a smooth wind.
\subsubsection{Sk -70 50}
(Fig. \ref{fig7050}).
Optical depth ratios indicate the presence of significant structure in the wind of this star. The optical depth of the wind increases substantially at velocities greater than $w$ = 0.4 (this deep absorption trough is a feature of the wind itself, within which the Milky Way ISM absorption line can just be discerned.
\subsubsection{Sk -69 140}
(Fig. \ref{fig69140}).
This star exhibits an unusually high terminal wind velocity for its spectral type (B4). A shallow, but clear, NAC feature is visible in both elements of the Si \textsc{iv} profile. There is evidence from the inferred oscillator strength ratios that there is structure within the wind.
\subsubsection{Sk -70 16}
(Fig. \ref{fig7016}).
An unusual wind profile with deep absorption at lower wind velocities and the wind becoming very optically thin at velocities above approximately $w$ = 0.6. There appears however to be a highly prominent absorption feature approaching the terminal wind velocity which is wider than a typical NAC. This may be evidence of a DAC feature which has migrated close to the terminal wind velocity. The plot of individual optical depth ratios shows considerable scatter.
\subsubsection{Sk -68 8}
(Fig. \ref{fig688}).
Optical depth ratios suggest that this B5 hypergiant exhibits a generally smooth wind with a low terminal wind velocity. The wind rapidly becomes optically thin however, meaning that meaningful data are not available for higher normalised velocities.
\subsubsection{NGC 2004 ELS 3}
(Fig. \ref{fig20043}).
The mean optical depth ratio suggests a smooth wind overall. The plot of individual optical depth ratios indicate however that there may be variability in the presence of structure at different velocities. There is some evidence of a shallow NAC feature.
\subsection{Notes on individual stellar spectra: SMC} \label{specSMC}
\subsubsection{AzV 215} \label{specav215}
(Fig. \ref{fig215}).
This star's spectrum exhibits the highest terminal wind velocity in the present sample. Very deep absorption at normalised velocities above $w$ = 0.6 and a prominent absorption feature, perhaps constituting a DAC approaching the terminal wind velocity and in the process of becoming a NAC, is observed in each element of the Si \textsc{iv} feature. This appears however to be a persistent, or recurring, feature since it is observed also in the STIS spectrum illustrated in \citet{Evans2004a} and \citet{Evans2004c}, who also note the unusual ``shelf-like'' appearance that this gives to the absorption profile. The wind appears to be relatively smooth, with a mean inferred optical depth ratio of 1.75 indicated by the model fitting process. A very prominent ISM feature is visible at the lower velocity end of the Si \textsc{iv} absorption profile.
\subsubsection{AzV 235}
(Fig. \ref{fig235}).
A relatively optically thin wind with only a short section of the Si \textsc{iv} absorption profile dropping below 0.5 of the normalised continuum flux. There is however prominent emission in that profile. A NAC feature is also apparent, the red element of which appears as an absorption feature within the slope of the blue element emission feature. There is some evidence of wind structure from the inferred mean optical depth ratio of 1.65.
\subsubsection{AzV 317}
(Fig. \ref{fig317}).
The shape of the absorption trough in the Si \textsc{iv} profile for this star is unusual. It shows a continuously increasing depth at higher radial wind velocities, but commences from a very shallow absorption at low velocities. The increase in absorption is in an almost linear relation with radial wind velocity. The red element of this doublet shows what appears to be a prominent NAC feature, which however cannot be confirmed due to saturation of the blue element profile very close to the terminal wind velocity. The inferred mean optical depth ratio from a combined SEI fit to the whole Si \textit{iv} feature suggests a relatively smooth, and moderately optically-thick, wind.
\subsubsection{AzV 104}
(Fig. \ref{fig104}).
The wind profile has an unusual appearance and is not very well-developed. Using published results for the terminal wind velocity suggests that the wind becomes extremely optically thin at higher velocities. the unusual nature of the spectrum of this star was first discussed by \citet{Evans2004a} who noted that in some respects, the star appears more akin to spectral type B0.5 Ib. It is noted that recent literature suggests widely-varying measures for terminal velocity depending upon how it is measured \citep{Hawcroft2023}. The rapid onset of a very optically thin regime greatly limits the data that may be obtained using the techniques applied here. This star is therefore excluded from the plots discussed in Section \ref{discussion}. The spectrum is however presented here as an unusual example of an early B supergiant with an uncharacteristically optically-thin wind. It is possible that the characteristics observed in this spectrum arise from the line of sight for this observation only containing an unstructured part of the stellar wind \citep[see the discussion in][]{Rickard2022}.
\subsubsection{AzV 488}
(Fig. \ref{fig488}).
The Si \textsc{iv} profile for this star shows deep, although not saturated, absorption as well as very strong emission peaks. A shallow NAC feature is apparent. The fitting process produces a large acceleration parameter ($\beta$ = 1.99) in order to fit the emission peaks, likely indicating a slowly accelerating, optically-thick wind. A considerable degree of structure appears to be present.
\subsubsection{AzV 242}
(Fig. \ref{fig242}).
There appears to be evidence for structure in a wide range of wind velocities, although there is considerable scatter in the plot of individual optical depth ratios. A significant NAC feature is present and the Milky Way and SMC ISM absorption lines are very prominent.
\subsubsection{AzV 264}
(Fig. \ref{fig264a}).
Optical depth ratios suggest the presence of wind structure throughout the velocity range. Only one ISM absorption feature (for the Milky Way) is prominent.
\subsubsection{AzV 266}
(Fig. \ref{fig266}).
Optical depth ratios suggest a mostly uniform or smooth wind, with some evidence for structure present in the higher velocity region. There is a very deep and significant NAC feature visible in both Si \textsc{iv} doublet elements.
\subsubsection{AzV 96}
(Figs. \ref{fig96} and \ref{fig96a}).
Initial attempts to produce model fits for the Si \textsc{iv} profiles for this star using the ULLYSES DR5 data gave very different optical depth profiles for each of the blue and red elements of the doublet. It is to be noted that the ULLYSES spectrum comprises an average of 3 separate observations, two of which are contiguous and the third of which was obtained at a remove of several days. Examination of the individual 1D spectra available in the MAST archive shows that there are significant variations in the shape of the absorption trough of the Si \textsc{iv} doublet elements between the first two spectra on the one hand and the third spectrum on the other, indicating blueward migration of a DAC feature. This in itself provides strong evidence of temporal variation in the conditions within the stellar wind of this star, consistent with the earlier results demonstrated by \citet{Prinja2002}. This variation in the central part of the absorption profile appears to produce contradictory results in the fitting process when the different spectra are averaged.

By model fitting against an average of only the two contiguous spectra, and separate fitting for the third spectrum, the results illustrated here were obtained. In addition to the foregoing issue, the effective temperature of this star contained in earlier literature appears to be too high. The lower temperature used here, which accords with results obtained (Bestenlehner, priv. comm.) using the procedure introduced by \citet{Bestenlehner2023}, appears to be necessary to produce a reasonable fit with a feasible $\beta$ value. There is a prominent NAC feature visible in both elements the Si \textsc{iv} profile, particularly in the later spectrum, the location of which also suggests a slightly lower terminal wind velocity than previously published results.
\subsubsection{AzV 78}
(Fig. \ref{fig78}).
The Al \textsc{iii} profile suggests a mostly uniform wind, although this feature becomes optically thin at higher velocities. The Si \textsc{iv} absorption for this hypergiant star is saturated.
\subsubsection{Sk 191}
(Fig. \ref{fig191a}).
The Al \textsc{iii} profile suggests a mostly uniform or smooth wind, except possibly in the lower velocity range. The Si \textsc{iv} absorption feature for this star is saturated so cannot be used for measuring fitted optical depth ratios. However, both the Si \textsc{iv} and C \textsc{iv} (also saturated) profiles indicate $\varv_\infty = $ 460 km s$^{-1}$, using the $\varv_{\rm{black}}$ measurement technique described in \citet{Prinja1990} and \citet{Hawcroft2023}. There is a possible NAC feature present close to the terminal wind velocity in both elements of the Al \textsc{iii} profile.
\subsubsection{AzV 210}
(Fig. \ref{fig210}).
The modelled optical depth ratios suggest a highly-structured wind across almost all of the wind velocity range. A shallow NAC is discernible in the Si \textsc{iv} profile.
\subsubsection{AzV 18}
(Fig. \ref{fig18}).
Optical depth ratios suggest a highly-structured wind throughout the velocity range. The mean optical depth ratio may however be affected by the blue element of the Si \textsc{iv} line profile beginning to saturate at higher wind velocities.
\subsubsection{NGC 330 ELS 4}
(Fig. \ref{fig3304}).
A generally smooth wind is indicated by the inferred optical depth ratios across most of the wind velocity range. A prominent ISM line affects the model fit at mid-lower velocities.
\subsubsection{AzV 187}
(Fig. \ref{fig187}).
Evidence for significant wind structure is apparent in mid-range velocities with limited usable data for this star. A prominent NAC feature is visible in the Si \textsc{iv} profile.
\subsubsection{AzV 234}
(Fig. \ref{fig234}).
A good fit for the observed spectrum is very difficult to obtain, due principally to the very low terminal wind velocity meaning that a large portion of the absorption troughs in the Si \textsc{iv}doublet profile is significantly affected by interstellar absorption lines. The limited useful data obtainable are however consistent with the presence of optically-thick structure in the stellar wind.
\subsubsection{NGC 330 ELS 2}
(Fig. \ref{fig3302}).
Limited data suggest a mostly smooth wind with some limited evidence of structure in the lower velocity region. What may be a prominent NAC feature is visible in the blue element of the Si \textsc{iv} profile, although the apparent optically thin regime pertaining at higher velocities makes this difficult to confirm in the red element.
\subsubsection{AzV 324}
(Fig. \ref{fig324}).
The very low terminal wind velocity for this star results in considerable intrusion upon the middle part of the Si \textsc{iv} absorption trough by the blue shifted Milky Way ISM line. The limited usable data do however indicate the presence of significant structure in the stellar wind. The limited data mean that the calculated mean inferred optical depth ratio is greater than the lower velocity part of the stellar wind would suggest.
\subsubsection{AzV 22}
(Fig. \ref{fig22}).
The Si \textit{iv} doublet profile suggests a wind that becomes optically thin at velocities well below the terminal velocity. The absorption profile is also significantly affected by the presence of deep interstellar absorption lines at a significant fraction of the relatively low terminal wind velocity. The limited data are however consistent with the presence of optically-thick structure.
\subsubsection{AzV 445}
(Fig. \ref{fig445}).
As with AzV 324, the amount of usable data is limited by the presence of a very significant Milky Way ISM feature in the mid- to higher-velocity range. Optical depth ratios at lower velocities suggest the presence of optically-thick structure.
\subsection{Notes on individual stellar spectra: low Z} \label{speclowZ}
\subsubsection{NGC 3109 EBU07}
(Fig. \ref{fig31097}).
SEI fitting has enabled a reliable estimate for stellar wind velocity to be obtained. There is a significant broad (DAC) type absorption feature visible at higher velocities in the Si \textsc{iv} doublet profile. The C \textsc{iv} doublet profile shows an apparent NAC feature close to the measured terminal wind velocity and this appears also to be present in the Si \textsc{iv} shown here. The derived optical depth ratio is consistent with the presence of significant degree of optically-thick structure in the stellar wind. 
\subsubsection{IC 1613 BUG2007 B3}
(Fig. \ref{fig20073}).
The wind absorption profile is very shallow, as may be expected at lower metallicity, and there is very little emission evident in the P Cygni profile of the Si \textsc{iv} resonance line doublet. The inferred optical depth ratios do however suggest the presence of a structured wind.
\subsubsection{IC 1613 BUG2007 A10}
(Fig. \ref{fig200710}).
The stellar wind appears to be relatively small based on the inferred optical depth ratios. The Si \textsc{iv} wind-formed profiles are however strong, with the wind exhibiting relatively high optical depth in the middle of its velocity range. The wind becomes optically tin towards the terminal velocity.
\subsubsection{IC 1613 BUG2007 B4}
(Fig. \ref{fig20074}).
As noted by \citet{Garcia2014}, the wind is relatively weak, as expected at this low metallicity. The inferred optical depth ratios suggest a relatively smooth wind. There is what may be a relatively shallow NAC approaching the terminal wind velocity of both components of the Si \textsc{iv} doublet.

\begin{figure*}
\begin{center}
 \includegraphics[scale=0.79]{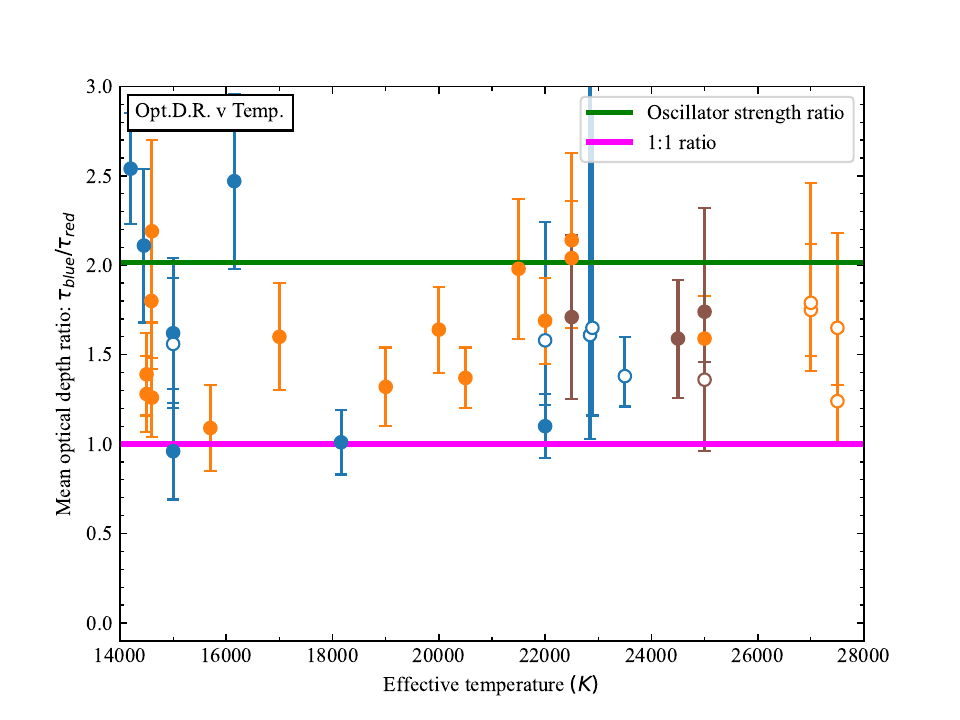}
 \caption{Plot of mean optical depth ratios against effective temperature for all target stars. LMC stars are plotted in blue, SMC stars in orange and ``low Z'' stars in brown. In each case, the filled circles show stars for which a separate fitting of each element of the Si \textsc{iv} or Al \textsc{iii} doublet has been carried out. The open circles show stars for which a single SEI model fitting has been made to the entire doublet feature, with the inferred oscillator strength ratio being allowed to float, as described in the text.}
 \label{fig:ratio_temp}
\end{center}
\end{figure*}

\begin{figure*}
\begin{center}
 \includegraphics[scale=0.79]{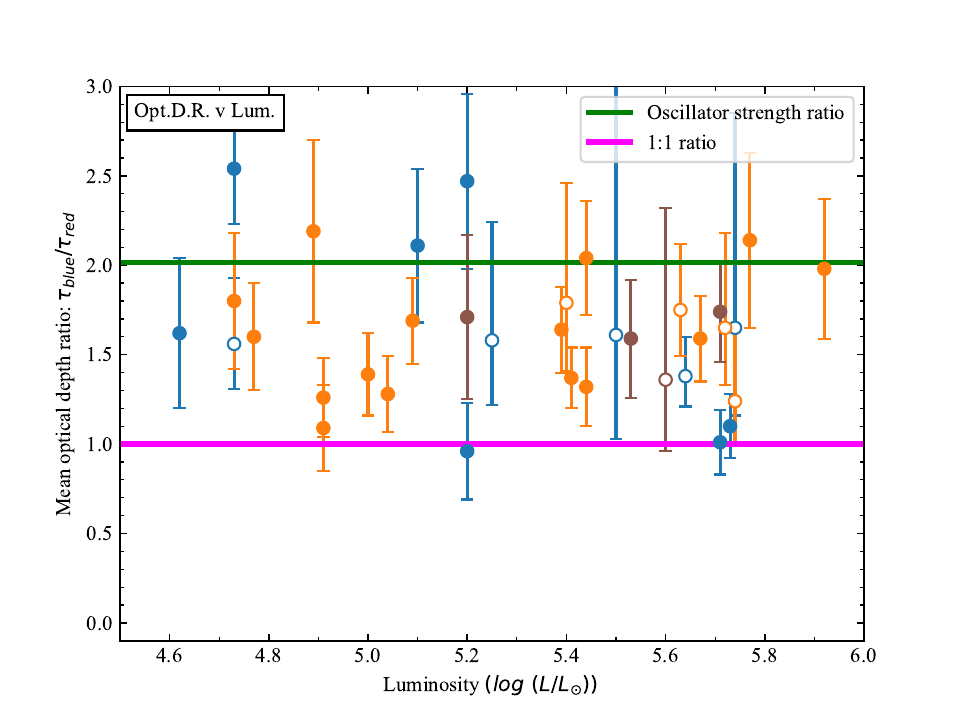}
 \caption{Plot of mean optical depth ratios against luminosity for all target stars. Plot colours and symbols are as described for Figure \ref{fig:ratio_temp} above.}
 \label{fig:ratio_lum}
\end{center}
\end{figure*}

\begin{figure*}
\begin{center}
 \includegraphics[scale=0.79]{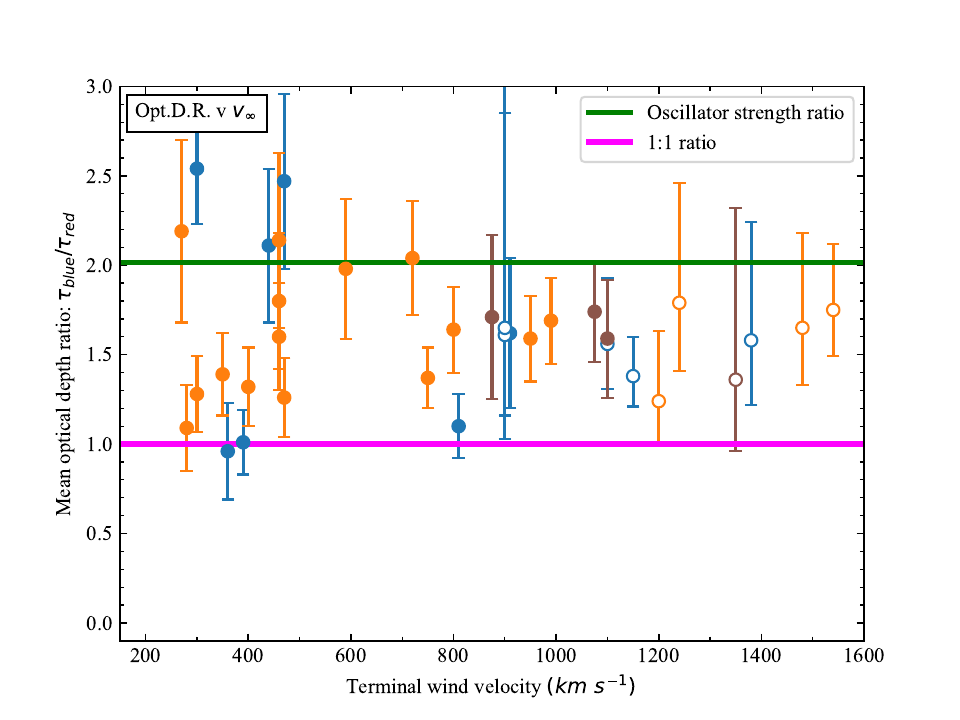}
 \caption{Plot of mean optical depth ratios against terminal wind velocity for all target stars. Plot colours and symbols are as described for Figure \ref{fig:ratio_temp} above.}
 \label{fig:ratio_vinf}
\end{center}
\end{figure*}

\begin{figure*}
\begin{center}
 \includegraphics[scale=0.79]{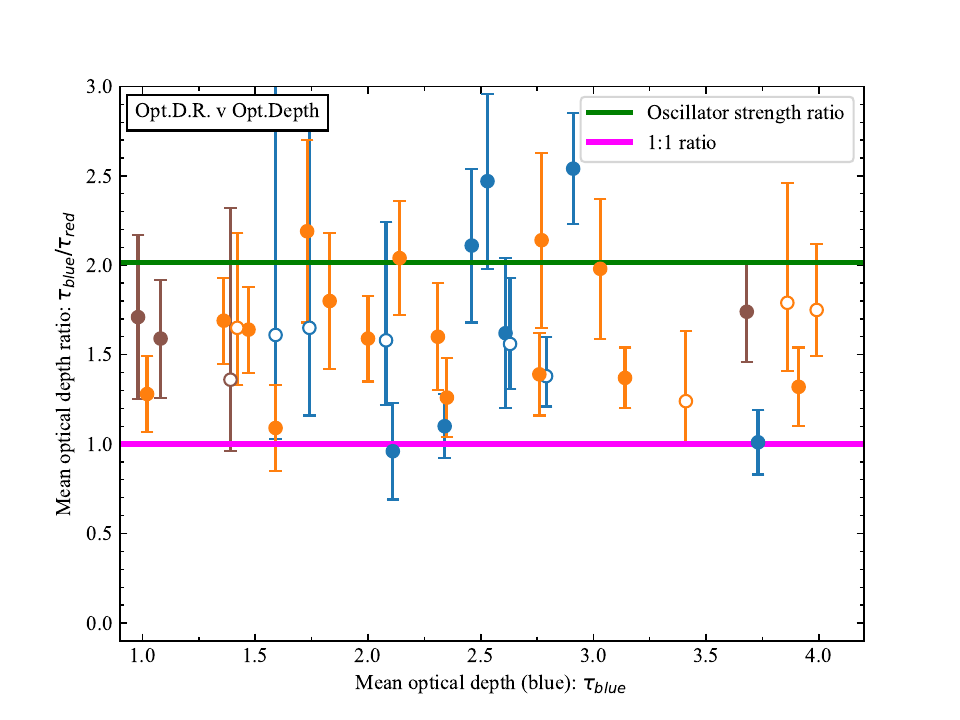}
 \caption{Plot of mean optical depth ratios against the mean optical depth of the blue doublet element for all target stars. Plot colours and symbols are as described for Figure \ref{fig:ratio_temp} above.}
 \label{fig:ratio_depth}
\end{center}
\end{figure*}

\begin{figure*}
\begin{center}
 \includegraphics[scale=0.79]{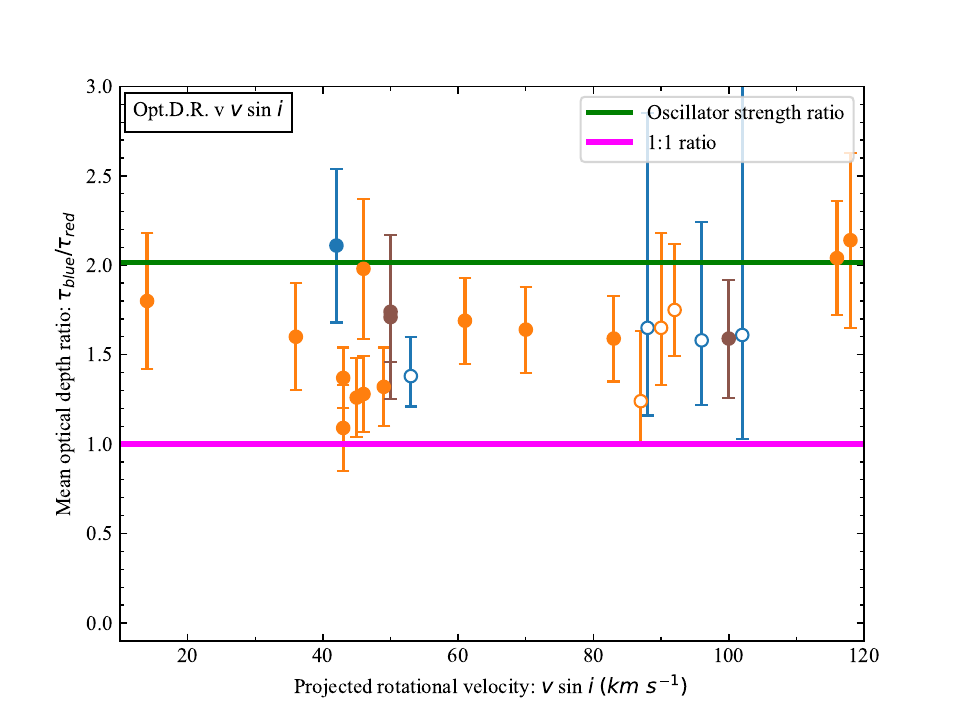}
 \caption{Plot of mean optical depth ratios against the rotational velocity of selected target stars. Plot colours and symbols are as described for Figure \ref{fig:ratio_temp} above.}
 \label{fig:ratio_vsini}
\end{center}
\end{figure*}

\section{Discussion} \label{discussion}
The data reductions undertaken as part of this work, in particular, the SEI-fitting processes carried out have produced some new results for stellar radial velocities and the terminal wind velocities of some early B supergiants in the LMC, SMC and in low-metallicity local group dwarf galaxies IC 1613 and NGC 3109. These new results are highlighted in Table \ref{table:Alltargets}. In addition, some amendments to existing literature values of terminal wind velocities and of stellar radial velocities are presented.

We provide plots of the derived mean radial optical depth ratios against a range of physical stellar parameters in order to determine whether there is evidence of any clear correlation between those ratios and any of those physical parameters.

Figure \ref{fig:ratio_temp} plots the derived mean optical depth ratios for all target stars against stellar effective temperature. Consistently with the findings for Galactic stars discussed in \citet{Prinja2010}, we do not observe evidence of correlation between the mean optical depth ratio and stellar effective temperature. We do however observe an apparent increase in the scatter of optical depth ratios at effective temperatures around 22,000 K, just above that at which a ``bi-stability jump'' in the ratio of terminal wind velocity to escape velocity is observed; from $\varv_\infty \simeq$ 2.6$\varv_{\rm{esc}}$ for supergiants of spectral type earlier than B1 to ${\varv_\infty \simeq}$ 1.3${\varv_{\rm{esc}}}$ for later-type supergiants \citep{Lamers1995, Markova2008}.

Whether this potential increased scatter in the prevalence of optically-thick wind structure affects the predictions of \citet{Vink1999, Vink2000, Vink2001} that mass loss rates increase across the bi-stability jump requires further investigation, particularly in the light of more recent studies which have suggested that such a jump in mass-loss rates may not be evident \citep[see, for example, ][]{Krticka2021, Bjorklund2023}. A similar increase in scatter of optical depth ratios can also be observed at effective temperatures around 14,500 K. It may therefore be the case that both observations are primarily the result of uneven sampling as a result of the limited data available at present.

Figures \ref{fig:ratio_lum}, \ref{fig:ratio_vinf} and \ref{fig:ratio_depth} respectively plot mean optical depth ratios for all target stars against stellar luminosity, measured terminal wind velocity and the derived mean optical depth of the blue component of the relevant doublet feature for each star. There does not appear to be clear evidence of correlation of the extent of the observed wind structure with any of these other features. We do observe that, at least at higher luminosities, lower metallicity stars generally exhibit, through the inferred mean optical depth ratios plotted, less evidence of optically-thick structure in their winds. This result is consistent with the theoretical predictions of \citet{Driessen2022a}. There is little evidence shown in our data at the lower end of the luminosity scale, however this may be a consequence of the limited data available.

With limited exceptions, we do observe in Figure \ref{fig:ratio_depth} that, for stellar winds of a given mean optical depth, the winds of stars in lower metallicity environments (i.e. SMC and ``low Z'' stars) appear to be smoother and disclose less evidence of structure. There is some evidence for a similar metallicity-based differential when we examine stars of a given luminosity (see Figure \ref{fig:ratio_lum}), i.e. that, at any given luminosity, the stars in a lower metallicity environment appear to be more likely to exhibit smoother stellar winds. If clumping in the stellar winds is indeed driven by the presence of sub-photospheric iron convection zones in the outer envelopes of hot massive stars, as proposed by \citet{Cantiello2009} and \citet{Cantiello2011}, some positive correlation between metallicity and the presence of wind structure is to be expected. The results shown here are consistent with that.

Figure \ref{fig:ratio_vsini} plots the derived mean optical depth ratios for selected target stars against rotational velocity of those stars for which such data are available in earlier literature. If observed large-scale structure in stellar winds arises from the existence, and persistence, of so-called co-rotating interaction regions \citep[``CIRs'', see][]{Owocki1995, Cranmer1996, Fullerton1997, deJong2001, Prinja2002, Prinja2007, David2017}, we may expect to see some relationship between the extent of observed wind structure and stellar rotational velocity. The recurrence and migration timescales of large-scale structure, evidenced by broad DACs observable in the troughs of UV resonance line P Cygni profiles in stellar spectra, have been demonstrated to be related to the rotation timescales of such stars \citep{Prinja2002}. CIRs are expected to form where differing conditions in regions of the stellar photosphere (potentially from features such as star spots or magnetic disturbances generally) result in the entrainment of material into the stellar wind at differing velocities and rates of acceleration \citep{Cranmer1996, Fullerton1997}. The higher velocity stream(s) will overtake those of lower velocity, producing an archimedean spiral structure in the wind as the star's rotation imparts transverse angular momentum of the stellar wind material. The ability to detect evidence of this will, of course, depend to a significant degree upon the viewing angle at which the system is observed.

It is expected that rapidly rotating stars exhibit a greater degree of detectable coherent macro-scale wind structure compared with slower rotators \citep{Cranmer1996}. We have limited the data set used in Figure \ref{fig:ratio_vsini} to those stars for which there are reasonably precise $\varv$ sin $i$ determinations in existing literature. We do not observe clear evidence of the postulated correlation from these limited data. It should be noted that the available sample size used here may be too small to permit $\varv$ sin $i$ to be used as an effective proxy for true rotational velocity \citep[c.f. the discussion of this issue in][]{Penny2009}, rather placing only a lower bound on that measure. The lack of correlation shown here in a relatively small available sample may be dominated by differing viewing angles to rotational axes of the stars observed \citep{Rickard2022}.

\section{Conclusions}
\subsection{Temporal variability}
The distinct variations observed in individual spectral observations of SMC star AzV 96 shows that, in this example at least, there is clear evidence of temporal variation in the structure and optical depth of the stellar wind, observable over a period of several days. Further analysis of individual 1D UV spectra underlying the ULLYSES data set is likely, therefore, to be of value in indicating suitable target stars for follow-up time series observations at low metallicity. AzV 96 is now established as a suitable candidate star for this purpose.

\subsection{Wind structure}
Using the SEI method applied to unsaturated P Cygni resonance line profiles in early B supergiant stellar spectra at sub-solar metallicity, we have demonstrated that single-epoch UV spectra may be used to find the presence of structure within the stellar winds of those stars. Most stars in our target sample show clear evidence of a significant degree of wind structure, consistent with earlier studies at solar metallicity. Our findings demonstrate that even at low metallicities, wind variability needs to be taken in to account when determining mass-loss rates of massive stars.

The results shown here confirm the conclusion reached by \citet{Prinja2010} for Galactic early B type supergiants that there is widespread evidence for macro-scale, optically-thick, clumping in the winds of B supergiants. We have extended this result to the lower metallicity environments of the Large and Small Magellanic Clouds and, to a limited degree, to very low metallicity environments elsewhere in the Local Group. We confirm, in the context of these other environments, the earlier conclusion of \citet{Prinja2010} that there does not appear to be evidence for a direct correlation between stellar effective temperature and the degree of such structure evident in those stellar winds. This conclusion also therefore suggests that, in these environments, there are likely to be time-variable factors at work, with transient events contributing to the scatter in the results. This conclusion further emphasises the desirability of obtaining time series UV observations particularly of selected SMC and lower metallicity stars from the ULLYSES sample.

\subsection{NACs at low metallicity}
The observable presence of NACs in wind profiles within the UV spectra of many early B supergiant stars at low metallicity, as illustrated in a significant proportion of the appended spectra, shows that a comprehensive investigation of this phenomenon at low metallicity is warranted. As shown by \citet{Prinja2013}, this is likely to play an important role in better constraining quantification of mass-loss rates from hot, massive stars at lower metallicity, leading to more accurate models of stellar and galactic evolution in the early Universe.

\subsection{Relationship with other stellar parameters}
We similarly find no evidence for correlation between the mean optical depth ratios inferred by the fitting process adopted here and each of stellar luminosity, terminal wind velocity and the mean optical depth of the wind itself. We do however find limited evidence suggesting a stratification of the prevalence of wind structure by reference to metallicity for any given stellar luminosity and for any given mean optical depth of the stellar wind, with some indication that a smoother wind is more likely to be found in lower metallicity environments based on these measures.

Finally, we have sought to test the hypothesis that faster rotating stars are likely to exhibit more coherent structures in their stellar winds when compared with slower rotators.

These results also suggest target stars, such as AzV 96, that are likely to reveal the most useful results from further detailed investigation by way of obtaining high-cadence time series UV spectra. This information may be contained within the underlying spectra forming the ULLYSES high level science products, from which further evidence of suitable candidates for time-series observation may be obtained. Such observations will be of key importance to establishing tight constraints upon variations in rates of mass-loss from stars in low-metallicity environments exhibiting significant structure in their stellar winds, thereby enabling more meaningful determination of the effects of metallicity upon variation in such mass-loss rates.

In conclusion, there is some evidence from the data presented here of the following relationships:
\begin{enumerate}
    \item Evidence of wind structure is generally clearer in cases where the mean optical depth of the blue doublet element is greater - this is perhaps to be expected as variations in optical depths are more clearly identifiable at greater average optical depths provided the medium is not reaching an optical depth that results in saturation;
    \item With some exceptions, evidence of the existence of macro-scale structure appears to be clearer, and to suggest a greater degree of clumping, at higher metallicities for a given effective temperature. At lower effective temperatures however the scatter in the LMC data is larger than at higher effective temperatures, so this conclusion is made only tentatively.
\end{enumerate}
We have, in addition, demonstrated evidence for the following:
\begin{enumerate}
    \item Temporal variation observable of a period of several days within the stellar winds of stars at low- (i.e. SMC-) metallicity;
    \item That separate fitting of each element of the P Cygni profiles of resonance-line doublet features is an appropriate and reliable tool for the discernment of structure within stellar winds of massive stars and, in particular, that treating these elements as radiatively decoupled in circumstances where the terminal velocity of the stellar wind is sufficiently low is a suitable technique for this purpose;
    \item The presence of NAC features, important as a diagnostic of mass-loss rates, in many early B supergiant stellar winds at low metallicity.
\end{enumerate}

\section*{Acknowledgements}

Based on observations obtained with the NASA/ESA Hubble Space Telescope, retrieved from the Mikulski Archive for Space Telescopes (MAST) at the Space Telescope Science Institute (STScI). STScI is operated by the Association of Universities for Research in Astronomy, Inc. under NASA contract NAS 5-26555.

The \texttt{STARLINK} software \citep{Currie2014} is currently supported by the East Asian Observatory.

AACS and MBP acknowledge support by the Deutsche Forschungsgemeinschaft (DFG, German Research Foundation) in the form of an Emmy Noether Research Group -- Project-ID 445674056 (SA4064/1-1, PI Sander). 
AACS further acknowledges support from the Federal Ministry of Education and Research (BMBF) and the Baden-W\"{u}rttemberg Ministry of Science as part of the Excellence Strategy of the German Federal and State Governments.

DP acknowledges financial support by the Deutsches Zentrum f\"ur Luft und Raumfahrt (DLR) grant FKZ 50OR2005.

The authors also wish to thank the anonymous referee for the review of, and the very helpful comments provided on, the draft of this paper.

\section*{Data Availability}

The spectral data used in this work may be obtained from the Mikulski Archive for Space Telescopes (MAST) at the Space Telescope Science Institute (STScI), accessible at: \texttt{https://mast.stsci.edu/portal/Mashup/Clients/Mast/\hspace{0pt}Portal.html}
in particular, the ULLYSES data contained therein.

\texttt{DIPSO} spectral analysis software is available as part of the \texttt{STARLINK} package, accessible at: \texttt{https://starlink.eao.hawaii.edu/starlink}.

\texttt{TLUSTY} model grids and associated materials are accessible at: \texttt{http://tlusty.oca.eu/}, with thanks to Ivan Hubeny and Thierry Lanz.

The SEI-fitting code and results therefrom will be made available upon reasonable request to the corresponding author of this work.


\bibliographystyle{mnras}
\bibliography{bibliography}


\clearpage

\appendix

\section{SEI model line fits for LMC B supergiants}
As described in the introduction to Section \ref{SEImethod}, the lower panels of each plot in this and the following Appendices respectively show the radial optical depths of the stellar wind at differing normalised velocities which produce the best-fit model line (lower left panels) and the photospheric contribution to the spectrum as derived from the model photospheric spectrum selected based upon the physical parameters of the star (lower right panels). A detailed explanation of each of the parameters derived and displayed with the plots is set out in Subsection \ref{datared}. Rest wavelengths of the blue and red elements of the doublet are highlighted in those colours. Note also that for those plots prepared using a decoupled treatment of each doublet element the model fit for each element ``ignores'' the presence of the other element. This can be seen most clearly in the blue element model fits, however it does not affect the model fit \textit{within} the relevant doublet element.

   \begin{figure*}
\begin{center}
 \includegraphics[width=5.45in]{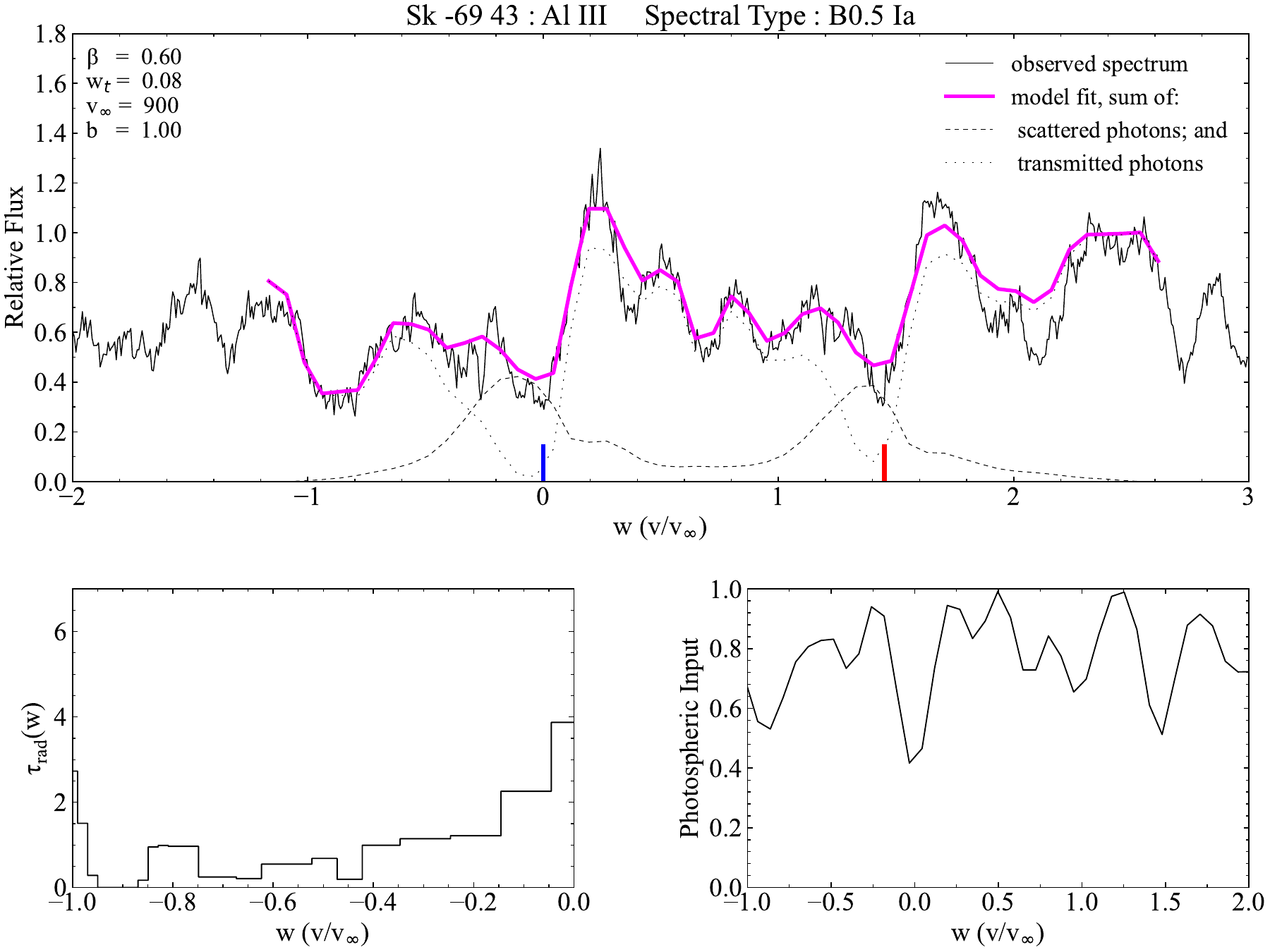}
 \caption{SEI-derived model combined fit for both components of the Al \textsc{iii} doublet feature in the UV spectrum of LMC star Sk -69 43 (spectral type B0.5 Ia). The rest wavelengths for each element of the doublet are highlighted in the relevant colours and a magenta colour is used to show a ``combined'' model fit.}
 \label{fig6943}
\end{center}
    \end{figure*}

   \begin{figure*}
\begin{center}
 \includegraphics[width=5.45in]{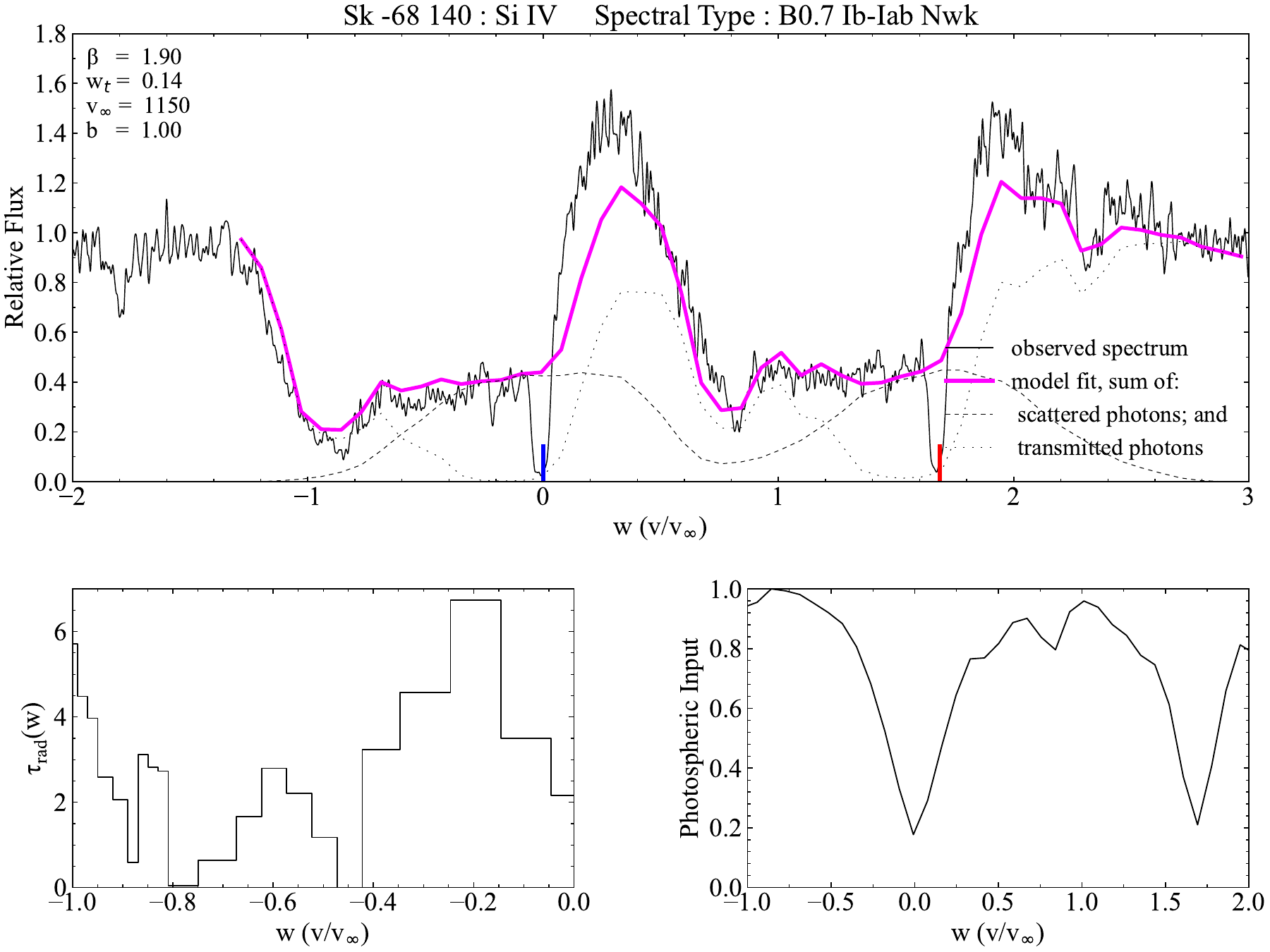}
 \caption{SEI-derived model combined fit for both components of the Si \textsc{iv} doublet feature in the UV spectrum of LMC star Sk -68 140 (spectral type B0.7 Ib-Iab Nwk).}
 \label{fig68140}
\end{center}
    \end{figure*}

   \begin{figure*}
\begin{center}
 \subfloat[ ]{\includegraphics[width=3.34in]{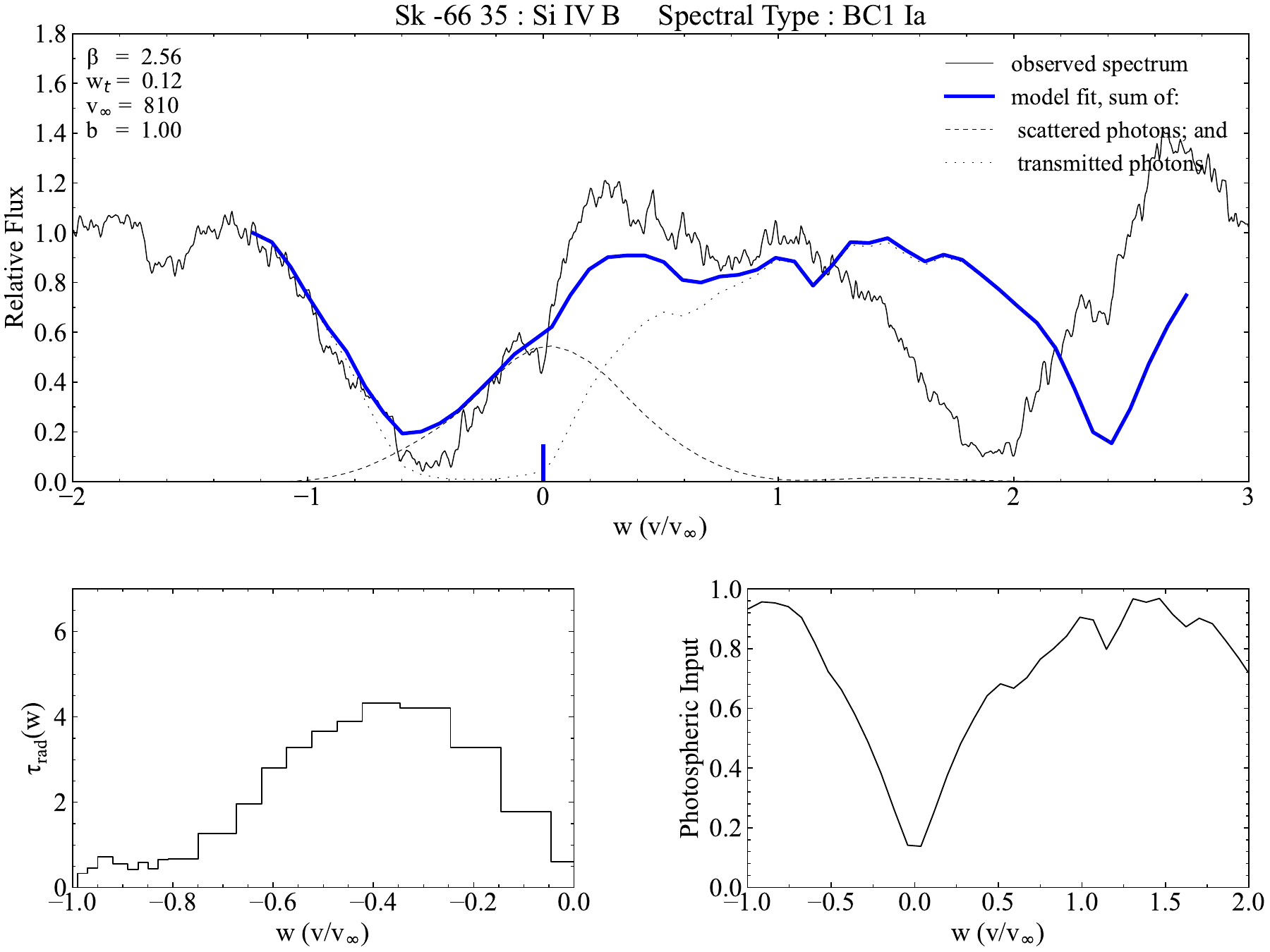} }
 \qquad
 \subfloat[ ]{\includegraphics[width=3.34in]{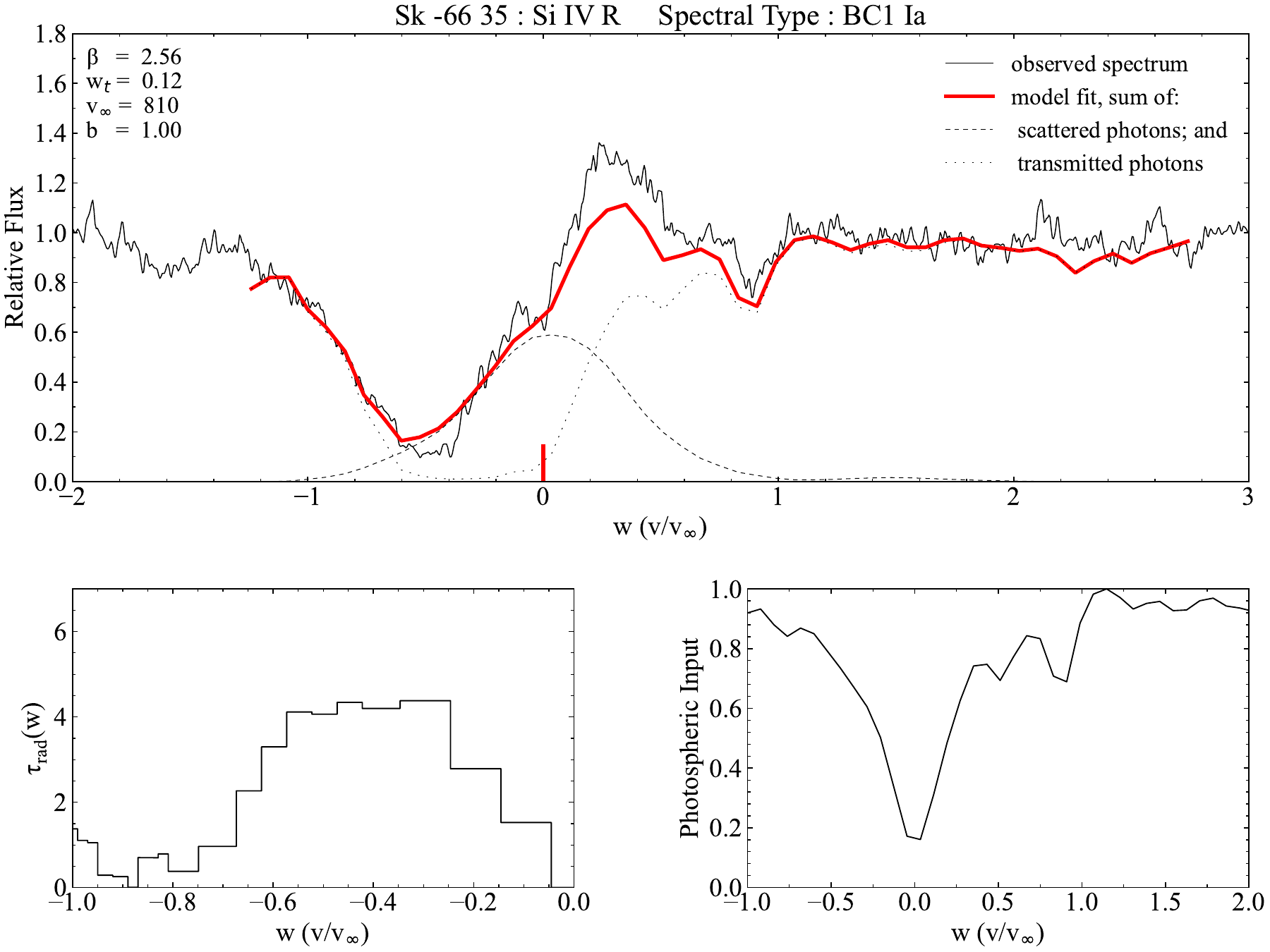} }
 \caption{SEI-derived model fits for (a) blue and (b) red components of the Si \textsc{iv} doublet feature in the UV spectrum of LMC star Sk -66 35 (spectral type BC1 Ia).}
 \label{fig6635}
\end{center}
    \end{figure*}

   \begin{figure*}
\begin{center}
 \includegraphics[width=5.52in]{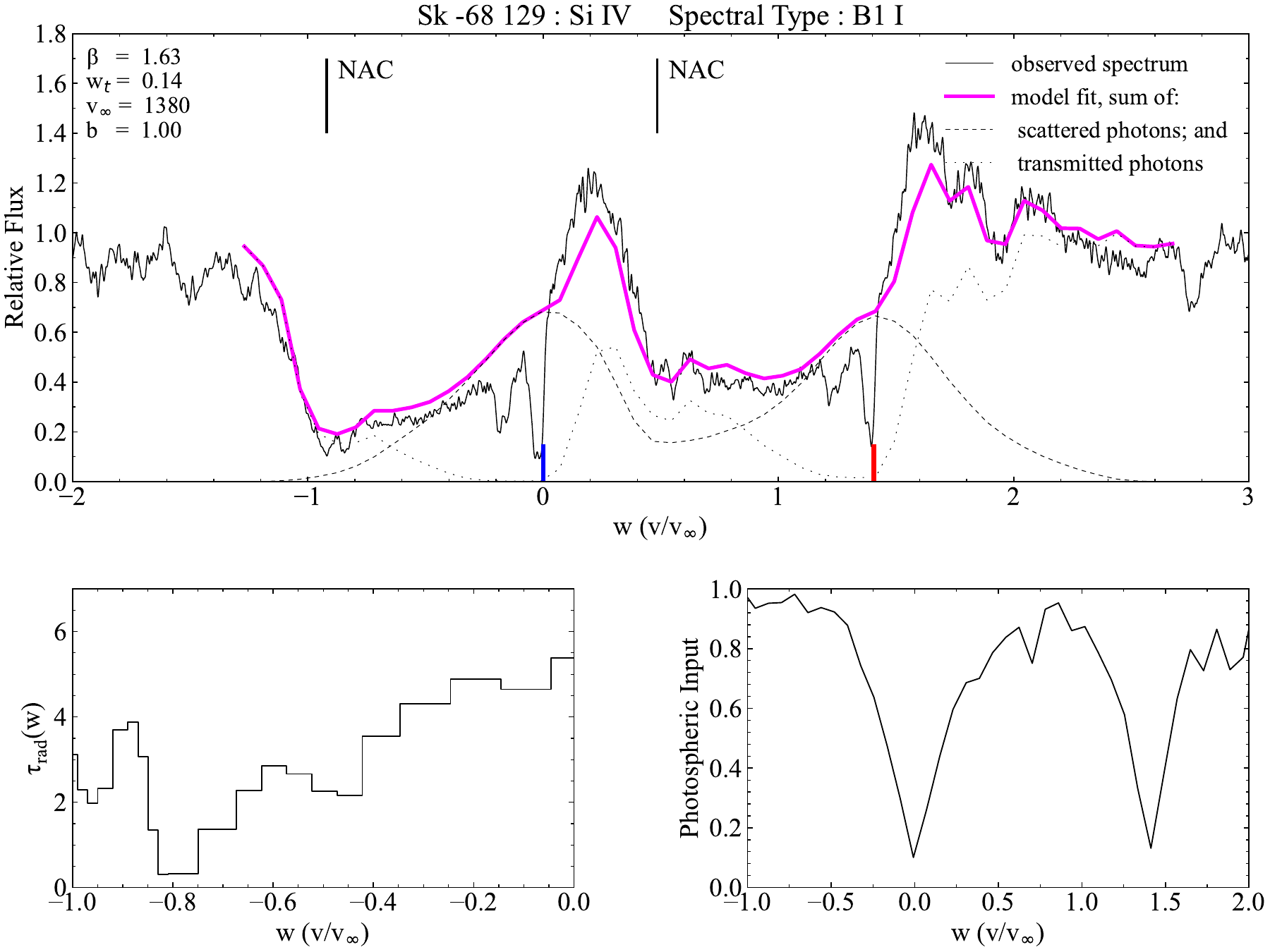}
 \caption{SEI-derived model combined fit for both components of the Si \textsc{iv} doublet feature in the UV spectrum of LMC star Sk -68 129 (spectral type B1 I). The location of likely narrow absorption components (NAC) in each doublet element are indicated, each at the same distance from the relevant rest wavelength.}
 \label{fig68129}
\end{center}
    \end{figure*}

   \begin{figure*}
\begin{center}
 \includegraphics[width=5.52in]{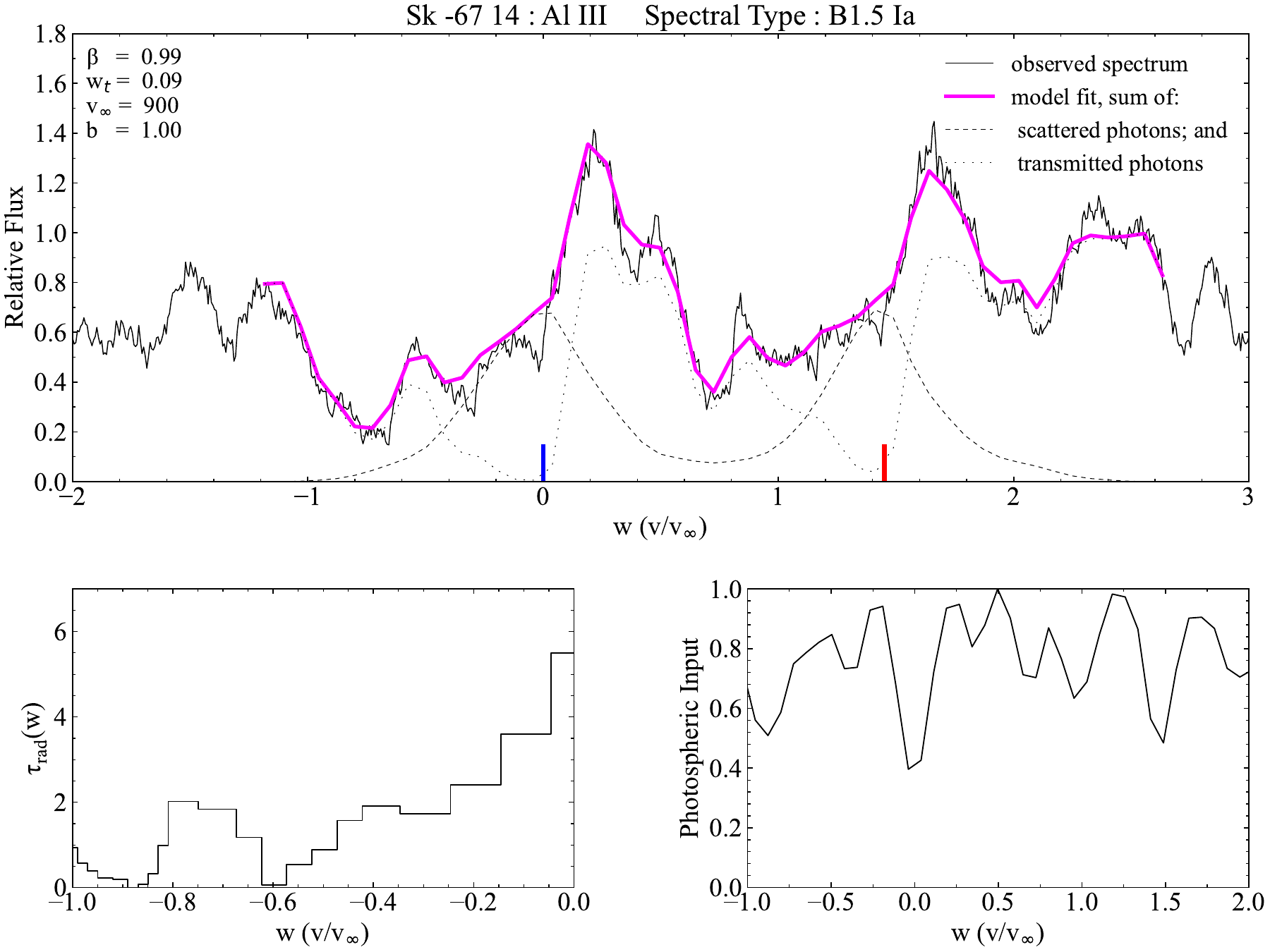}
 \caption{SEI-derived model combined fit for both components of the Al \textsc{iii} doublet feature in the UV spectrum of LMC star Sk -67 14 (spectral type B1.5 Ia).}
 \label{fig6714}
\end{center}
    \end{figure*}

   \begin{figure*}
\begin{center}
 \subfloat[ ]{\includegraphics[width=3.34in]{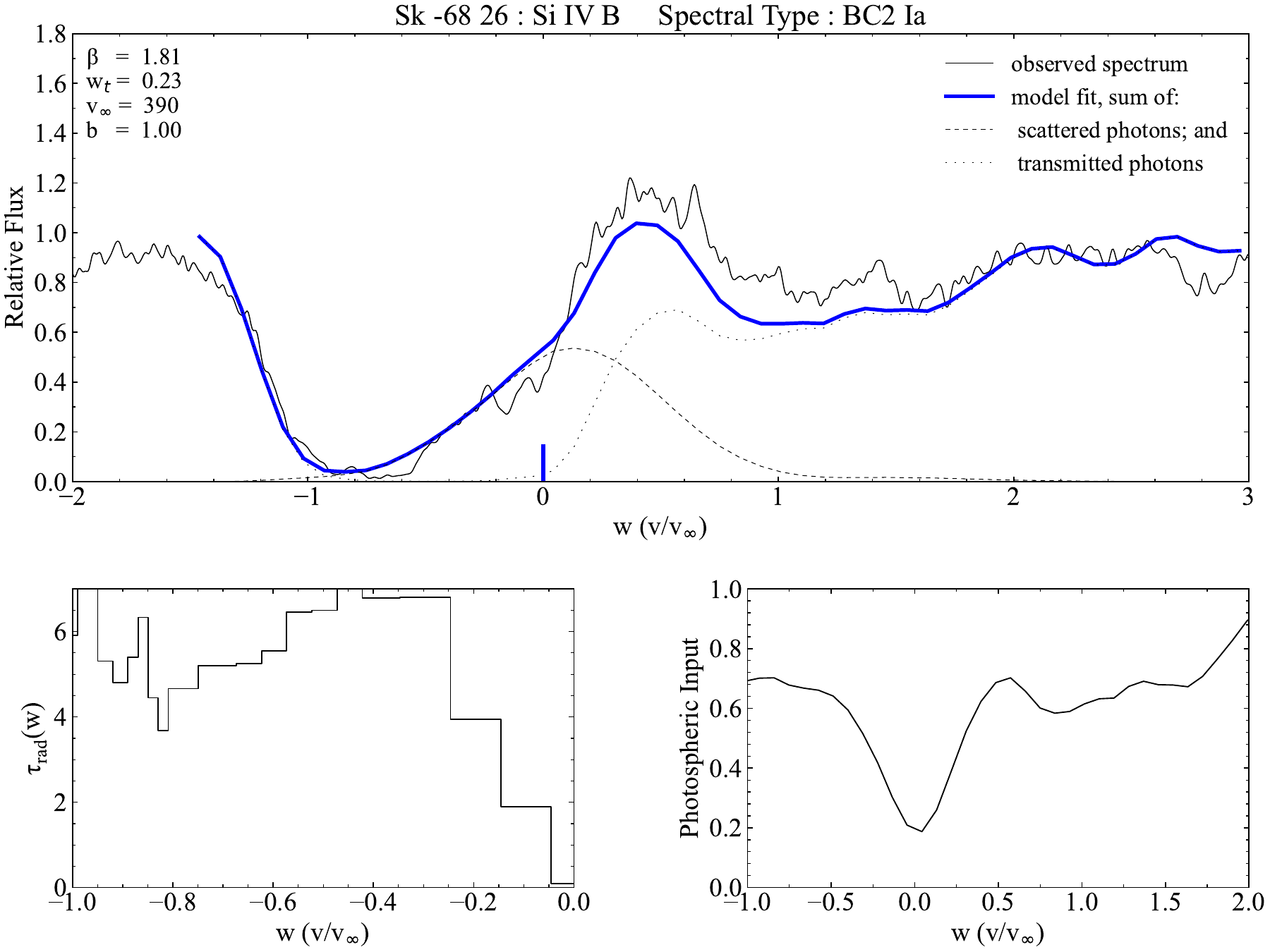} }
 \qquad
 \subfloat[ ]{\includegraphics[width=3.34in]{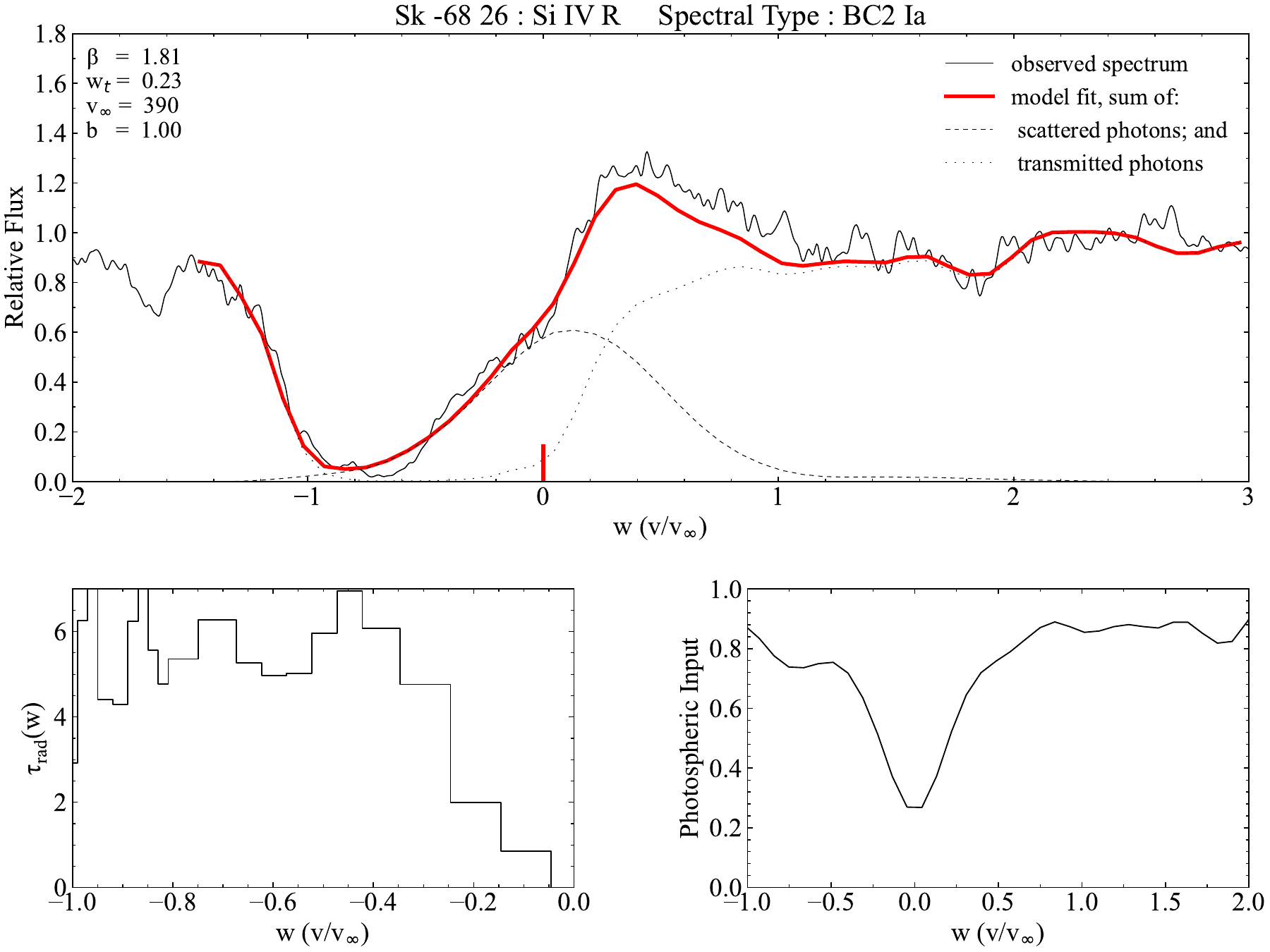} }
 \caption{SEI-derived model fits for (a) blue and (b) red components of the Si \textsc{iv} doublet feature in the UV spectrum of LMC star Sk -68 26 (spectral type BC2 Ia).}
 \label{fig6826}
\end{center}
    \end{figure*}

   \begin{figure*}
\begin{center}
 \subfloat[ ]{\includegraphics[width=3.34in]{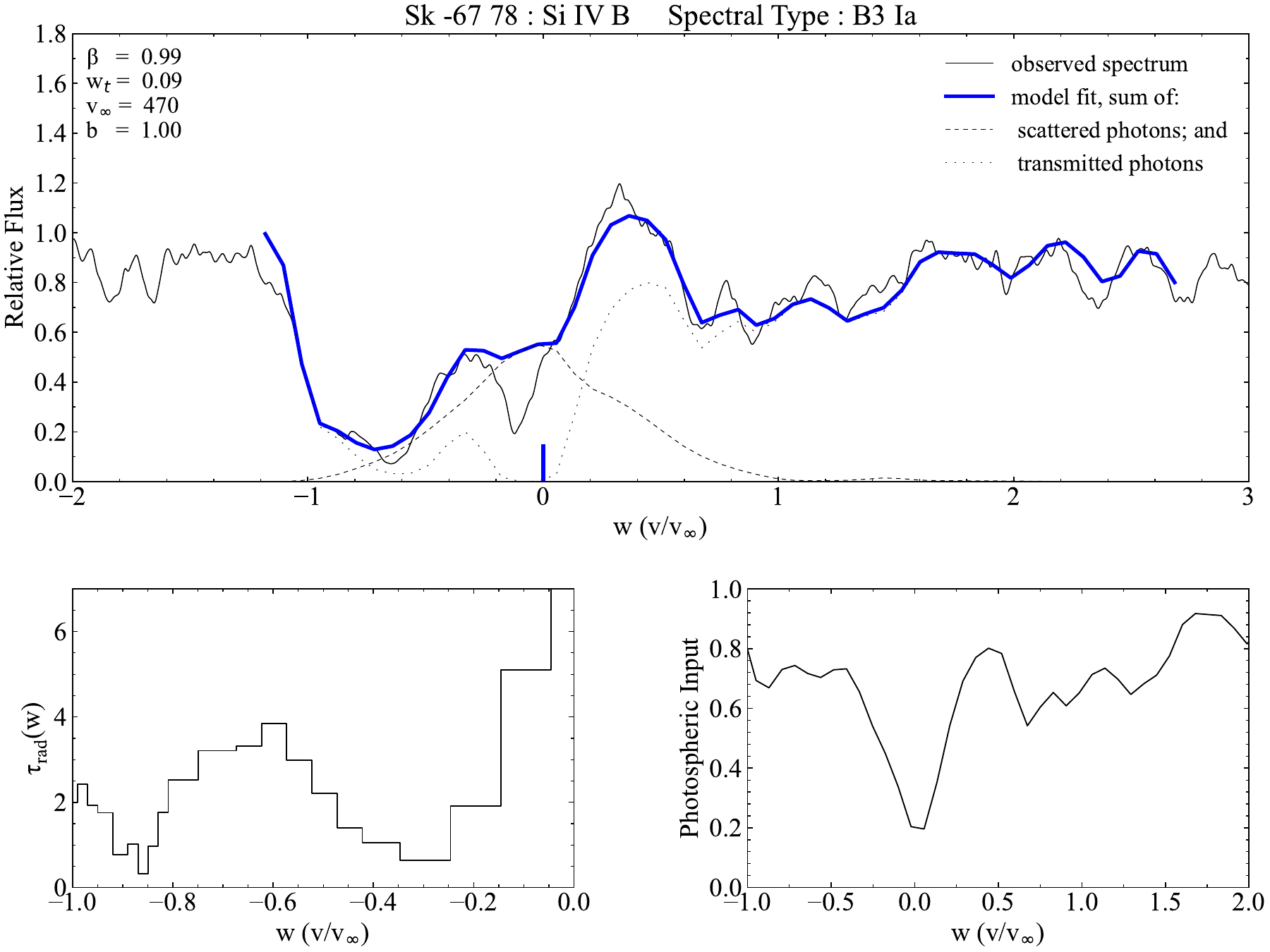} }
 \qquad
 \subfloat[ ]{\includegraphics[width=3.34in]{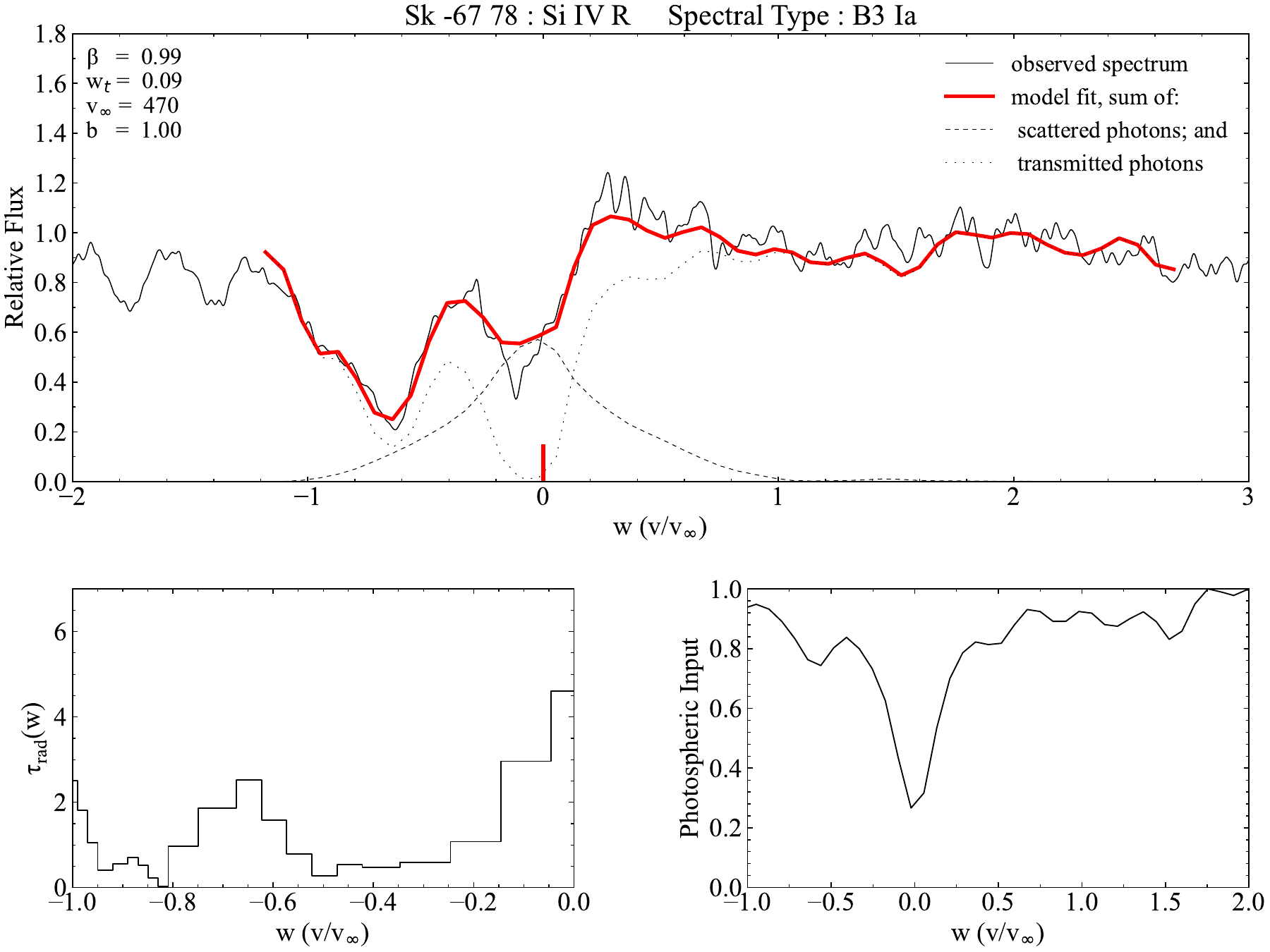} }
 \caption{SEI-derived model fits for (a) blue and (b) red components of the Si \textsc{iv} doublet feature in the UV spectrum of LMC star Sk -67 78 (spectral type B3 Ia).}
 \label{fig6778}
\end{center}
    \end{figure*}
%
   \begin{figure*}
\begin{center}
 \subfloat[ ]{\includegraphics[width=3.34in]{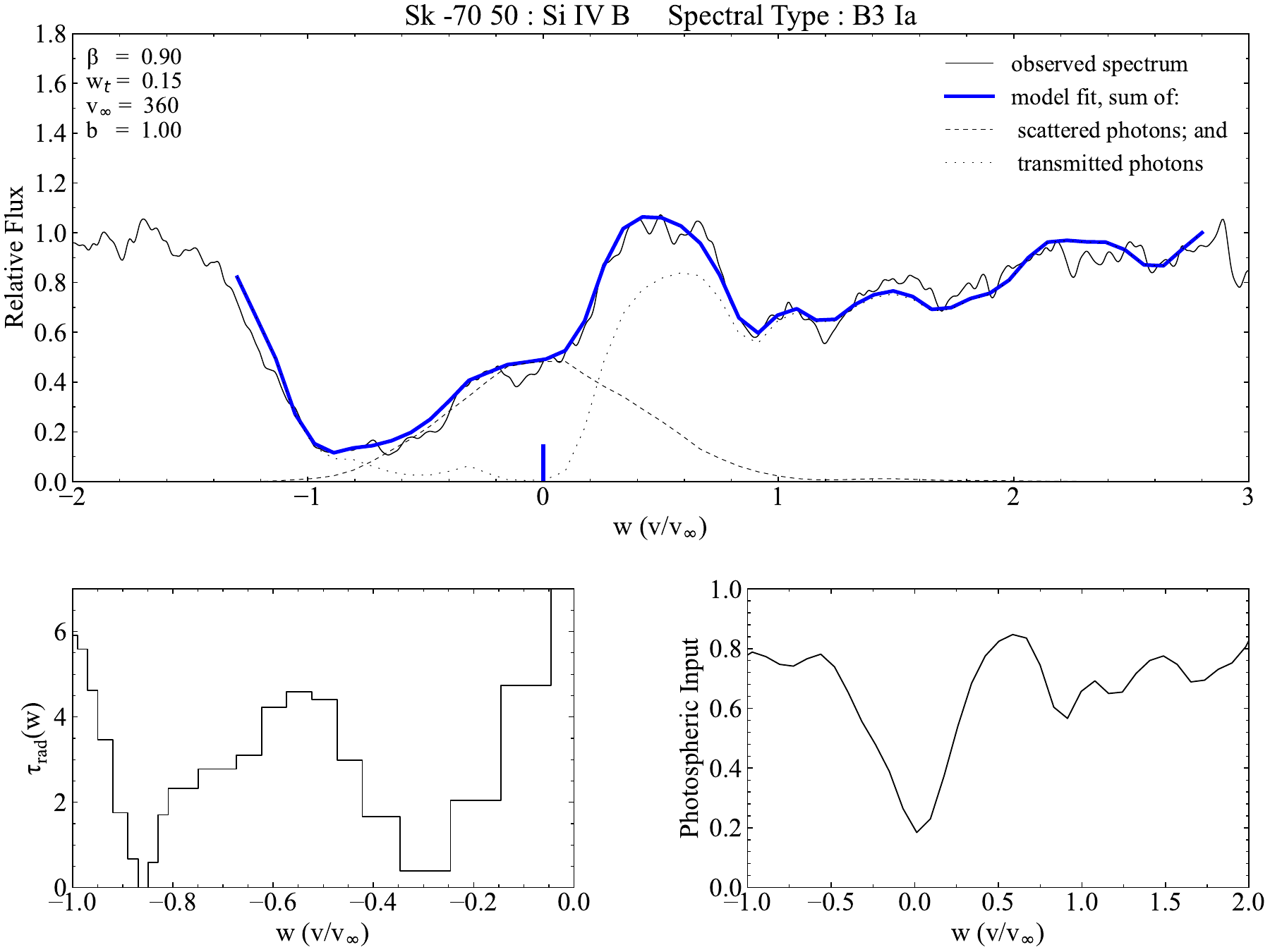} }
 \qquad
 \subfloat[ ]{\includegraphics[width=3.34in]{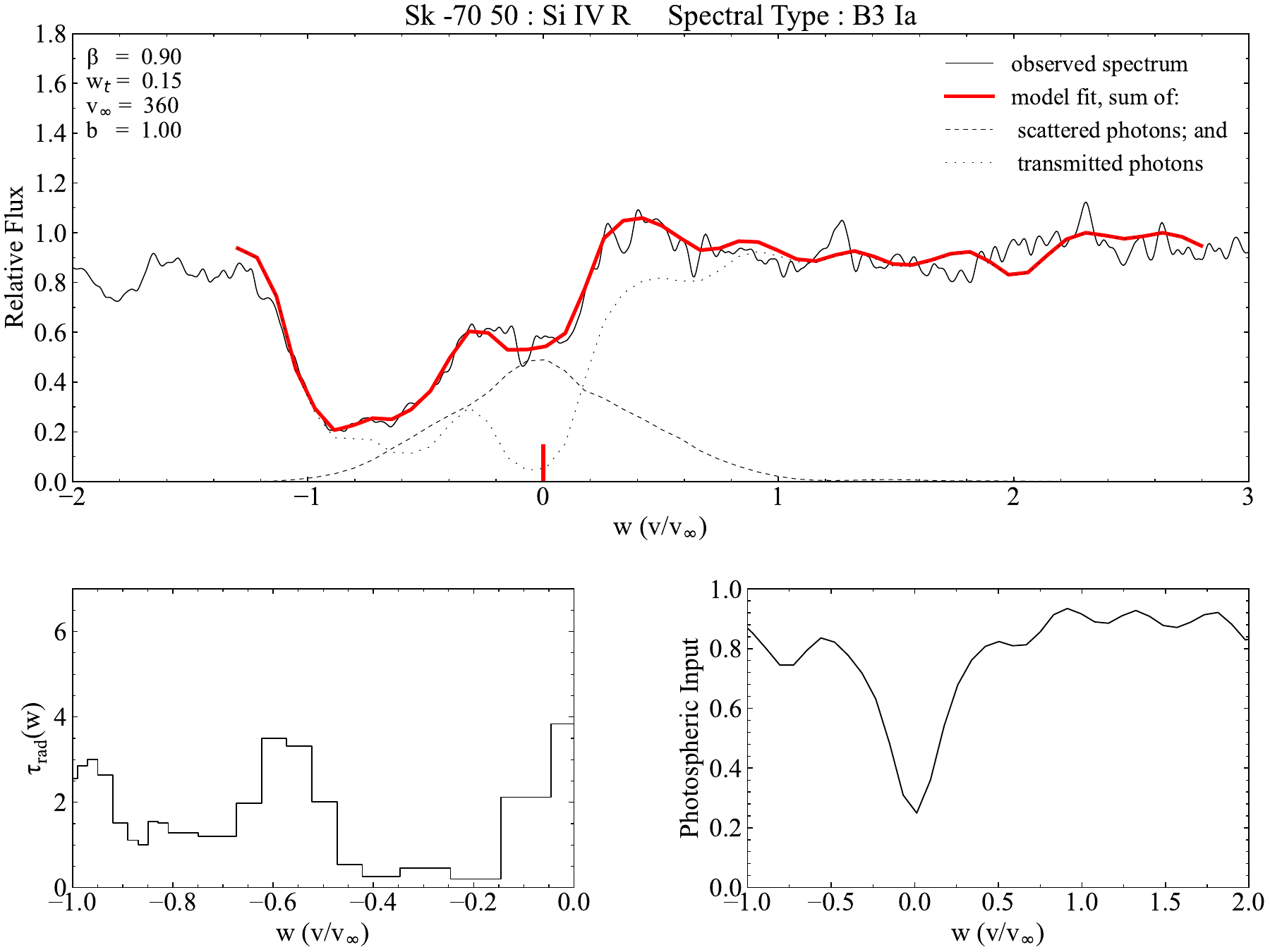} }
 \caption{SEI-derived model fits for (a) blue and (b) red components of the Si \textsc{iv} doublet feature in the UV spectrum of LMC star Sk -70 50 (spectral type B3 Ia).}
 \label{fig7050}
\end{center}
    \end{figure*}

   \begin{figure*}
\begin{center}
 \includegraphics[width=5.52in]{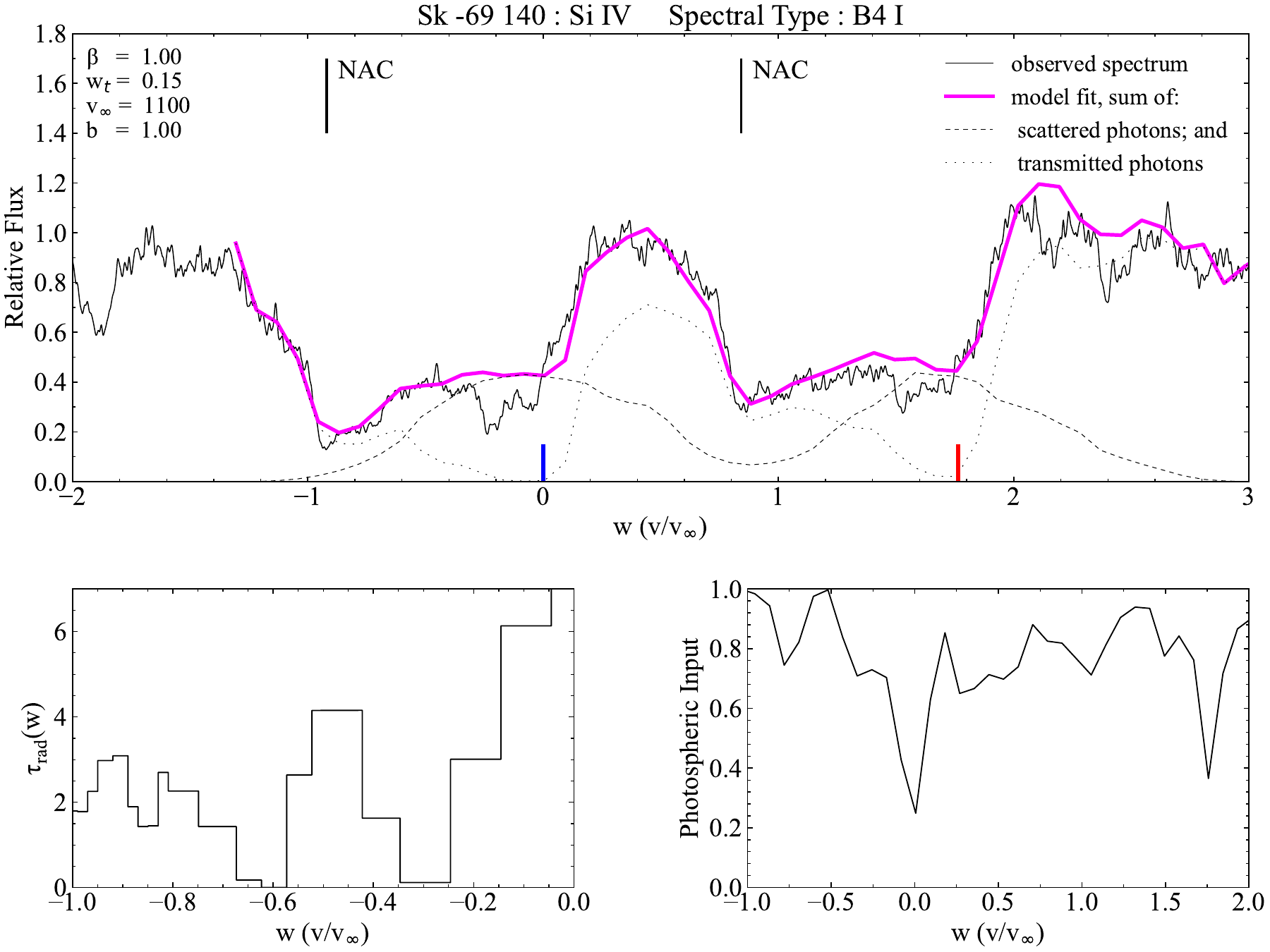}
 \caption{SEI-derived model combined fit for both components of the Si \textsc{iv} doublet feature in the UV spectrum of LMC star Sk -69 140 (spectral type B4 I). The location of likely narrow absorption components (NAC) in each doublet element are indicated, each at the same distance from the relevant rest wavelength.}
 \label{fig69140}
\end{center}
    \end{figure*}

   \begin{figure*}
\begin{center}
 \subfloat[ ]{\includegraphics[width=3.34in]{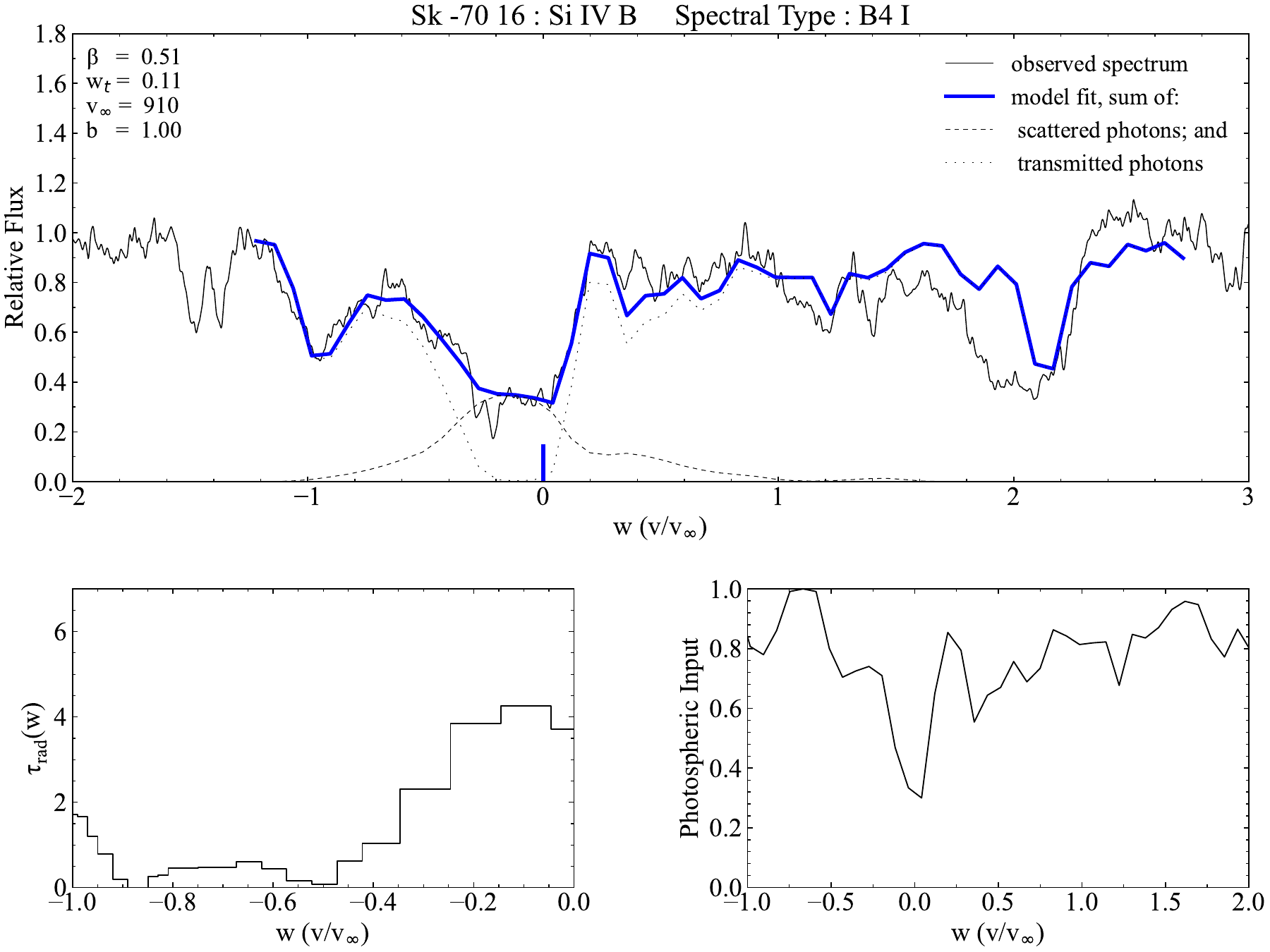} }
 \qquad
 \subfloat[ ]{\includegraphics[width=3.34in]{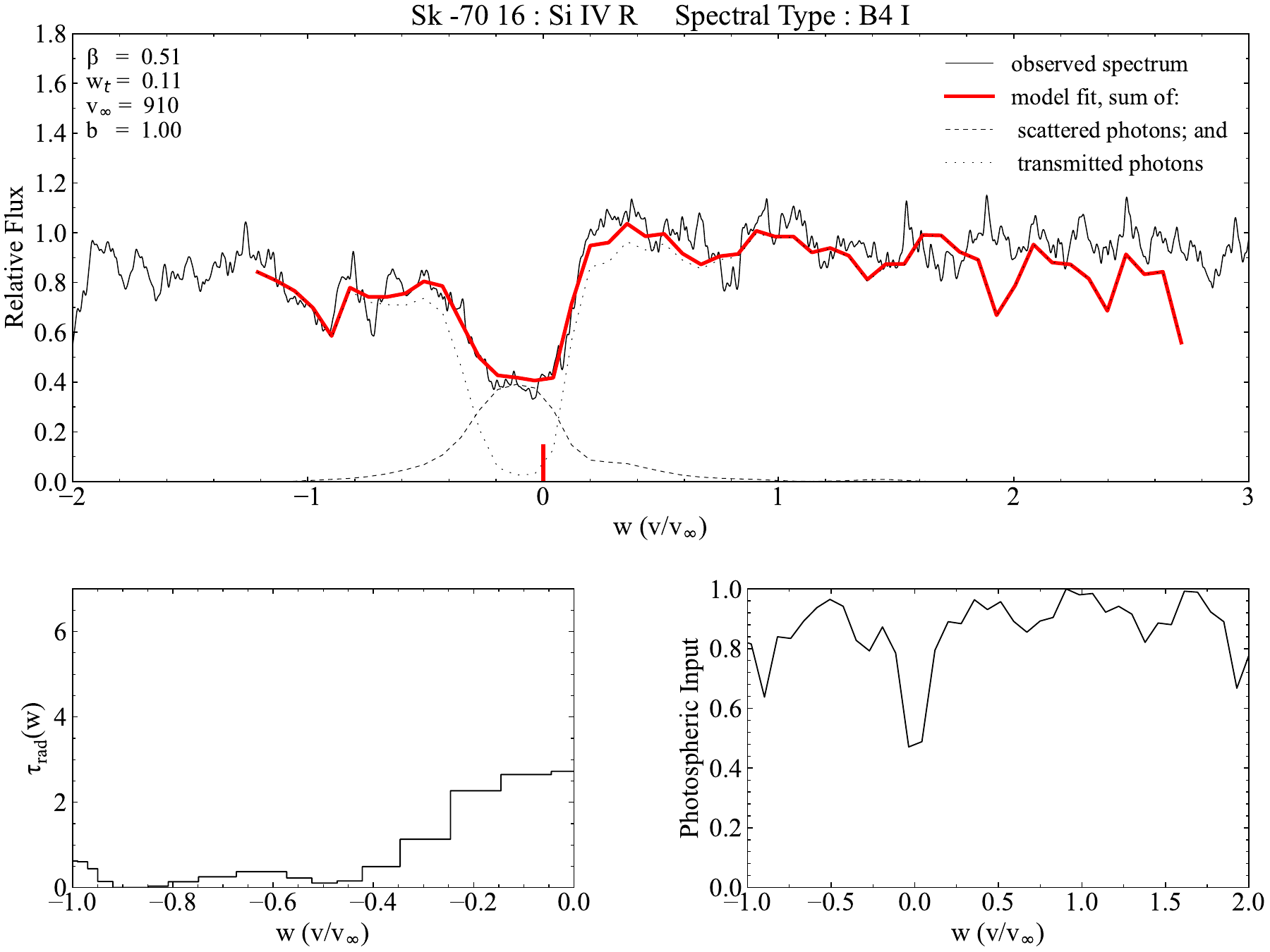} }
 \caption{SEI-derived model fits for (a) blue and (b) red components of the Si \textsc{iv} doublet feature in the UV spectrum of LMC star Sk -70 16 (spectral type B4 I).}
 \label{fig7016}
\end{center}
    \end{figure*}

   \begin{figure*}
\begin{center}
 \subfloat[ ]{\includegraphics[width=3.34in]{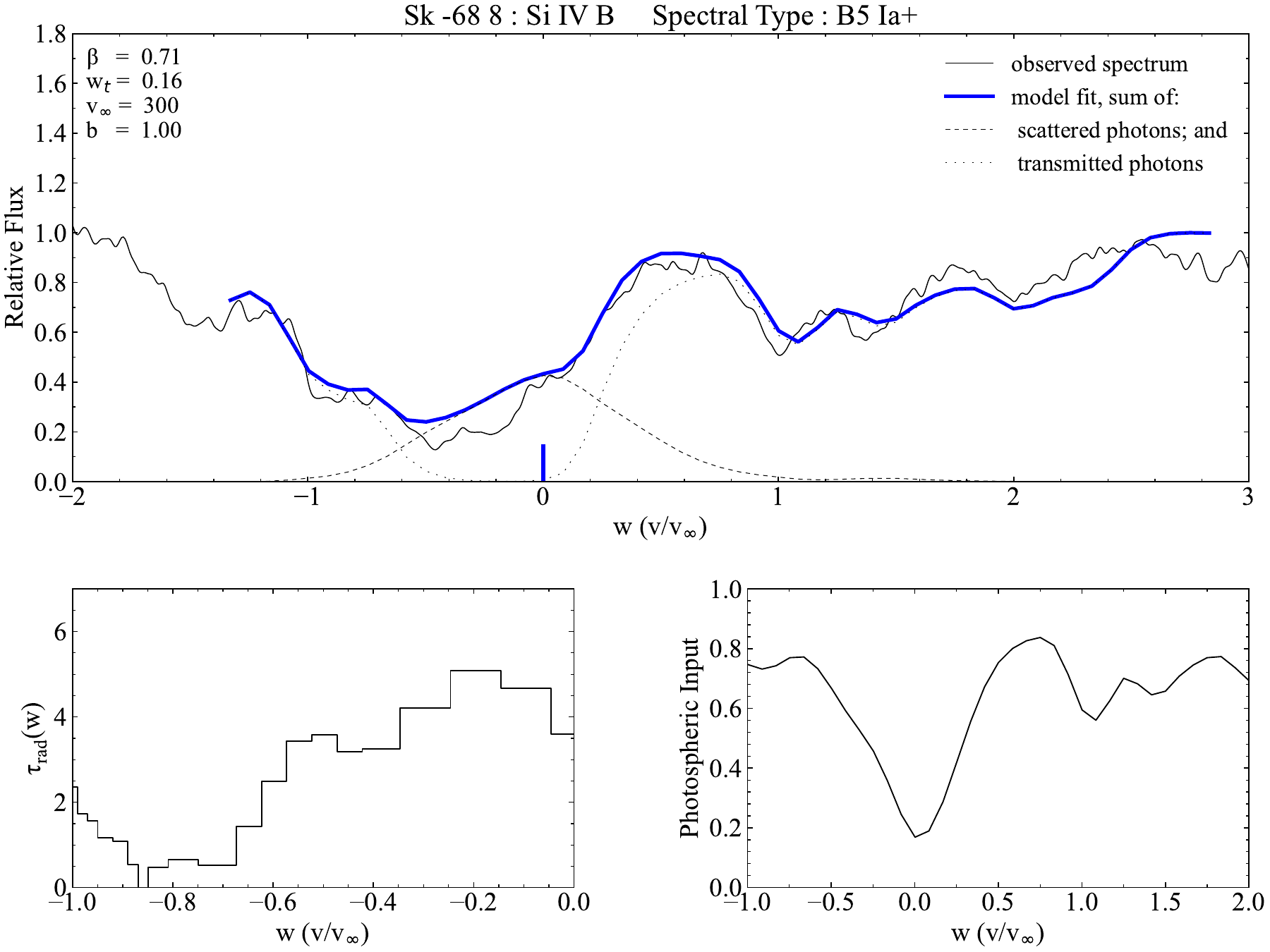} }
 \qquad
 \subfloat[ ]{\includegraphics[width=3.34in]{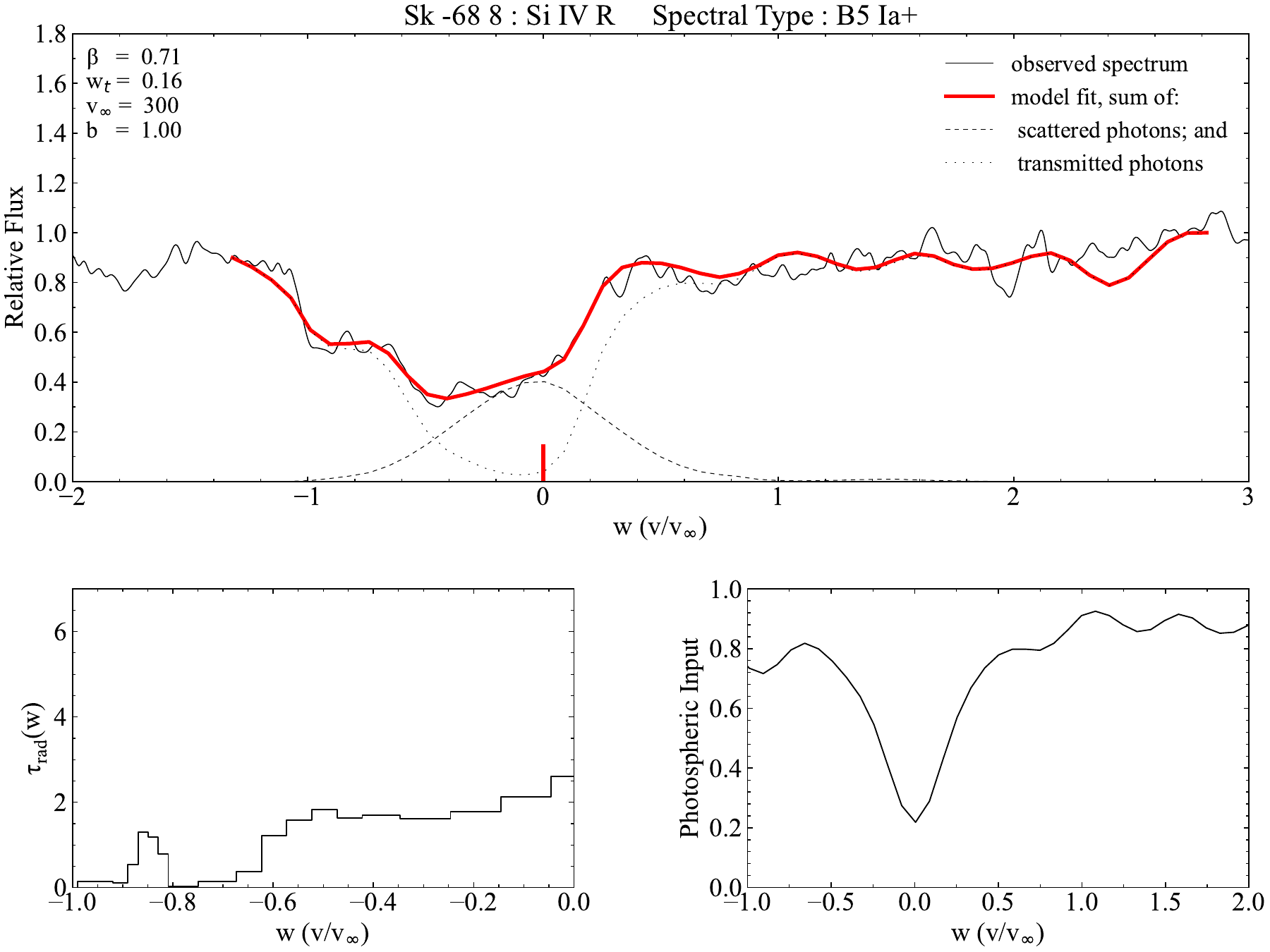} }
 \caption{SEI-derived model fits for (a) blue and (b) red components of the Si \textsc{iv} doublet feature in the UV spectrum of LMC star Sk -68 8 (spectral type B5 Ia+).}
 \label{fig688}
\end{center}
    \end{figure*}
    
   \begin{figure*}
\begin{center}
 \subfloat[ ]{\includegraphics[width=3.34in]{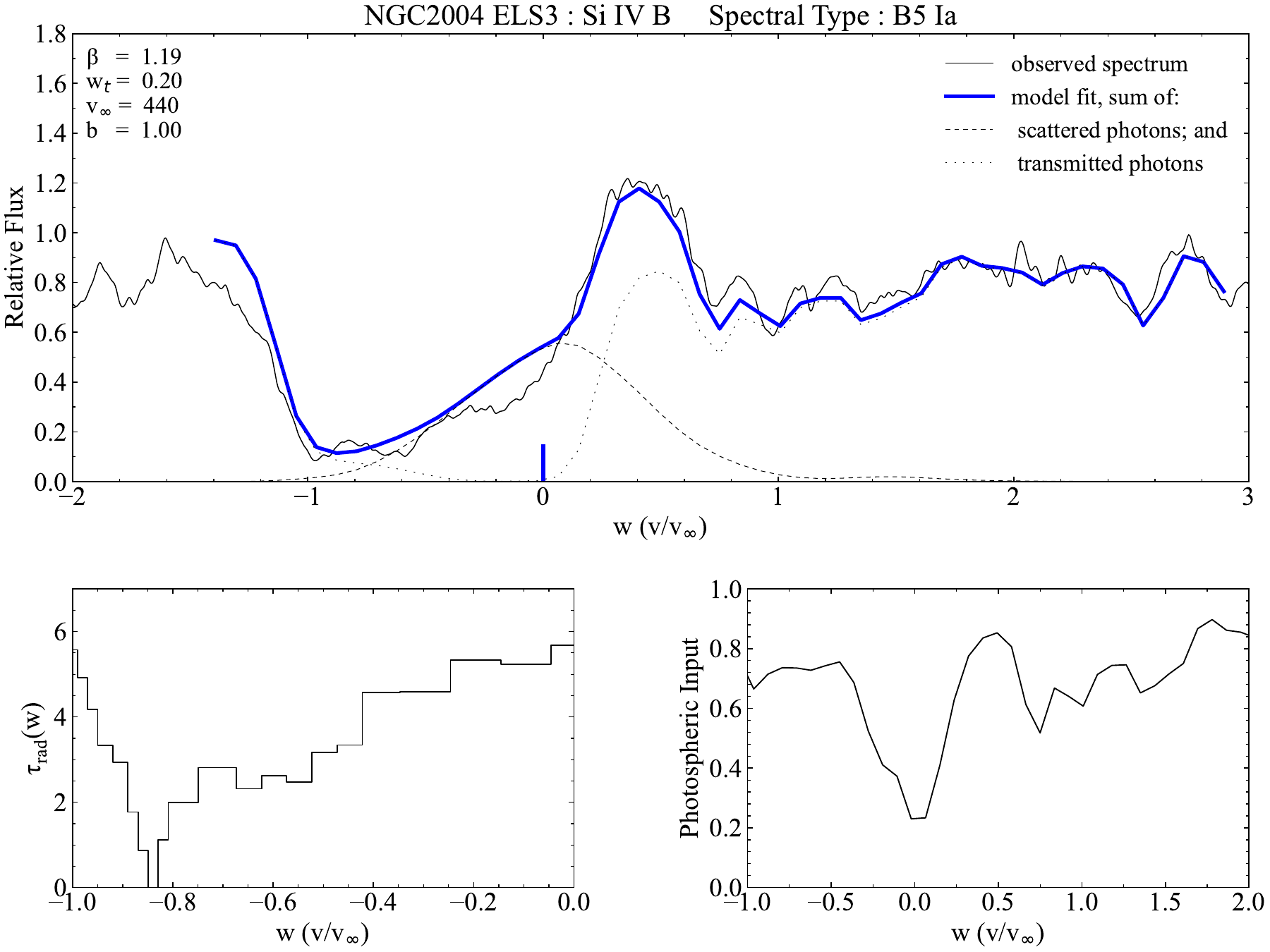} }
 \qquad
 \subfloat[ ]{\includegraphics[width=3.34in]{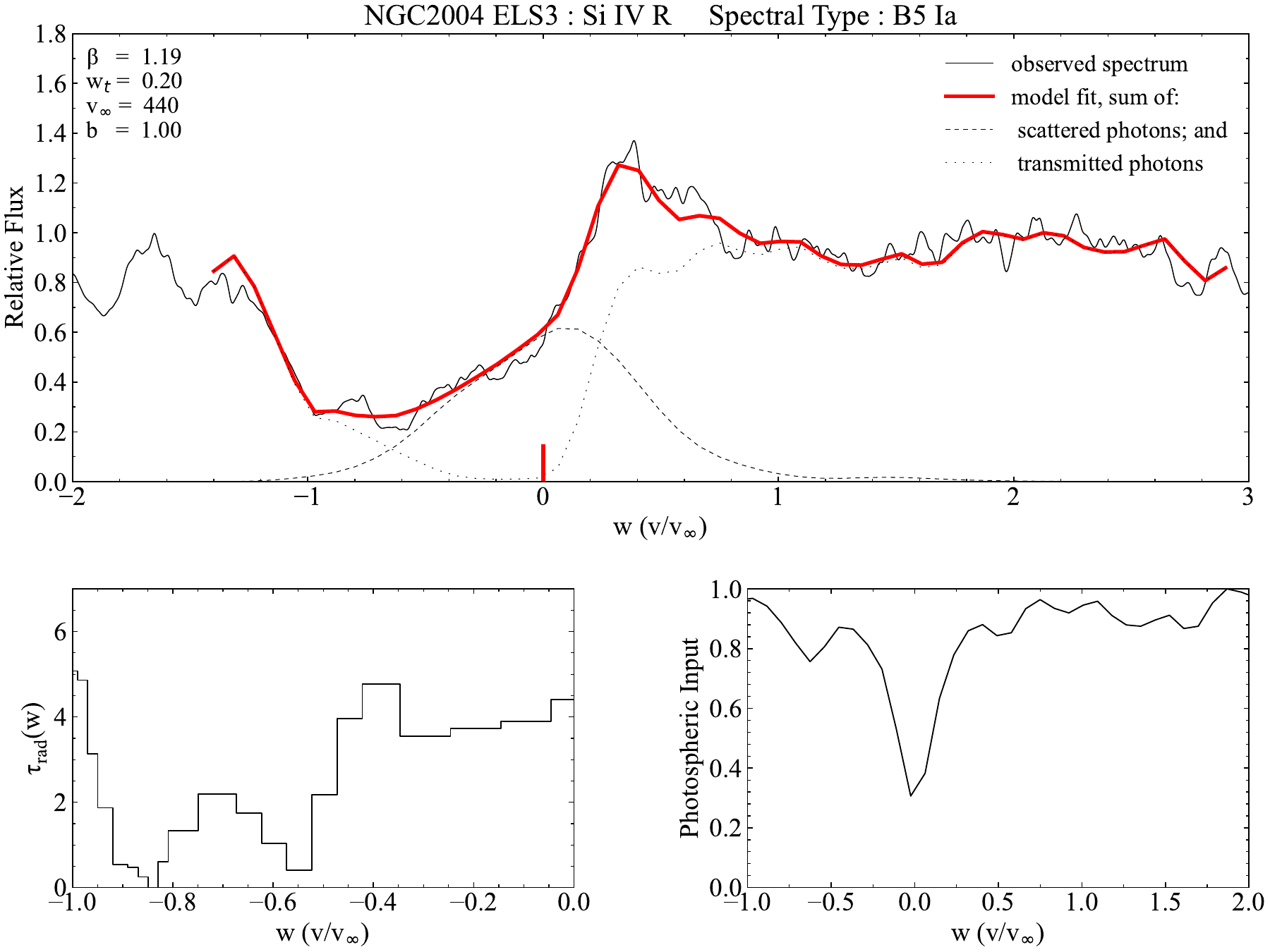} }
 \caption{SEI-derived model fits for (a) blue and (b) red components of the Si \textsc{iv} doublet feature in the UV spectrum of LMC star NGC 2004 ELS 3 (spectral type B5 Ia).}
 \label{fig20043}
\end{center}
    \end{figure*}

\clearpage
\newpage
\section{SEI model line fits for SMC B supergiants} \label{AppB}

   \begin{figure*}
\begin{center}
 \includegraphics[width=5.52in]{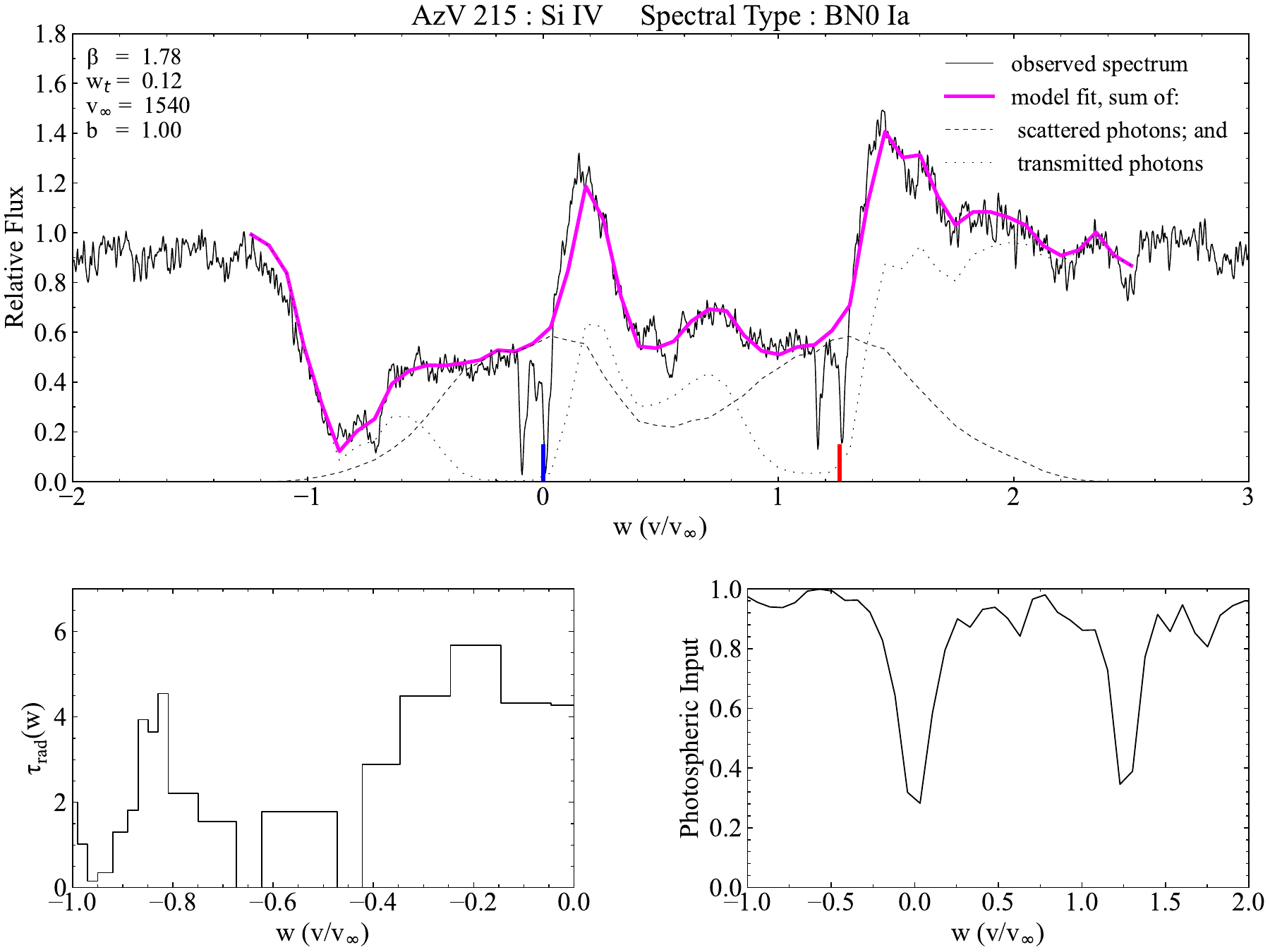}

 \caption{SEI-derived model combined fit for both components of the Si \textsc{iv} doublet feature in the UV spectrum of SMC star AzV 215 (spectral type B0 I(b)/BN0 Ia).}
 \label{fig215}
\end{center}
    \end{figure*}

   \begin{figure*}
\begin{center}
 \includegraphics[width=5.52in]{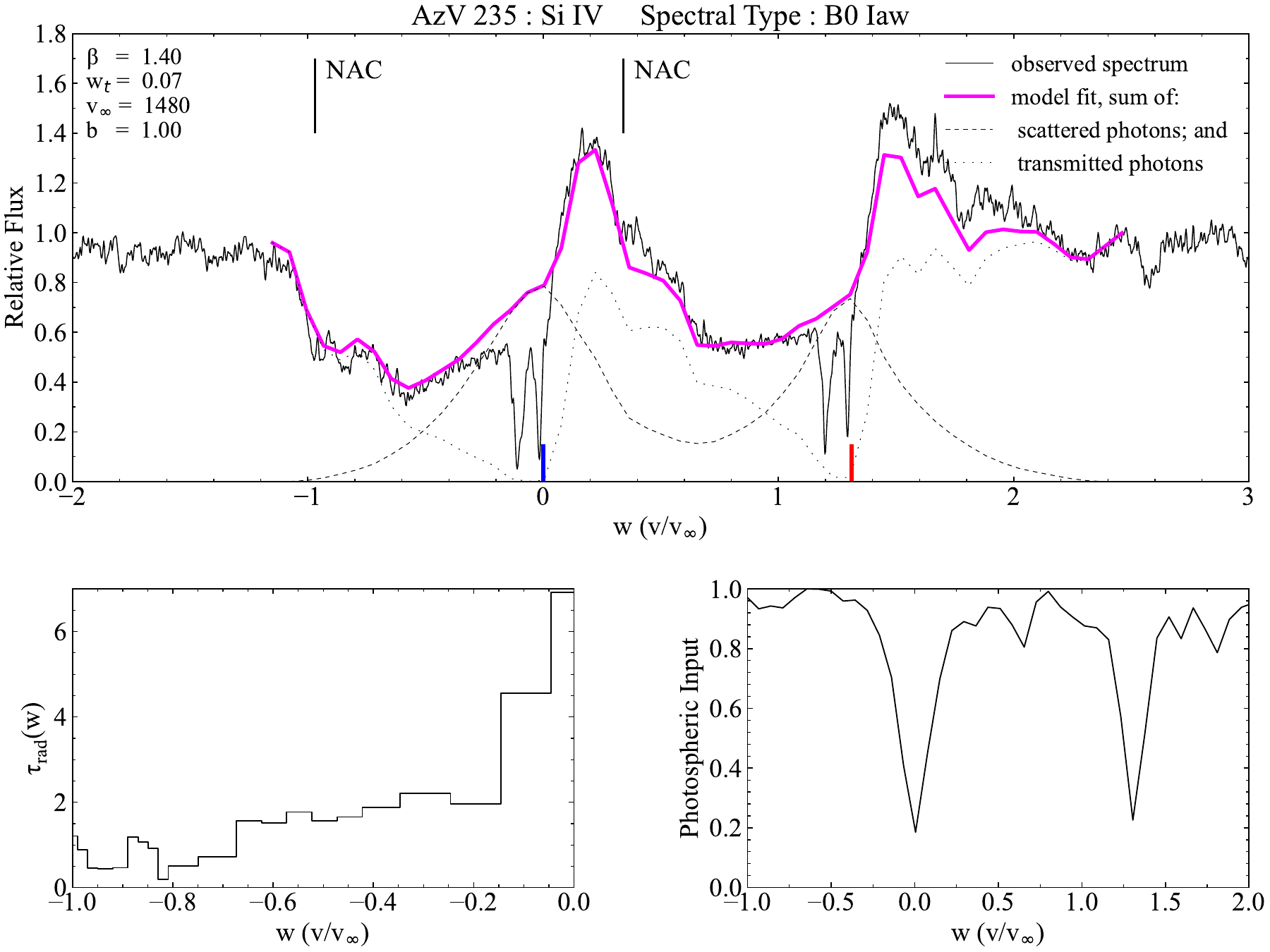}

 \caption{SEI-derived model combined fit for both components of the Si \textsc{iv} doublet feature in the UV spectrum of SMC star AzV 235 (spectral type B0 Iaw). The location of likely narrow absorption components (NAC) in each doublet element are indicated, each at the same distance from the relevant rest wavelength.}
 \label{fig235}
\end{center}
    \end{figure*}

   \begin{figure*}
\begin{center}
 \includegraphics[width=5.52in]{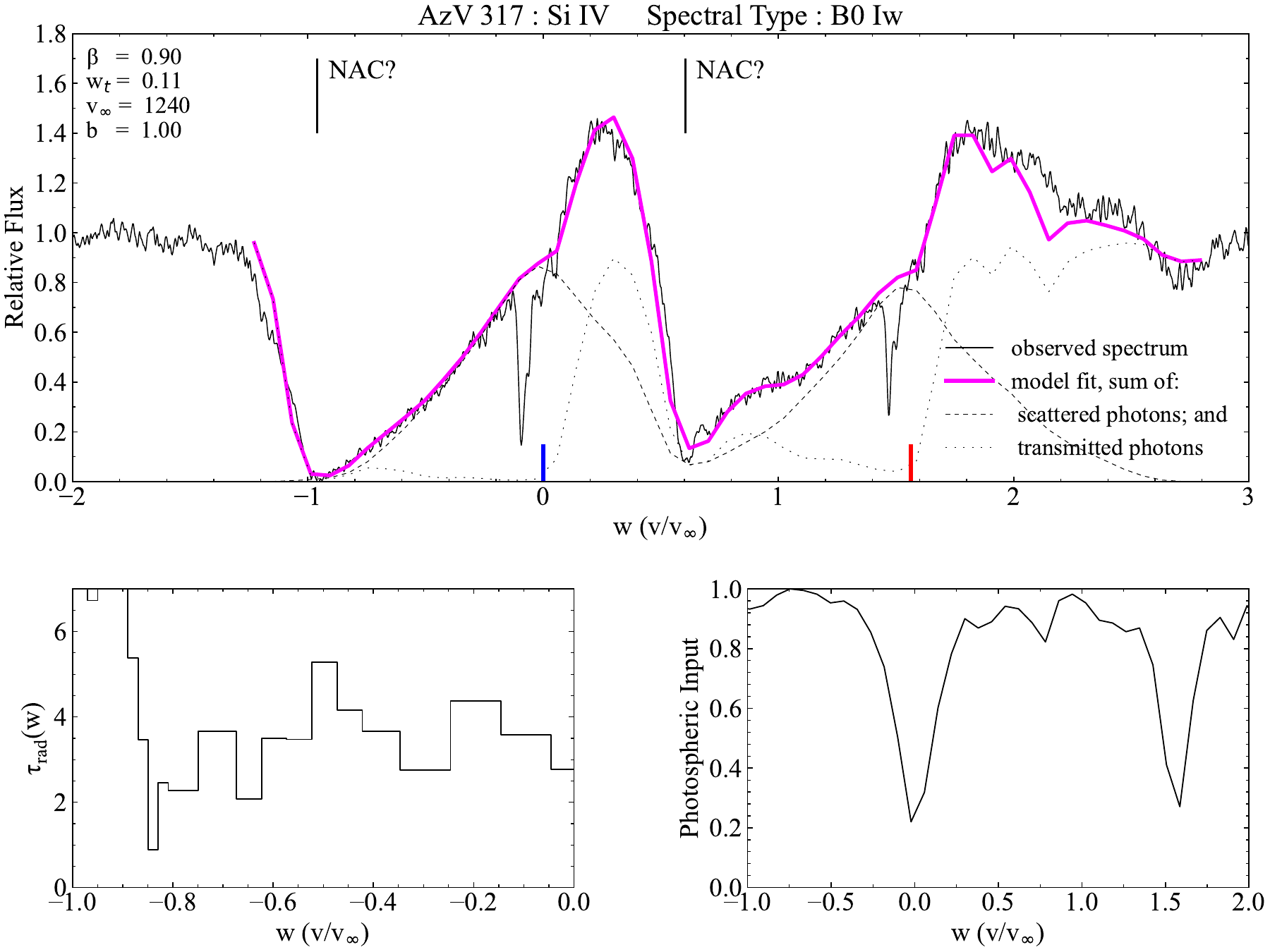}

 \caption{SEI-derived model combined fit for both components of the Si \textsc{iv} doublet feature in the UV spectrum of SMC star AzV 317 (spectral type B0 Iw). Note that the apparent narrow absorption component (NAC) visible in the red component profile is, unfortunately, lost due to the part saturation of the highest velocity part of the blue profile.}
 \label{fig317}
\end{center}
    \end{figure*}

   \begin{figure*}
\begin{center}
 \subfloat[ ]{\includegraphics[width=3.34in]{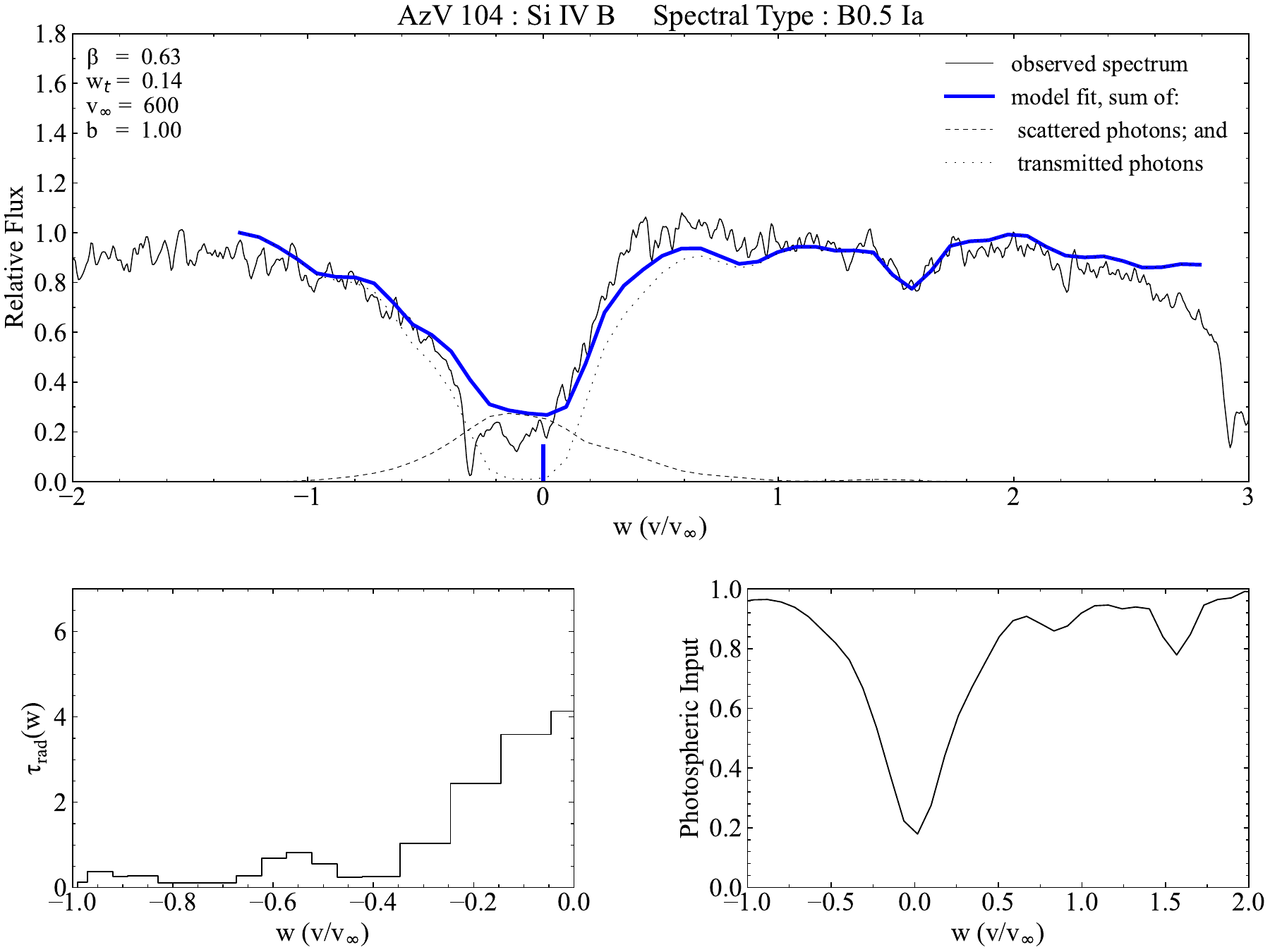} }
 \qquad
 \subfloat[ ]{\includegraphics[width=3.34in]{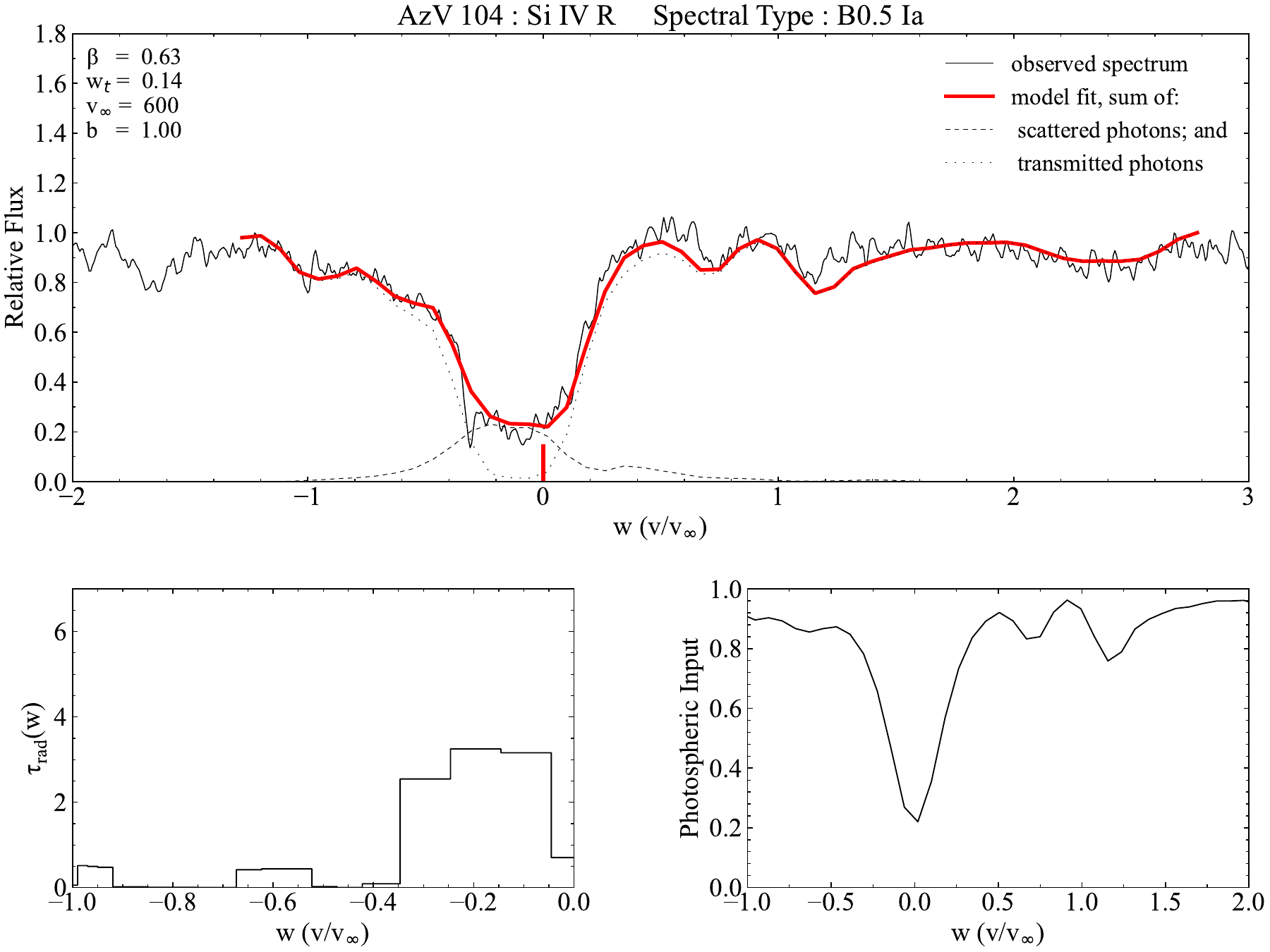} }
 \caption{SEI-derived model fits for (a) blue and (b) red components of the Si \textsc{iv} doublet feature in the UV spectrum of SMC star AzV 104 (spectral type B0.5 Ia). The deep absorption feature is the Milky Way ISM line, blue-shifted by an amount equal to the radial velocity of the star. Note that the wind appears to become very optically thin at higher velocities.}
 \label{fig104}
\end{center}
    \end{figure*}
    
   \begin{figure*}
\begin{center}
 \includegraphics[width=5.52in]{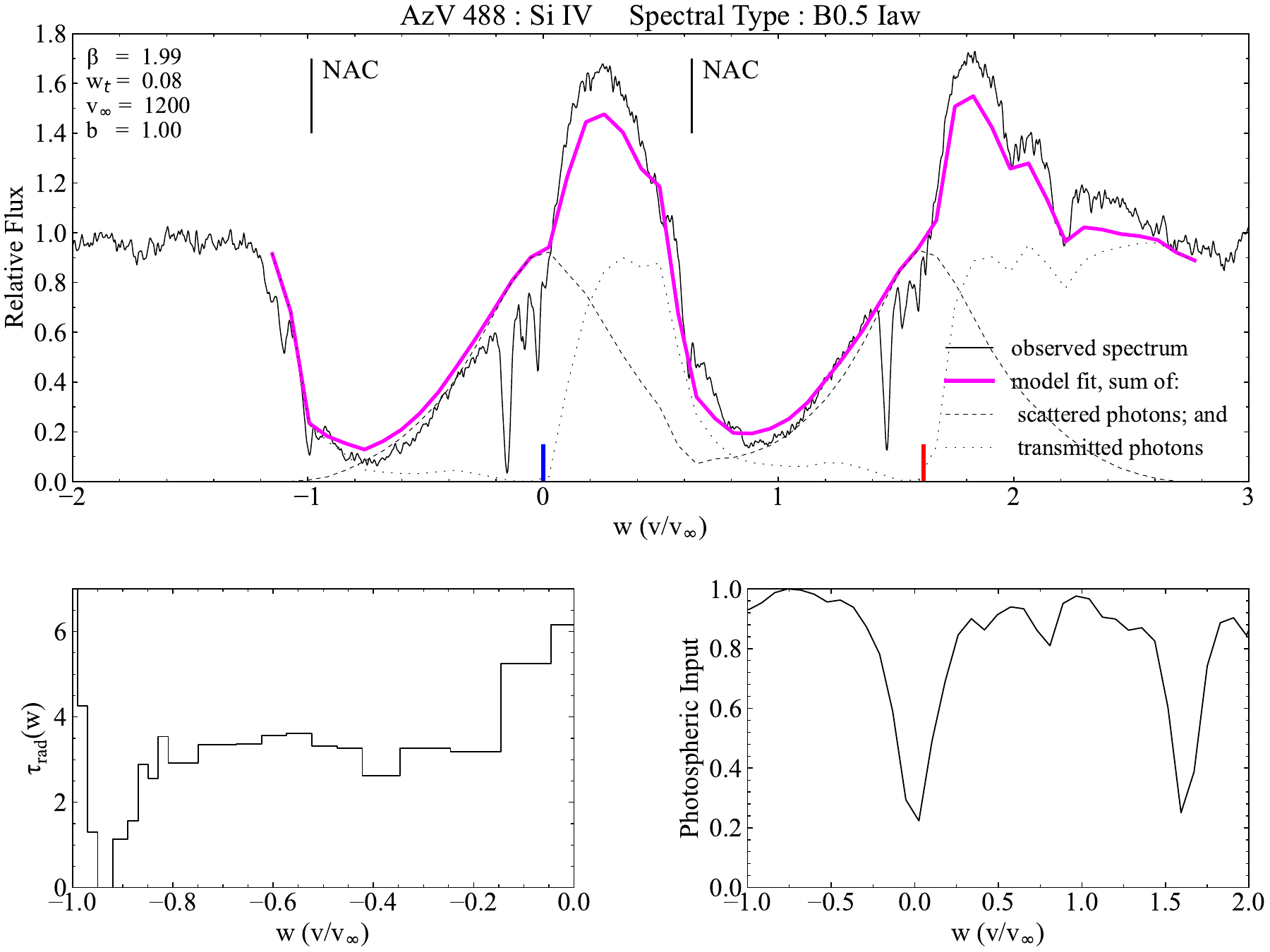}
 \caption{SEI-derived model combined fit for both components of the Si \textsc{iv} doublet feature in the UV spectrum of SMC star AzV 488 (spectral type B0.5 Iaw). The location of likely narrow absorption components (NAC) in each doublet element are indicated, each at the same distance from the relevant rest wavelength.}
 \label{fig488}
\end{center}
    \end{figure*}

   \begin{figure*}
\begin{center}
 \subfloat[ ]{\includegraphics[width=3.34in]{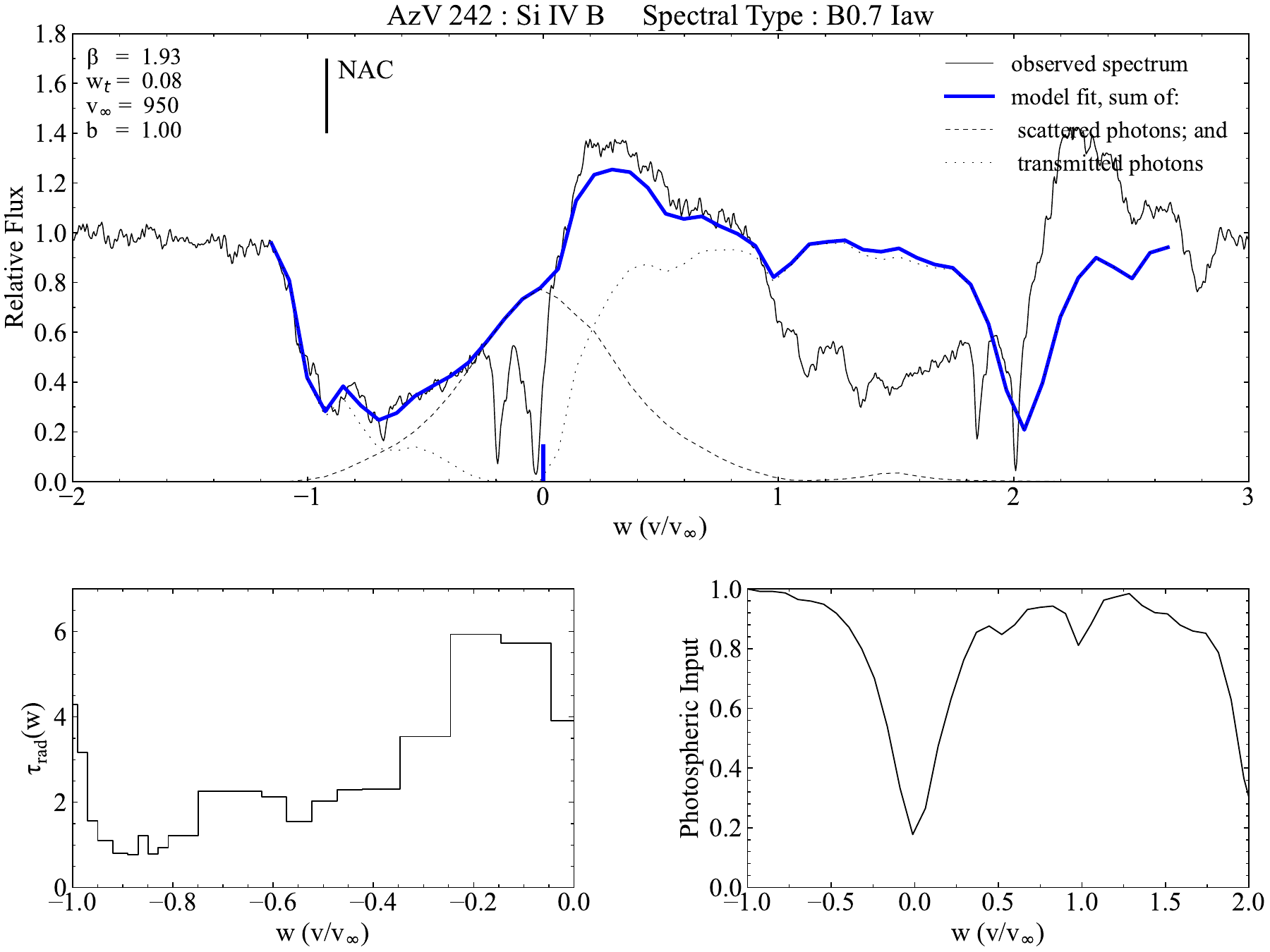} }
 \qquad
 \subfloat[ ]{\includegraphics[width=3.34in]{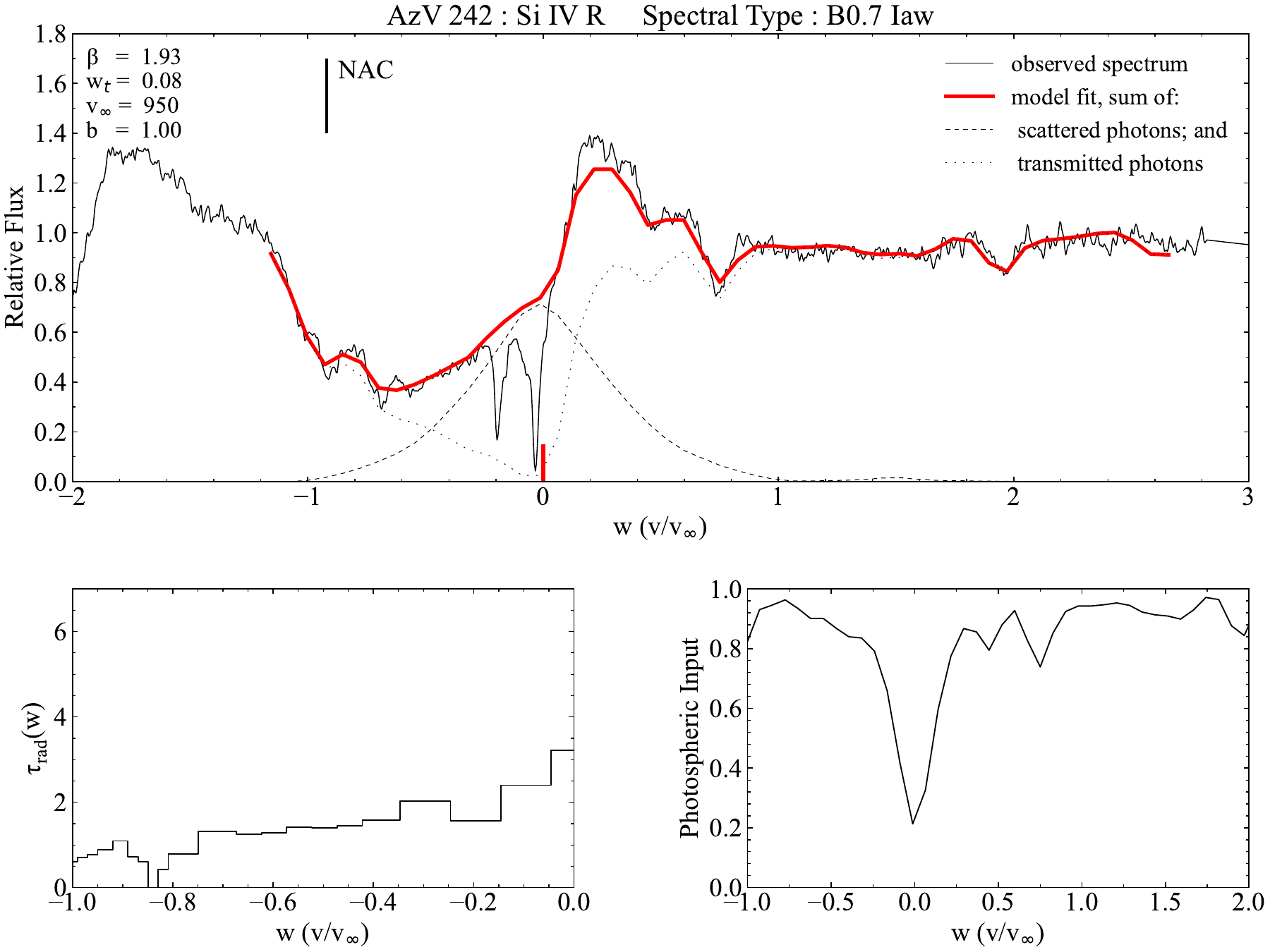} }
 \caption{SEI-derived model fits for (a) blue and (b) red components of the Si \textsc{iv} doublet feature in the UV spectrum of SMC star AzV 242 (spectral type B0.7 Iaw). The location of likely narrow absorption components (NAC) in each doublet element are indicated, each at the same distance from the relevant rest wavelength.}
 \label{fig242}
\end{center}
    \end{figure*}

   \begin{figure*}
\begin{center}
 \subfloat[ ]{\includegraphics[width=3.34in]{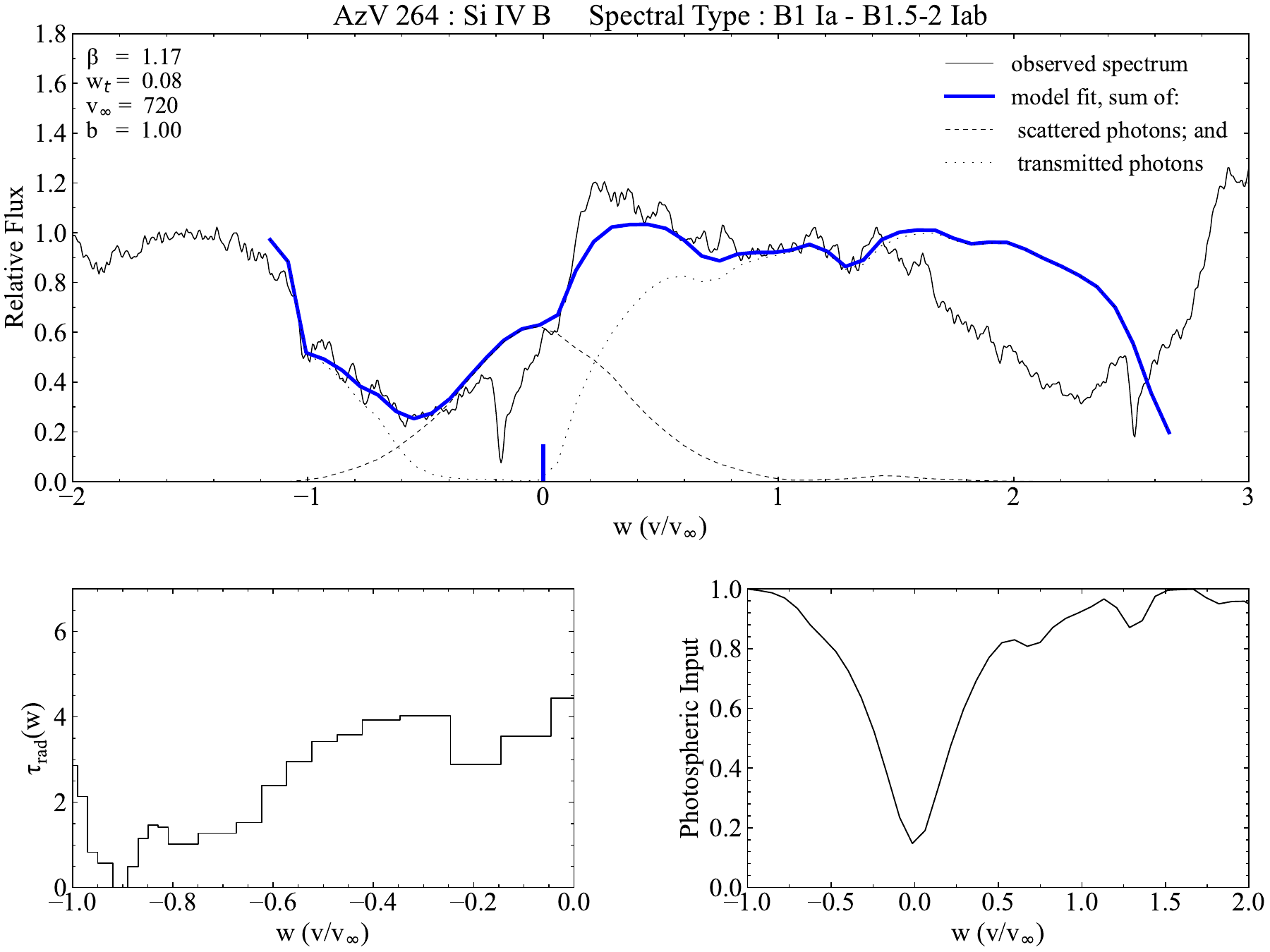} }
 \qquad
 \subfloat[ ]{\includegraphics[width=3.34in]{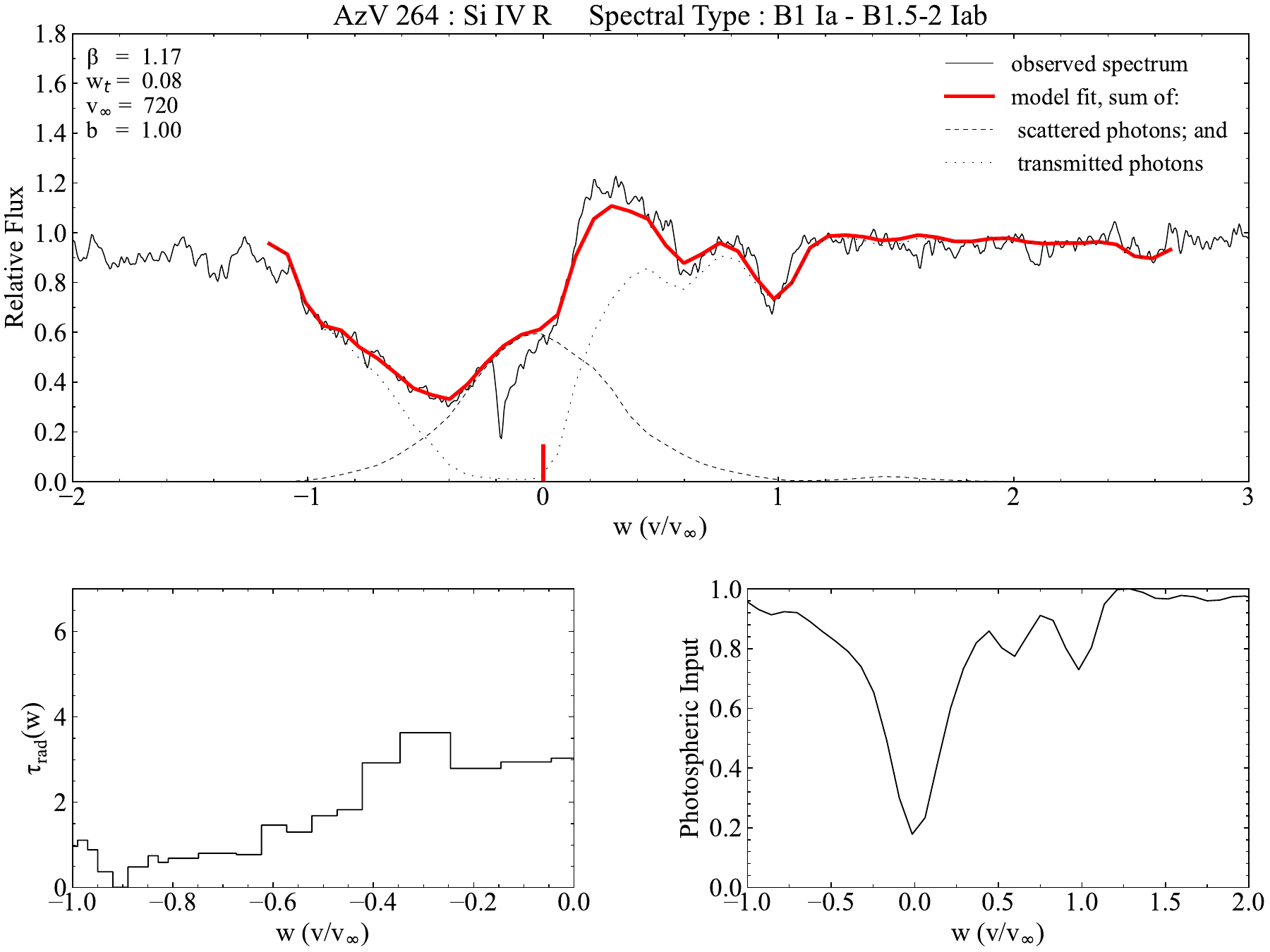} }
 \caption{SEI-derived model fits for (a) blue and (b) red components of the Si \textsc{iv} doublet feature in the UV spectrum of SMC star AzV 264 (spectral type B1 Ia).}
 \label{fig264a}
\end{center}
    \end{figure*}

   \begin{figure*}
\begin{center}
 \subfloat[ ]{\includegraphics[width=3.34in]{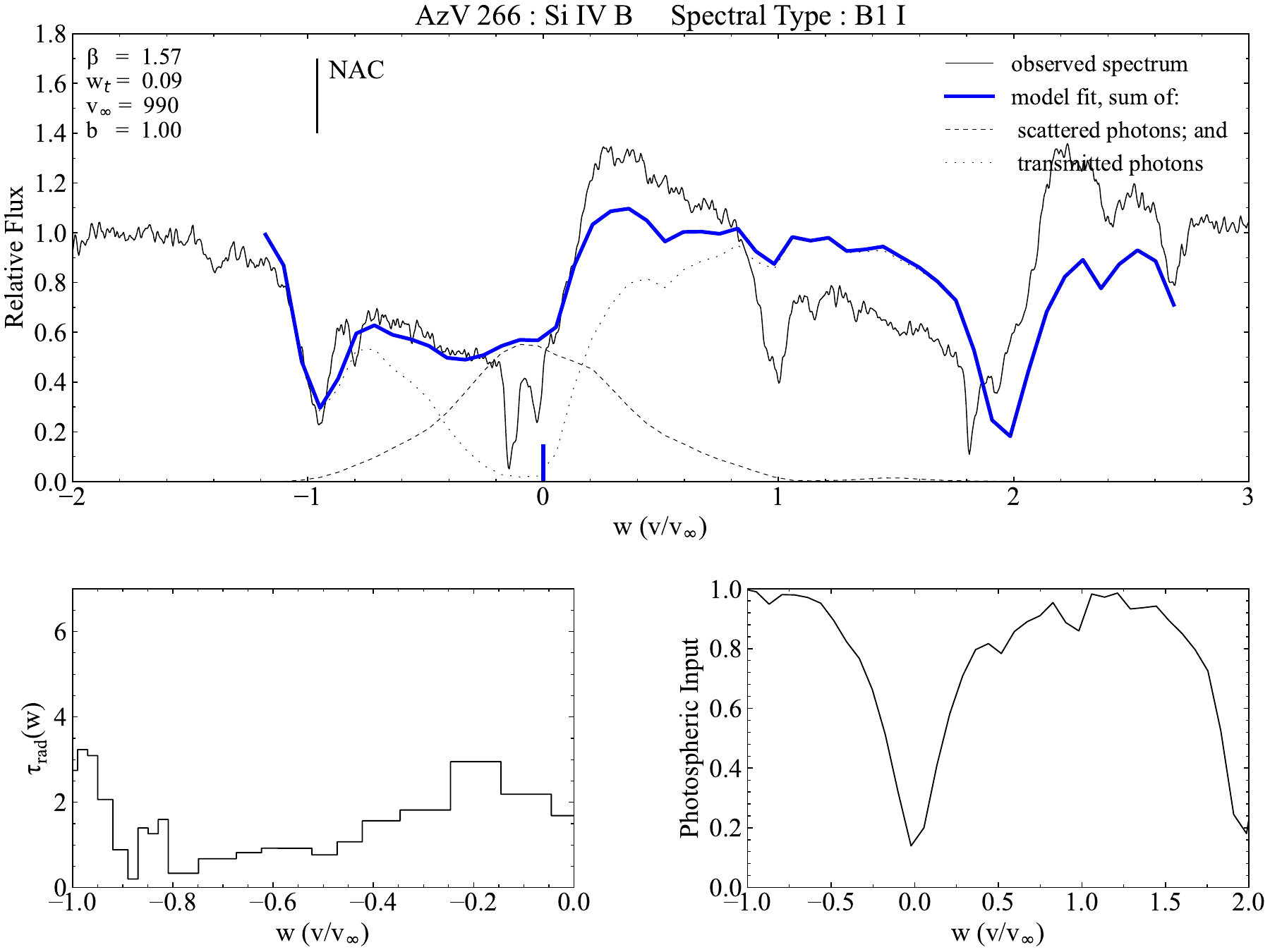} }
 \qquad
 \subfloat[ ]{\includegraphics[width=3.34in]{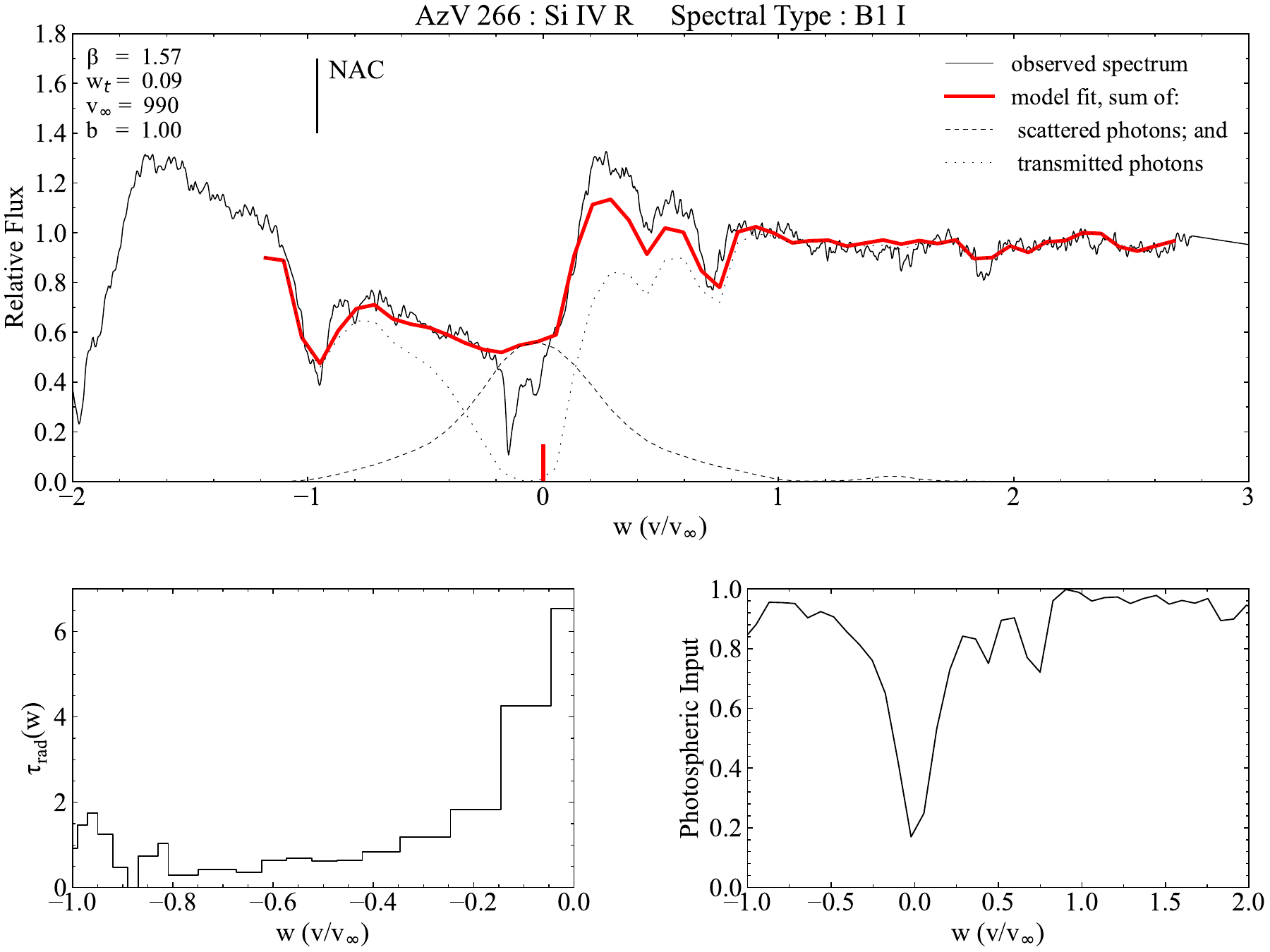} }
 \caption{SEI-derived model fits for (a) blue and (b) red components of the Si \textsc{iv} doublet feature in the UV spectrum of SMC star AzV 266 (spectral type B1 I). There appears to be a very prominent, and deep, narrow absorption component (NAC) feature in both elements of this doublet.}
 \label{fig266}
\end{center}
    \end{figure*}

   \begin{figure*}
\begin{center}
 \subfloat[ ]{\includegraphics[width=3.34in]{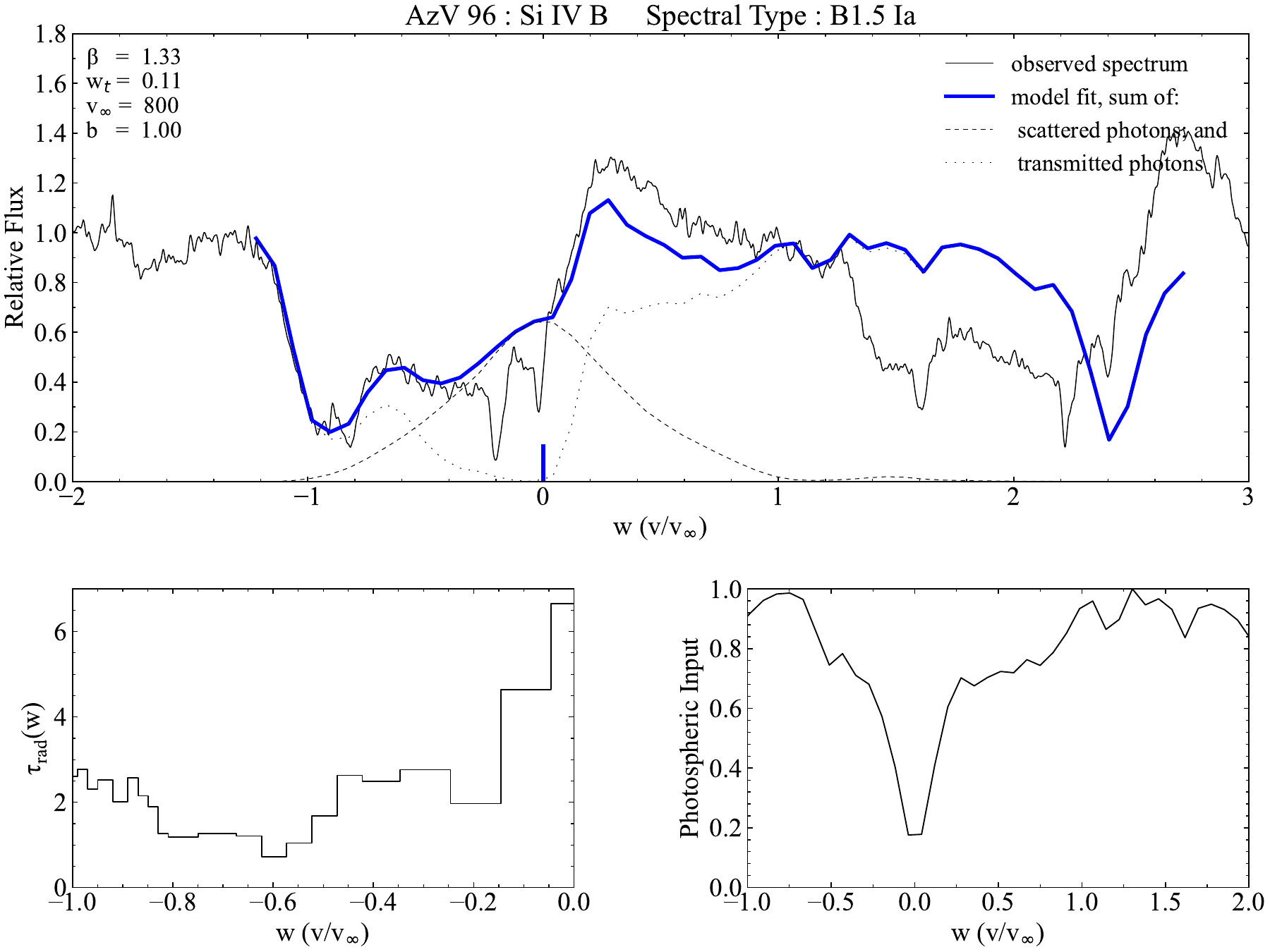} }
 \qquad
 \subfloat[ ]{\includegraphics[width=3.34in]{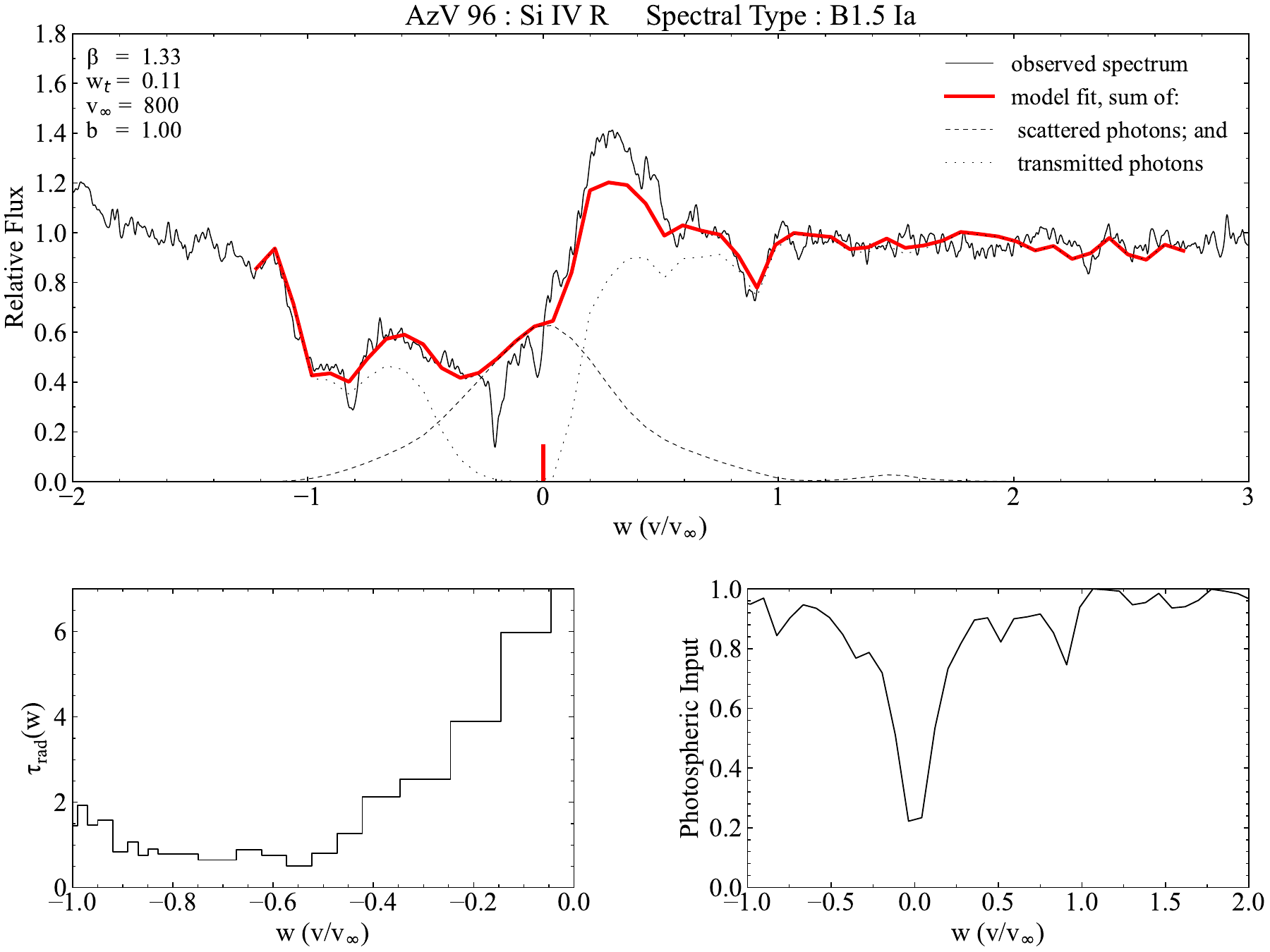} }
 \caption{SEI-derived model fits for (a) blue and (b) red components of the Si \textsc{iv} doublet feature in the UV spectrum of SMC star AzV 96 (spectral type B1 I/B1.5 Ia). Note that this spectrum is taken from 2 contiguous observation of this star and excludes a third observation which forms part of the ULLYSES spectrum, as discussed in the main text. The third spectrum is presented in Fig. \ref{fig96a} for comparison.}
 \label{fig96}
\end{center}
    \end{figure*}

   \begin{figure*}
\begin{center}
 \subfloat[ ]{\includegraphics[width=3.34in]{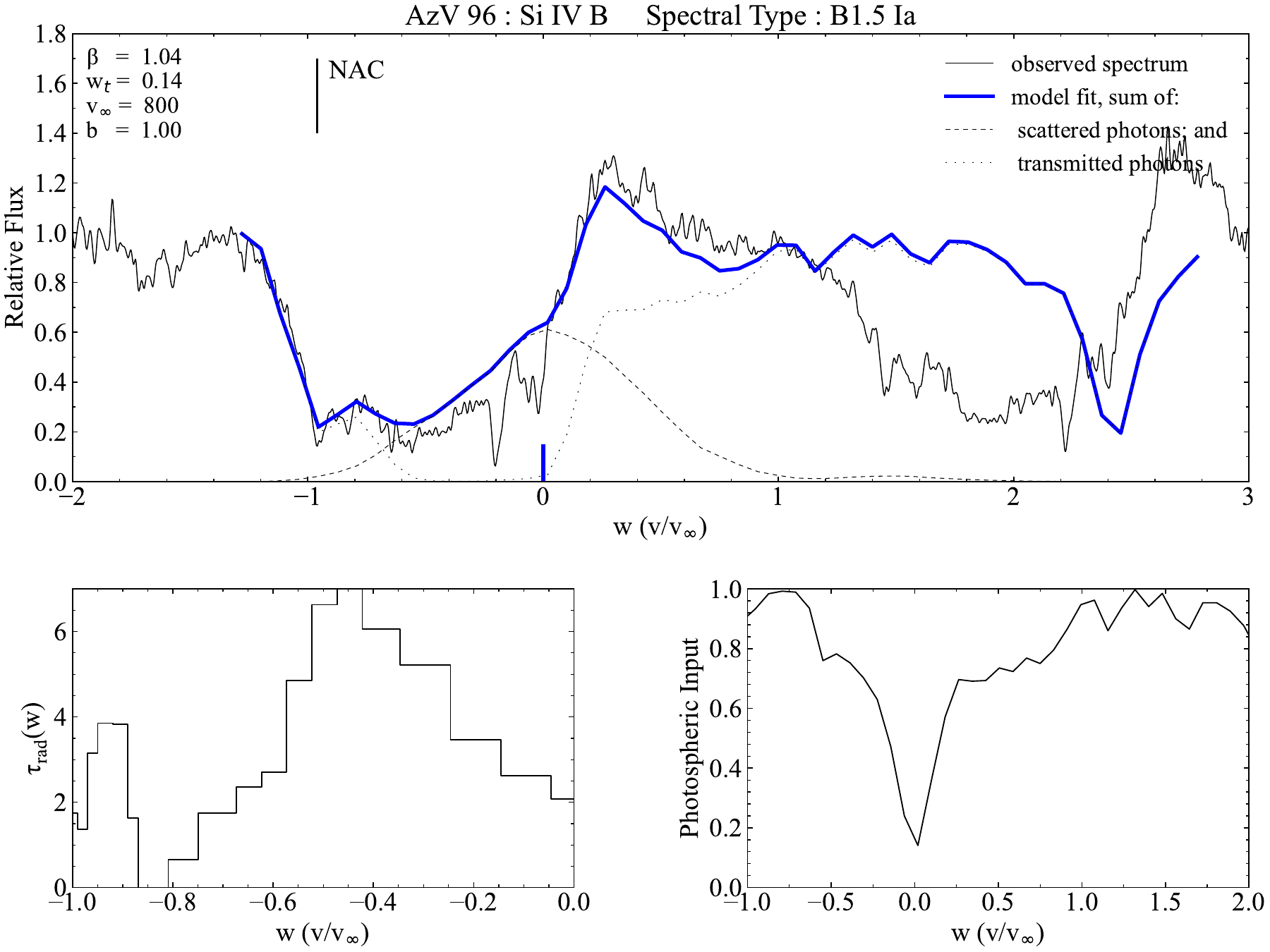} }
 \qquad
 \subfloat[ ]{\includegraphics[width=3.34in]{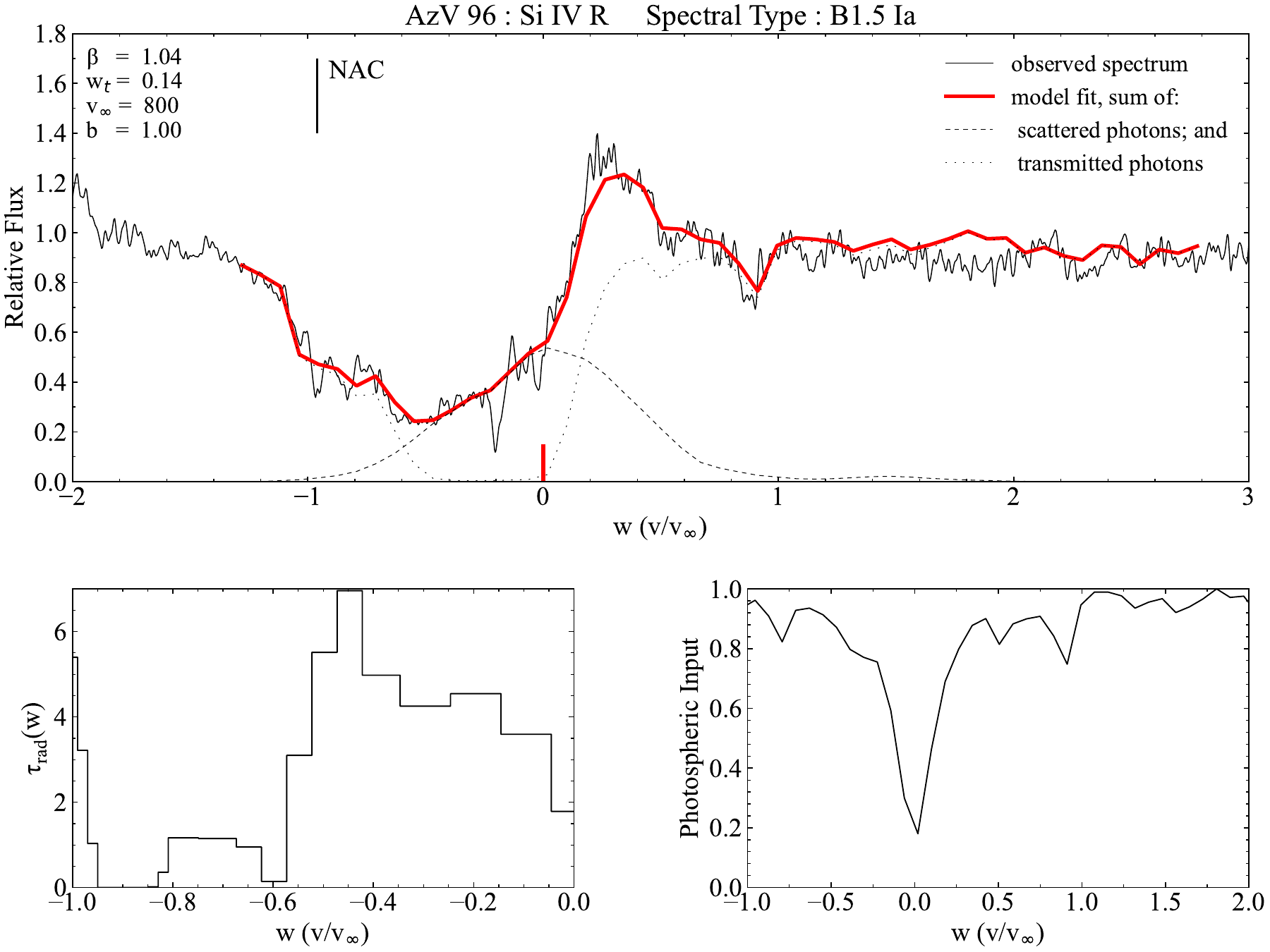} }
 \caption{SEI-derived model fits for (a) blue and (b) red components of the Si \textsc{iv} doublet feature in the UV spectrum of SMC star AzV 96 (spectral type B1 I/B1.5 Ia). Note that this spectrum is taken from a separate observation obtained several days after those shown in Fig. \ref{fig96}. The blueward migration of an apparent discrete absorption component (DAC) feature may be observed in both doublet elements. The location of likely narrow absorption components (NAC) in each doublet element are indicated, each at the same distance from the relevant rest wavelength. This feature appears in the blue, but not the red component of the preceding plots for this star.}
 \label{fig96a}
\end{center}
    \end{figure*}

   \begin{figure*}
\begin{center}
 \subfloat[ ]{\includegraphics[width=3.34in]{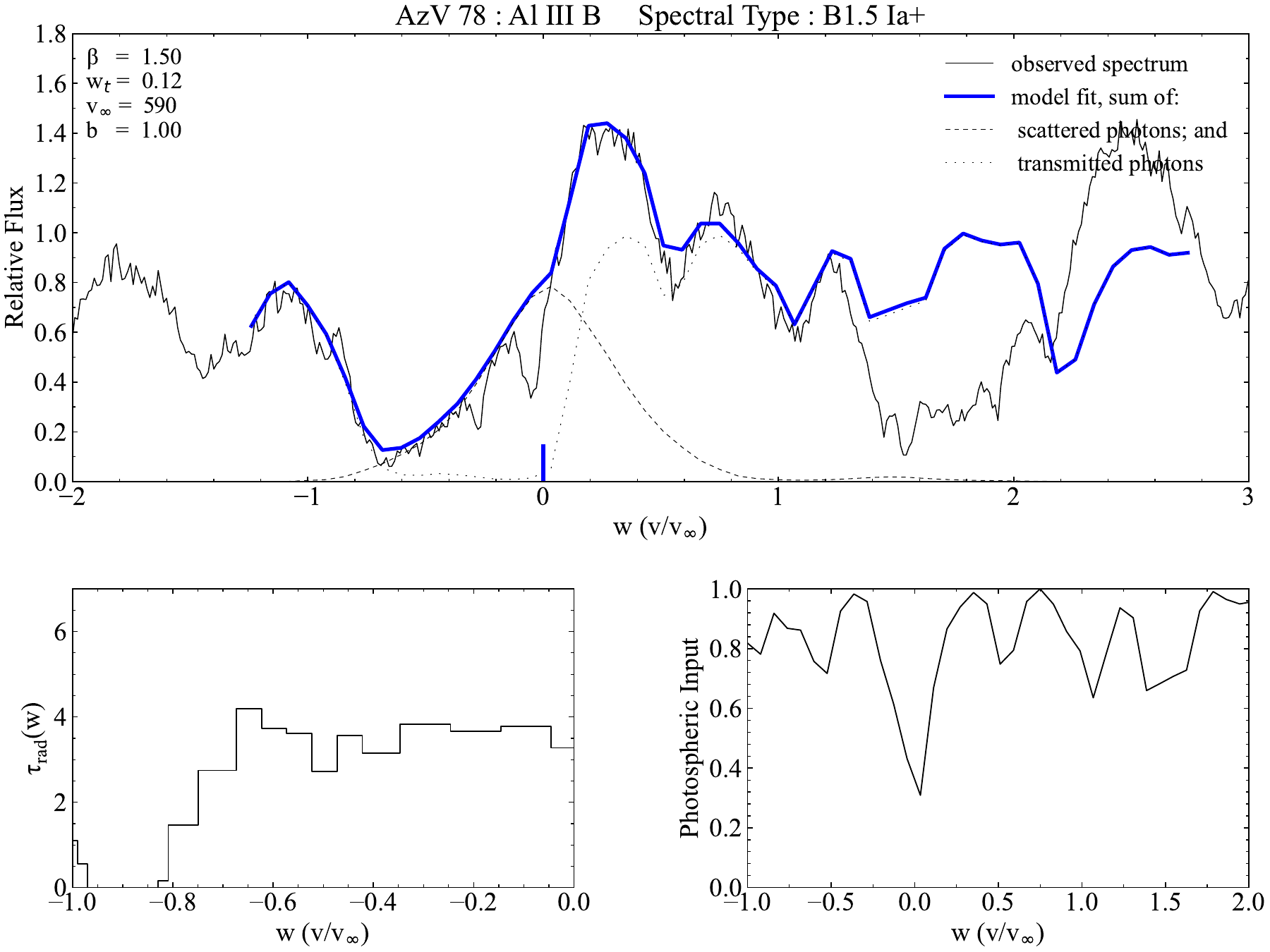} }
 \qquad
 \subfloat[ ]{\includegraphics[width=3.34in]{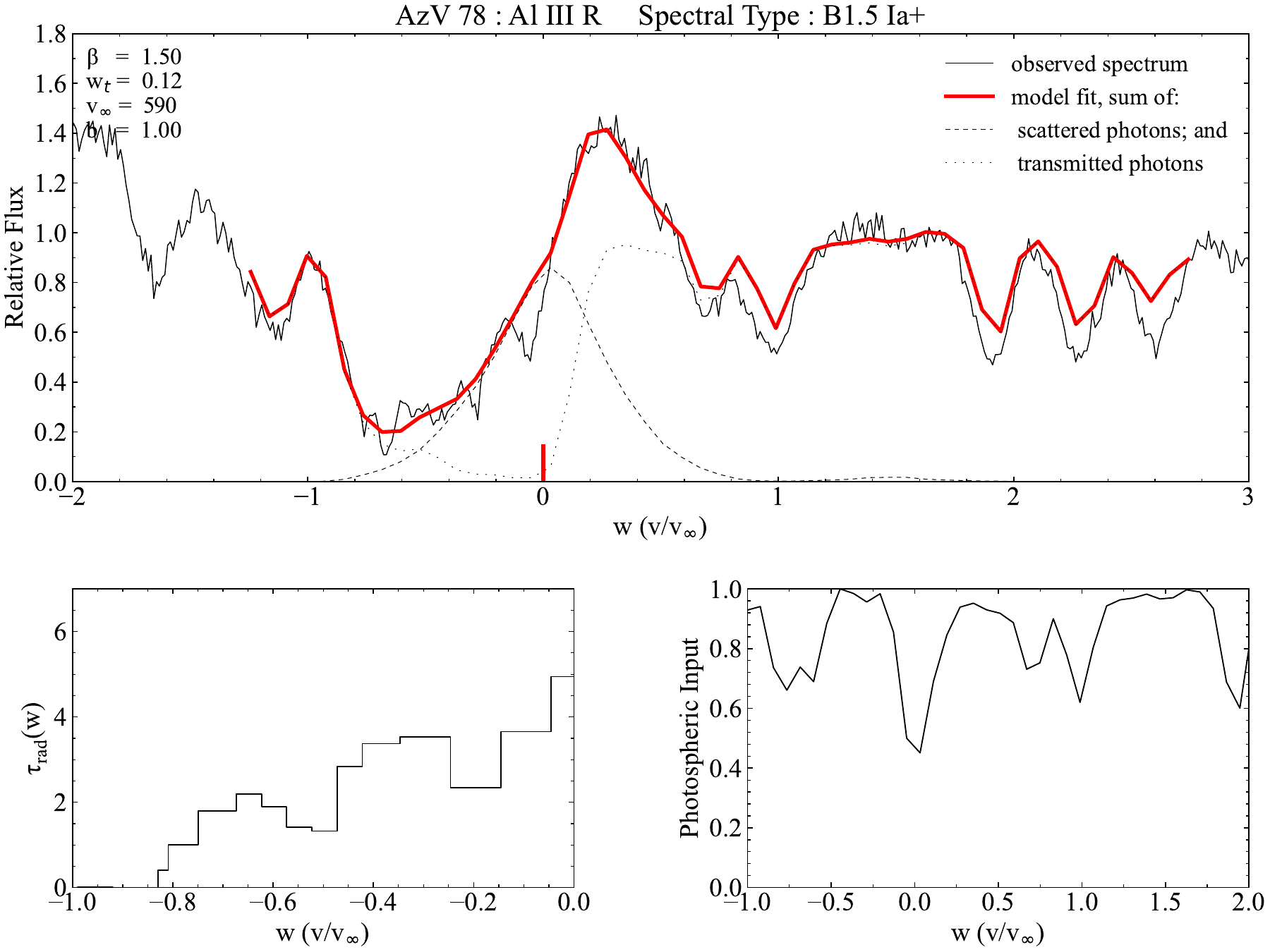} }
 \caption{SEI-derived model fits for (a) blue and (b) red components of the Al \textsc{iii} doublet feature in the UV spectrum of SMC star AzV 78 (spectral type B1.5 Ia+).}
 \label{fig78}
\end{center}
    \end{figure*}

   \begin{figure*}
\begin{center}
 \subfloat[ ]{\includegraphics[width=3.34in]{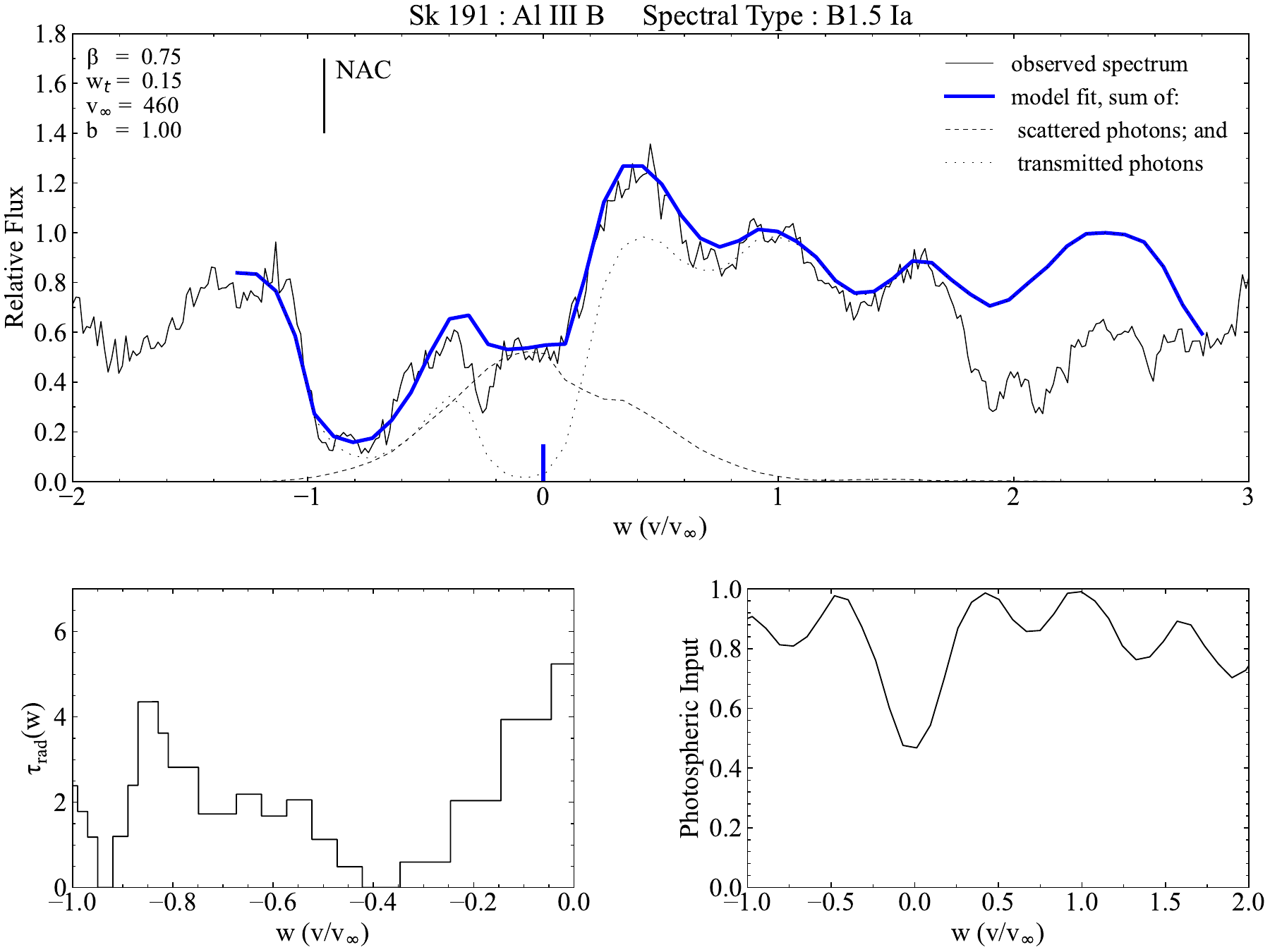} }
 \qquad
 \subfloat[ ]{\includegraphics[width=3.34in]{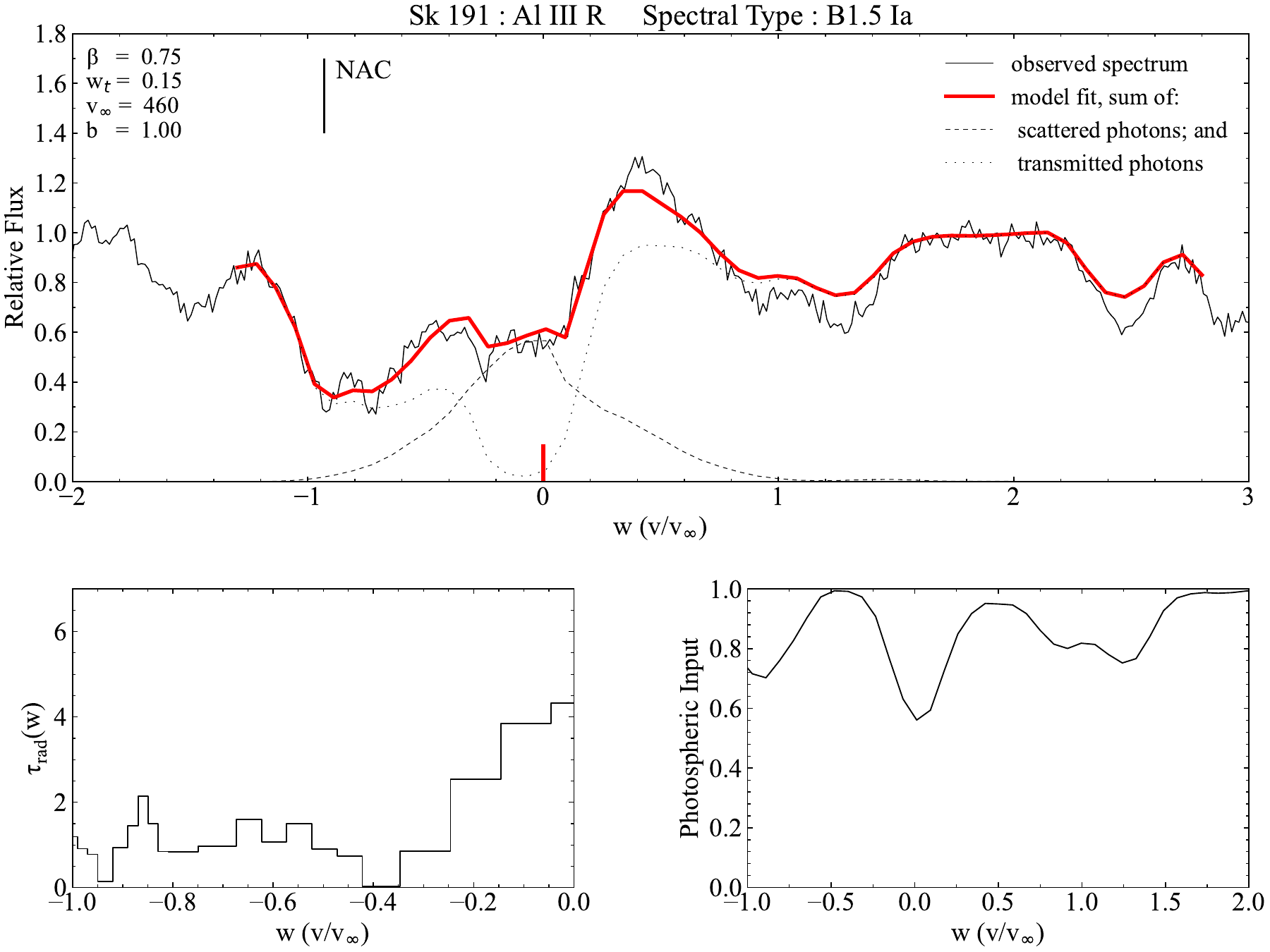} }
 \caption{SEI-derived model fits for (a) blue and (b) red components of the Al \textsc{iii} doublet feature in the UV spectrum of SMC star Sk 191 (spectral type B1.5 Ia). The location of likely narrow absorption components (NAC) in each doublet element are indicated, each at the same distance from the relevant rest wavelength.}
 \label{fig191a}
\end{center}
    \end{figure*}

   \begin{figure*}
\begin{center}
 \subfloat[ ]{\includegraphics[width=3.34in]{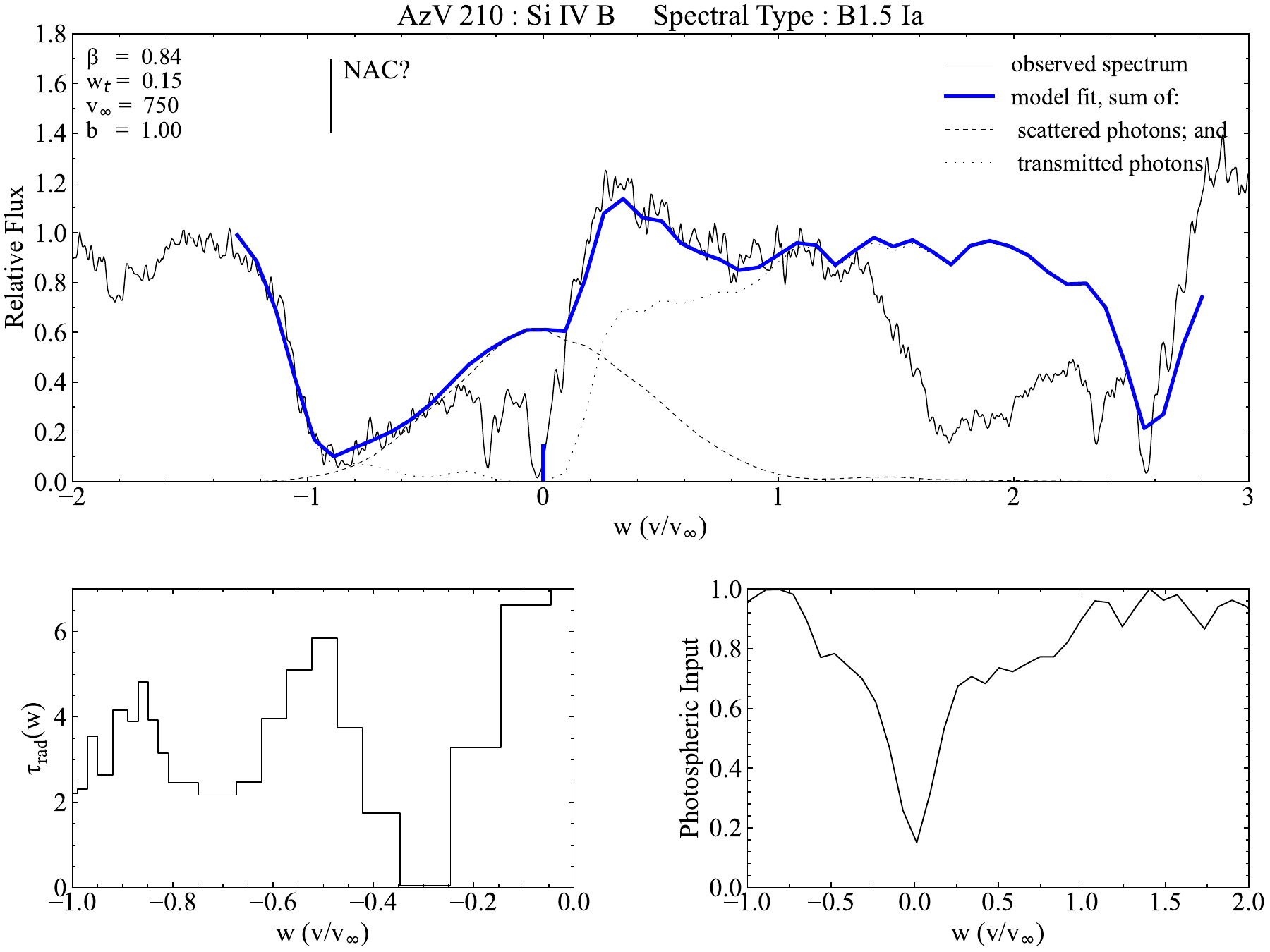} }
 \qquad
 \subfloat[ ]{\includegraphics[width=3.34in]{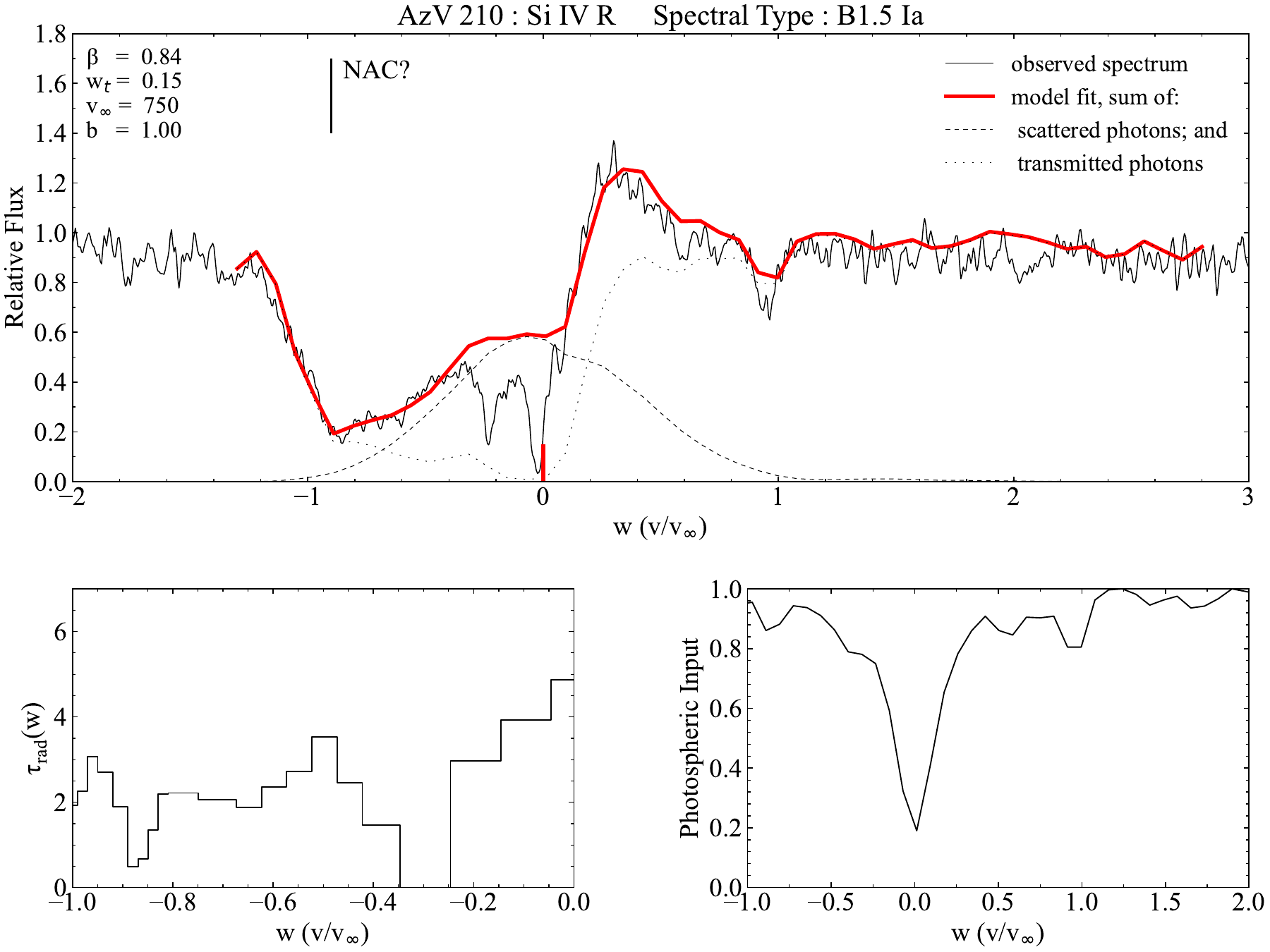} }
 \caption{SEI-derived model fits for (a) blue and (b) red components of the Si \textsc{iv} doublet feature in the UV spectrum of SMC star AzV 210 (spectral type B1.5 Ia). The location of possible narrow absorption components (NAC) in each doublet element are indicated, each at the same distance from the relevant rest wavelength.}
 \label{fig210}
\end{center}
    \end{figure*}

   \begin{figure*}
\begin{center}
 \subfloat[ ]{\includegraphics[width=3.34in]{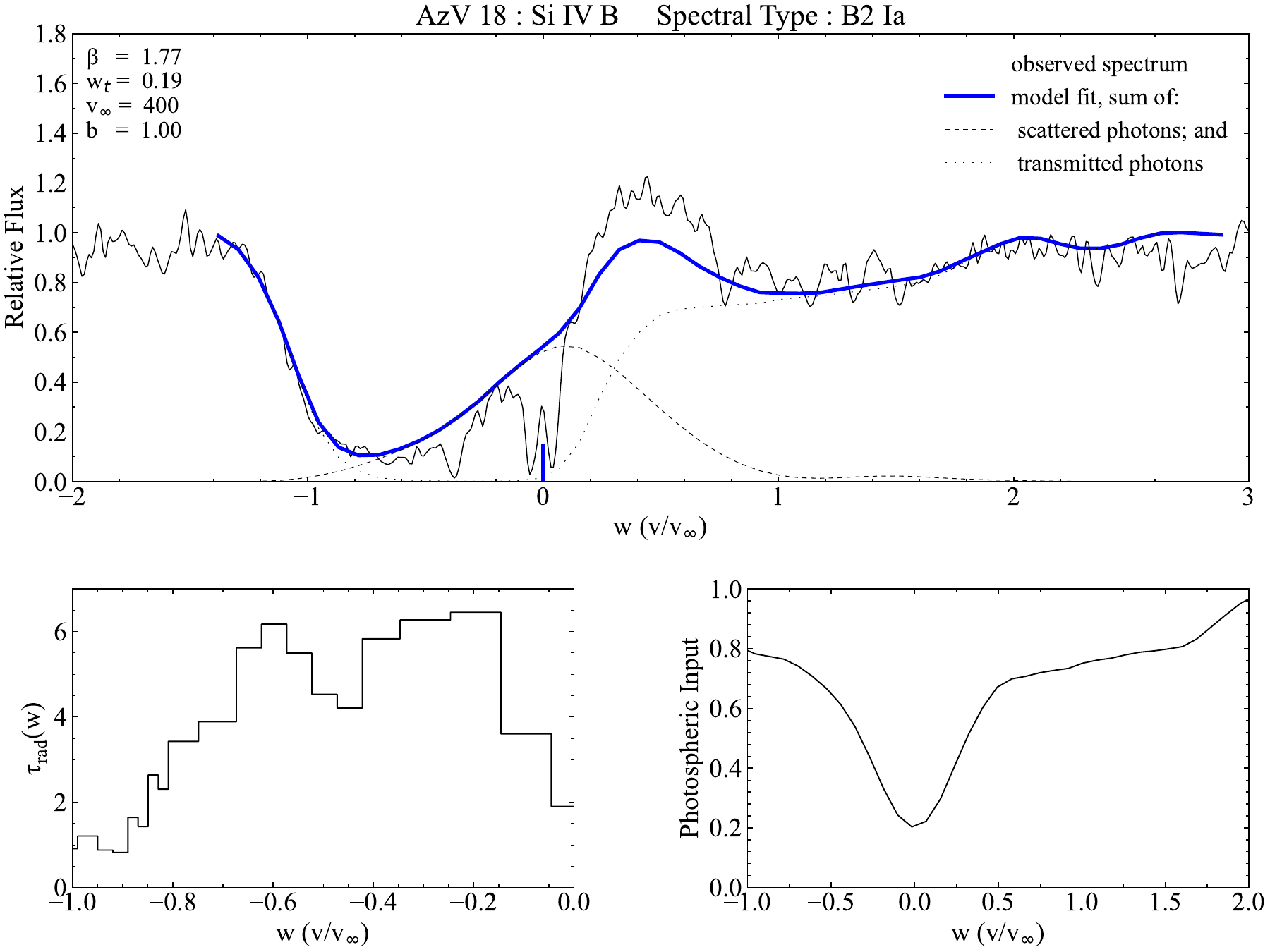} }
 \qquad
 \subfloat[ ]{\includegraphics[width=3.34in]{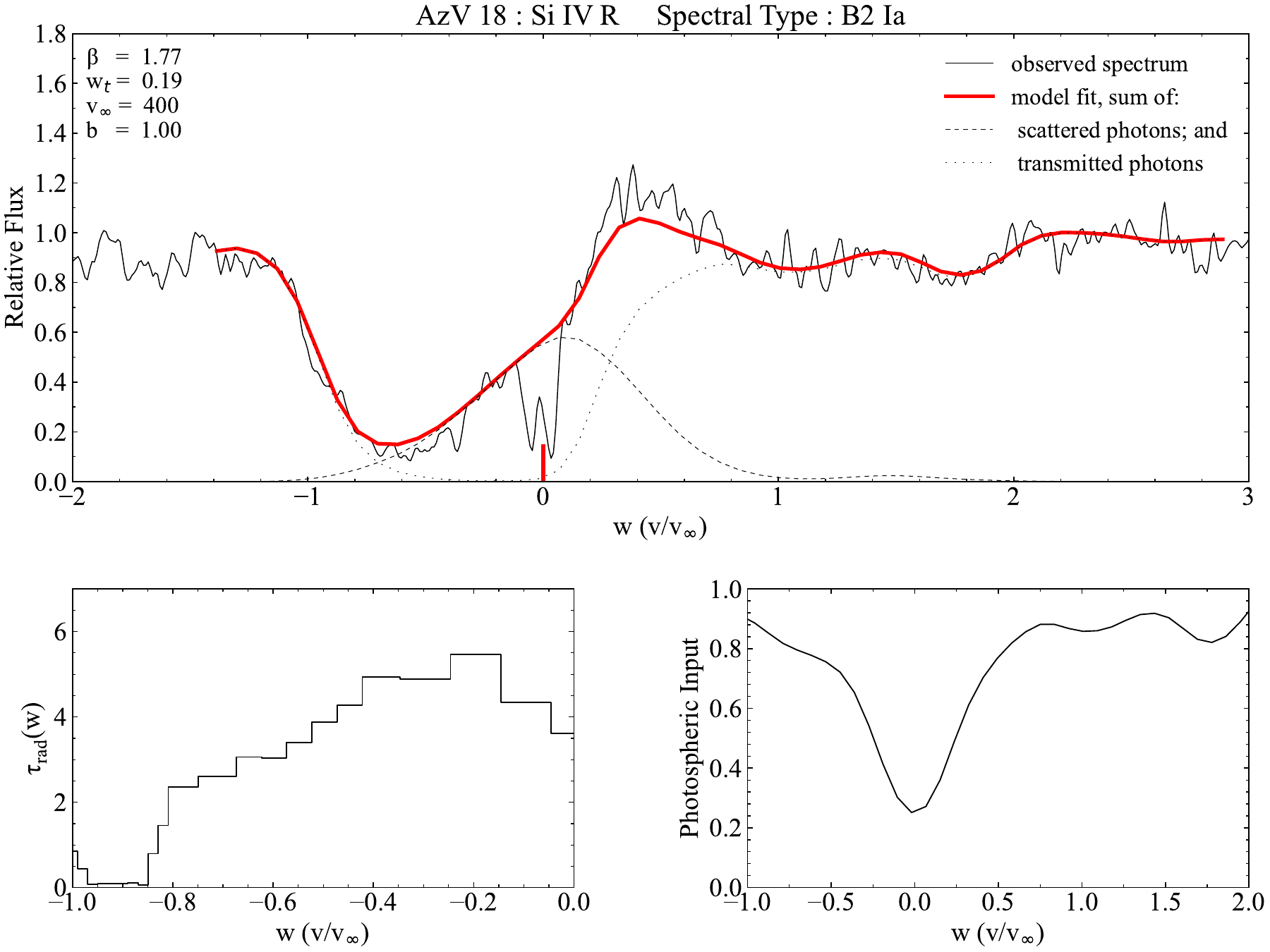} }
 \caption{SEI-derived model fits for (a) blue and (b) red components of the Si \textsc{iv} doublet feature in the UV spectrum of SMC star AzV 18 (spectral type B2 Ia).}
 \label{fig18}
\end{center}
    \end{figure*}

   \begin{figure*}
\begin{center}
 \subfloat[ ]{\includegraphics[width=3.34in]{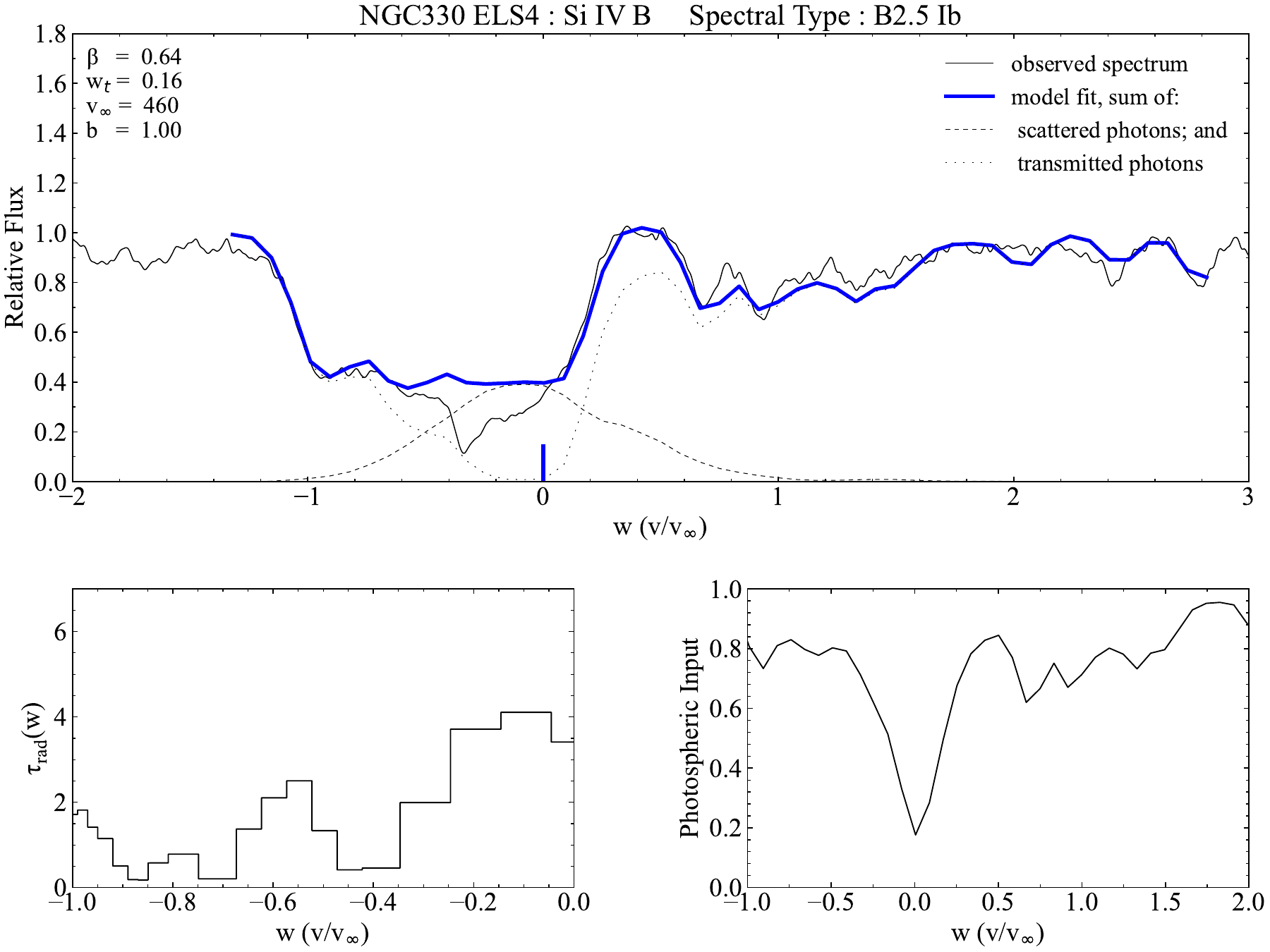} }
 \qquad
 \subfloat[ ]{\includegraphics[width=3.34in]{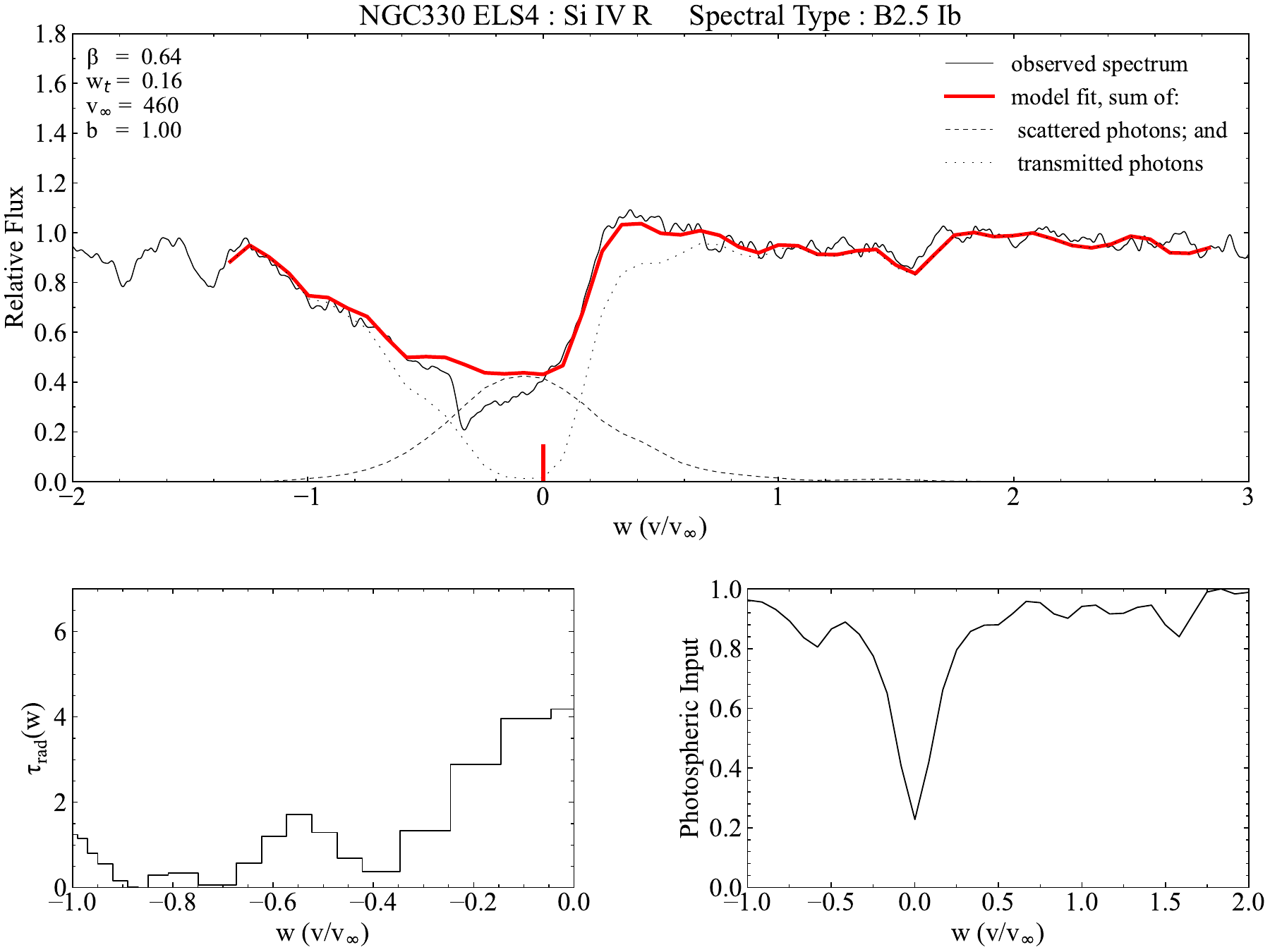} }
 \caption{SEI-derived model fits for (a) blue and (b) red components of the Si \textsc{iv} doublet feature in the UV spectrum of SMC star NGC 330 ELS 4 (spectral type B2.5 Ib).}
 \label{fig3304}
\end{center}
    \end{figure*}
    
   \begin{figure*}
\begin{center}
 \subfloat[ ]{\includegraphics[width=3.34in]{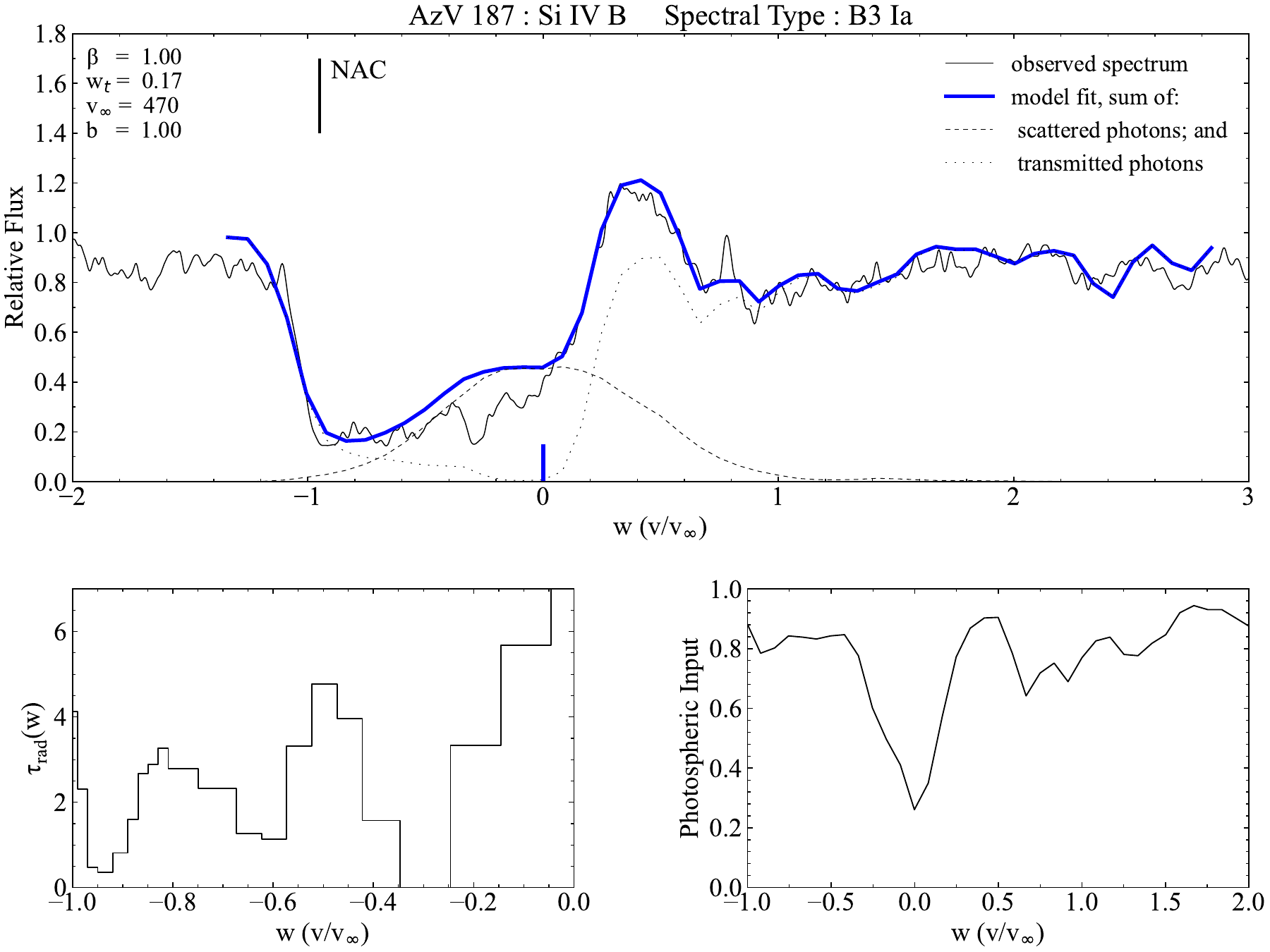} }
 \qquad
 \subfloat[ ]{\includegraphics[width=3.34in]{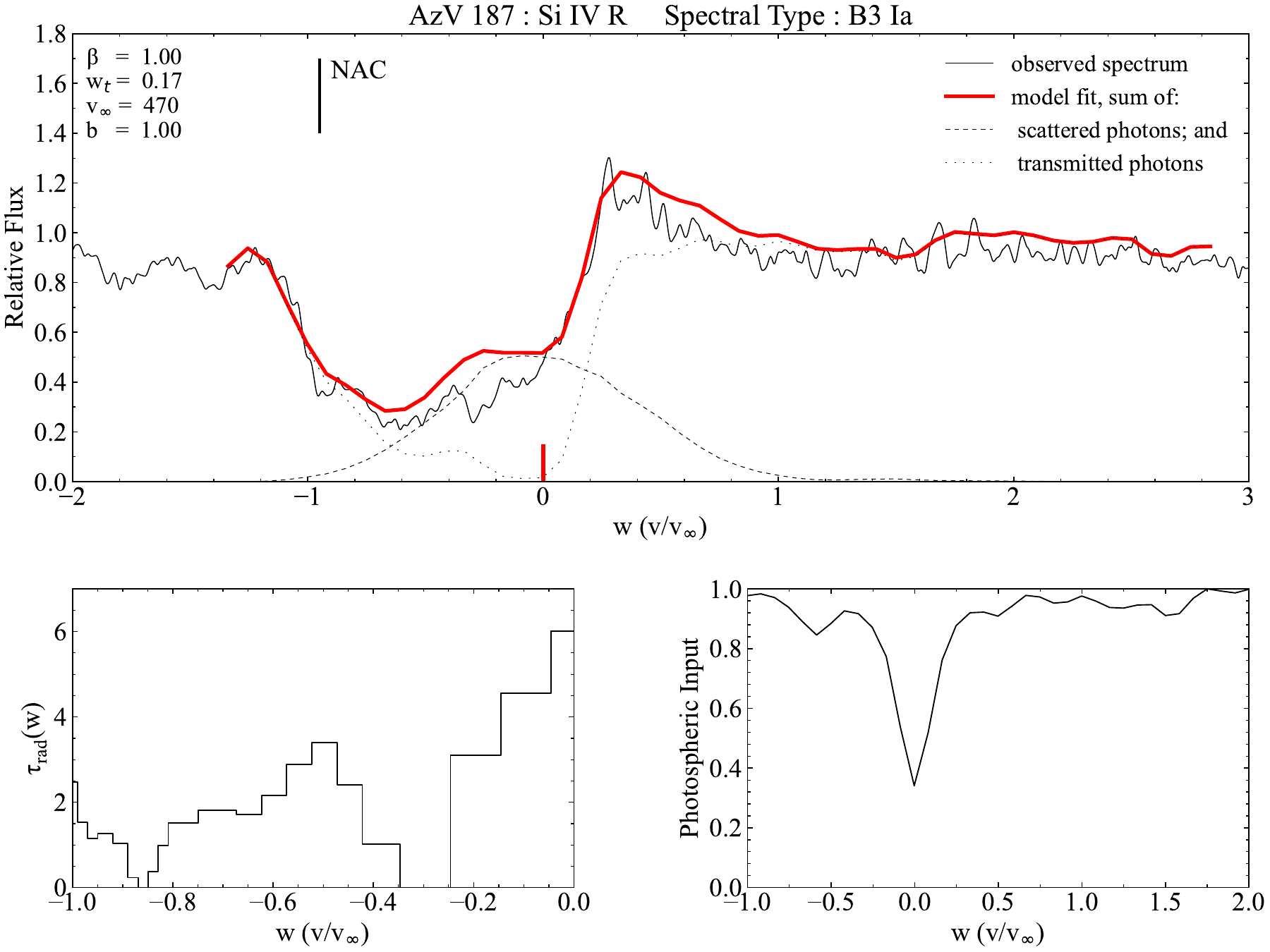} }
 \caption{SEI-derived model fits for (a) blue and (b) red components of the Si \textsc{iv} doublet feature in the UV spectrum of SMC star AzV 187 (spectral type B3 Ia). The location of likely narrow absorption components (NAC) in each doublet element are indicated, each at the same distance from the relevant rest wavelength.}
 \label{fig187}
\end{center}
    \end{figure*}

   \begin{figure*}
\begin{center}
 \subfloat[ ]{\includegraphics[width=3.34in]{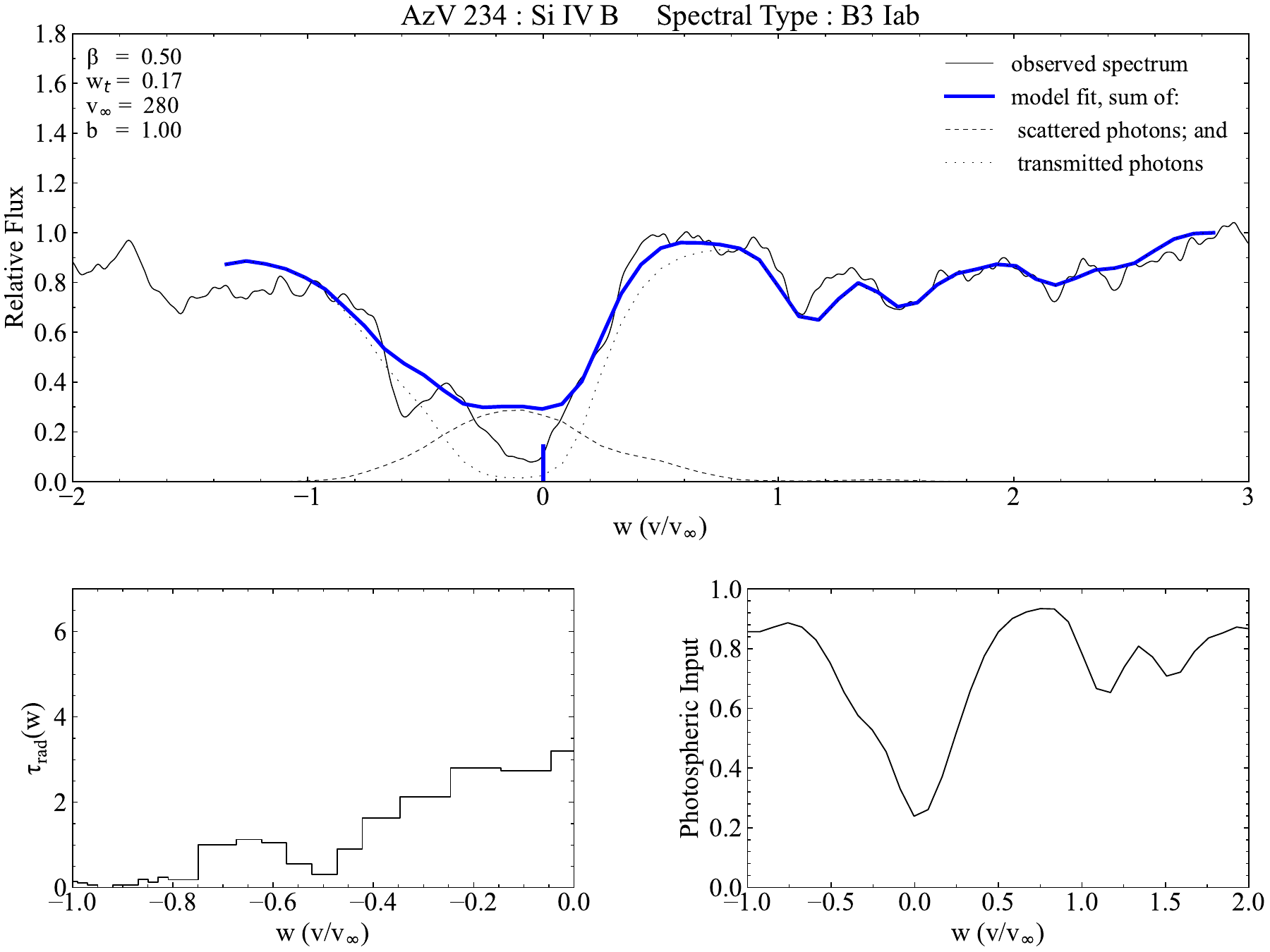} }
 \qquad
 \subfloat[ ]{\includegraphics[width=3.34in]{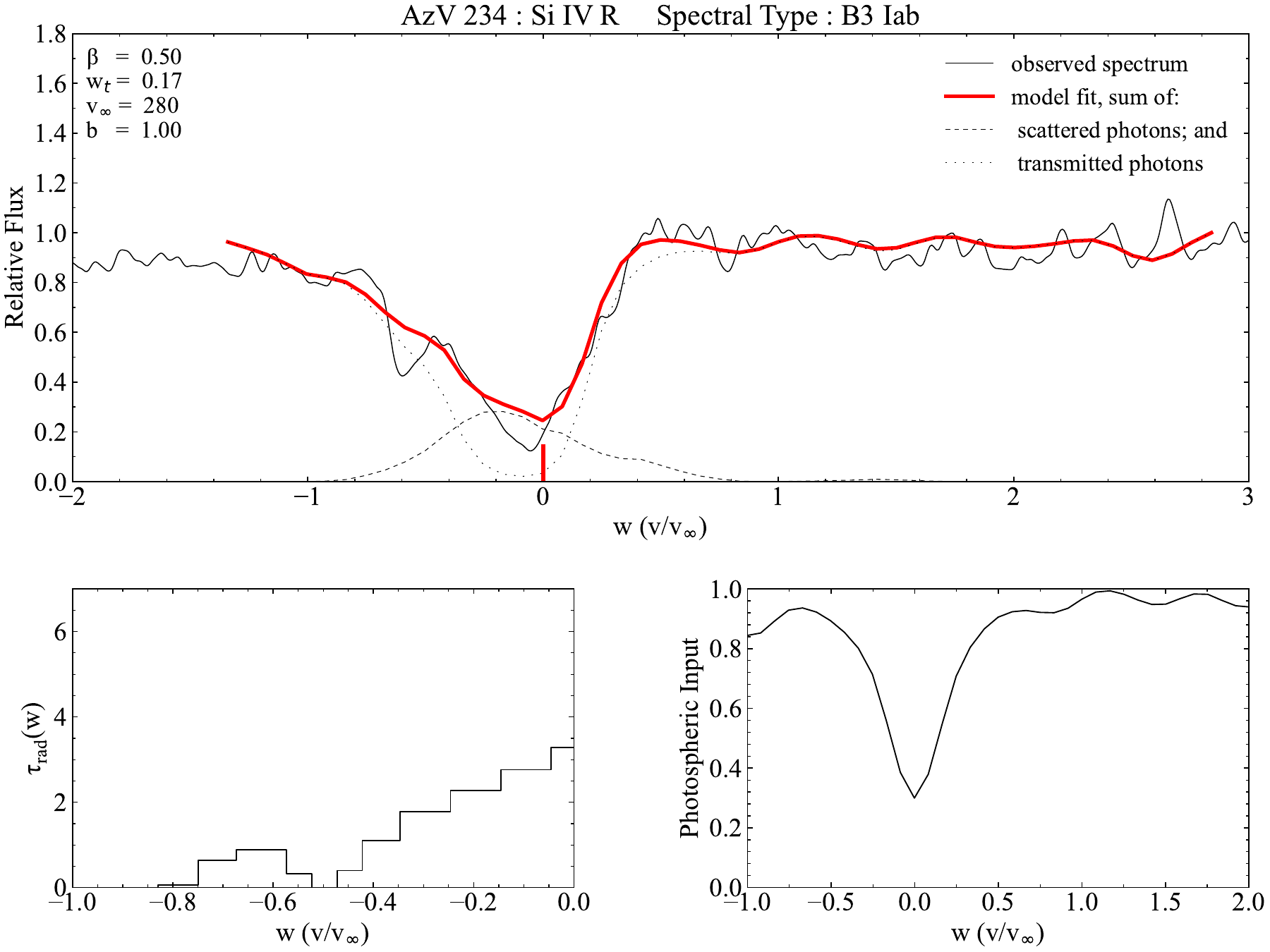} }
 \caption{SEI-derived model fits for (a) blue and (b) red components of the Si \textsc{iv} doublet feature in the UV spectrum of SMC star AzV 234 (spectral type B3 Iab).}
 \label{fig234}
\end{center}
    \end{figure*}

   \begin{figure*}
\begin{center}
 \subfloat[ ]{\includegraphics[width=3.34in]{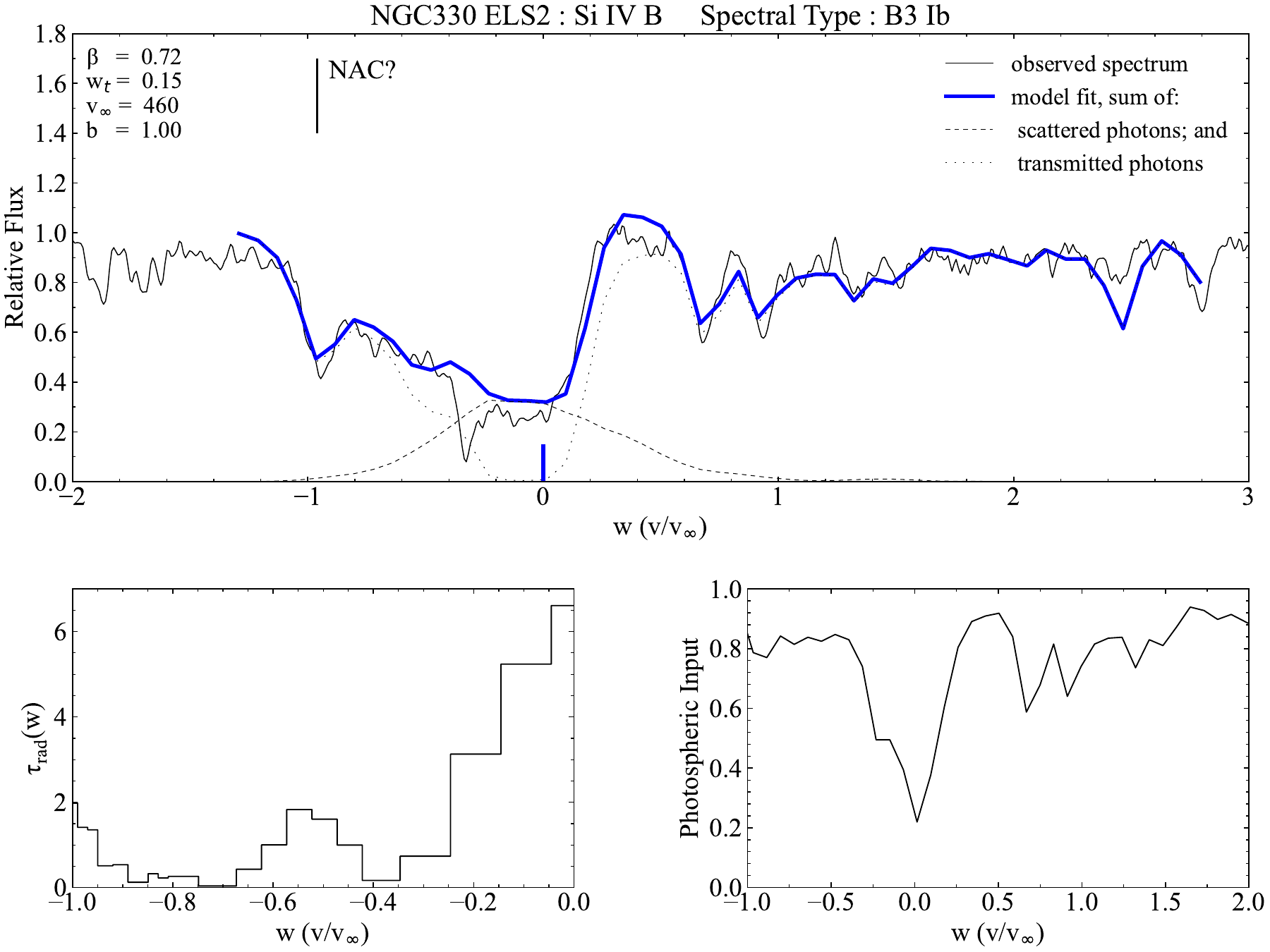} }
 \qquad
 \subfloat[ ]{\includegraphics[width=3.34in]{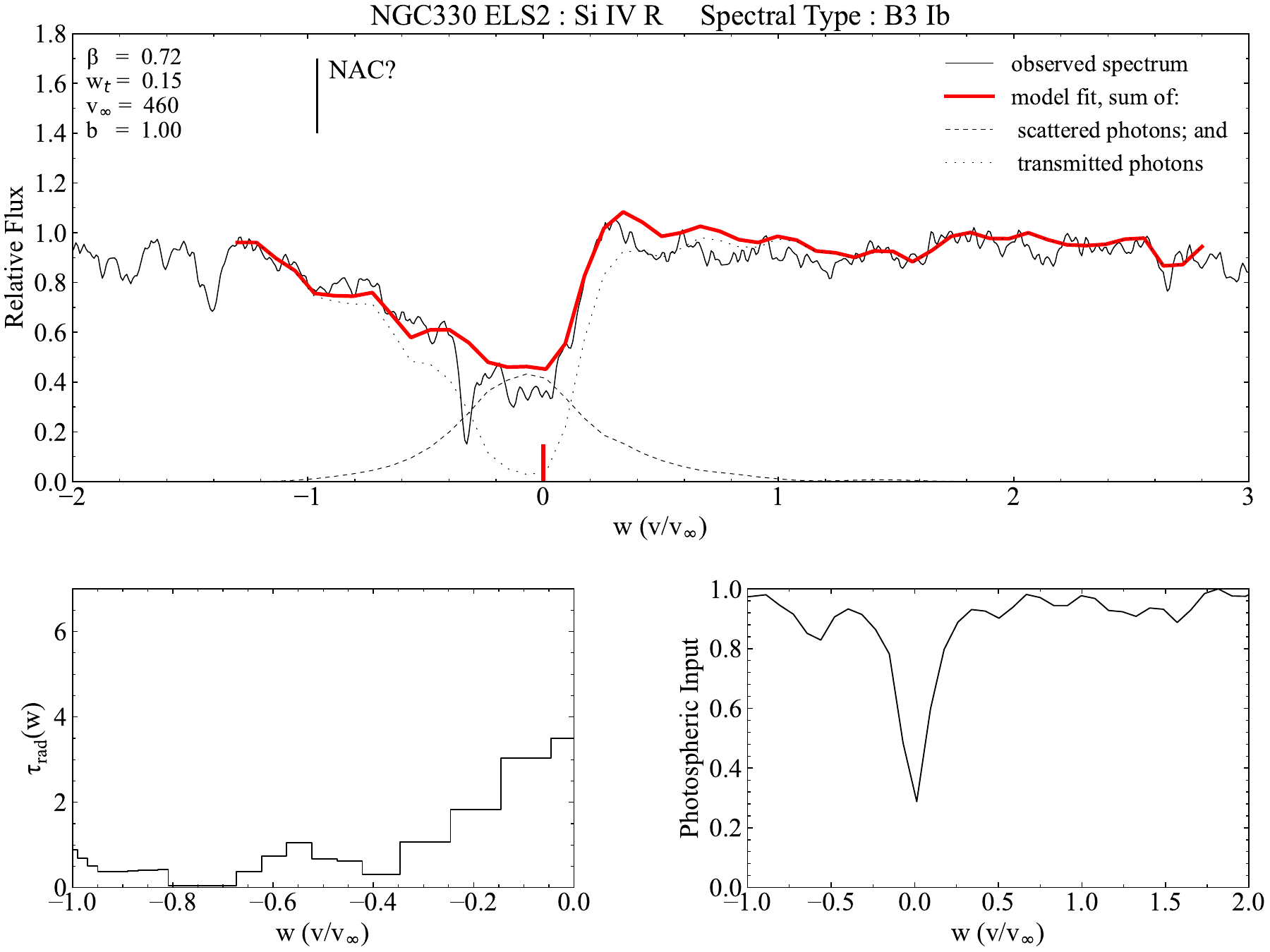} }
 \caption{SEI-derived model fits for (a) blue and (b) red components of the Si \textsc{iv} doublet feature in the UV spectrum of SMC star NGC 330 ELS 2 (spectral type B3 Ib). The location of possible narrow absorption components (NAC) in each doublet element are indicated, each at the same distance from the relevant rest wavelength.}
 \label{fig3302}
\end{center}
    \end{figure*}

   \begin{figure*}
\begin{center}
 \subfloat[ ]{\includegraphics[width=3.34in]{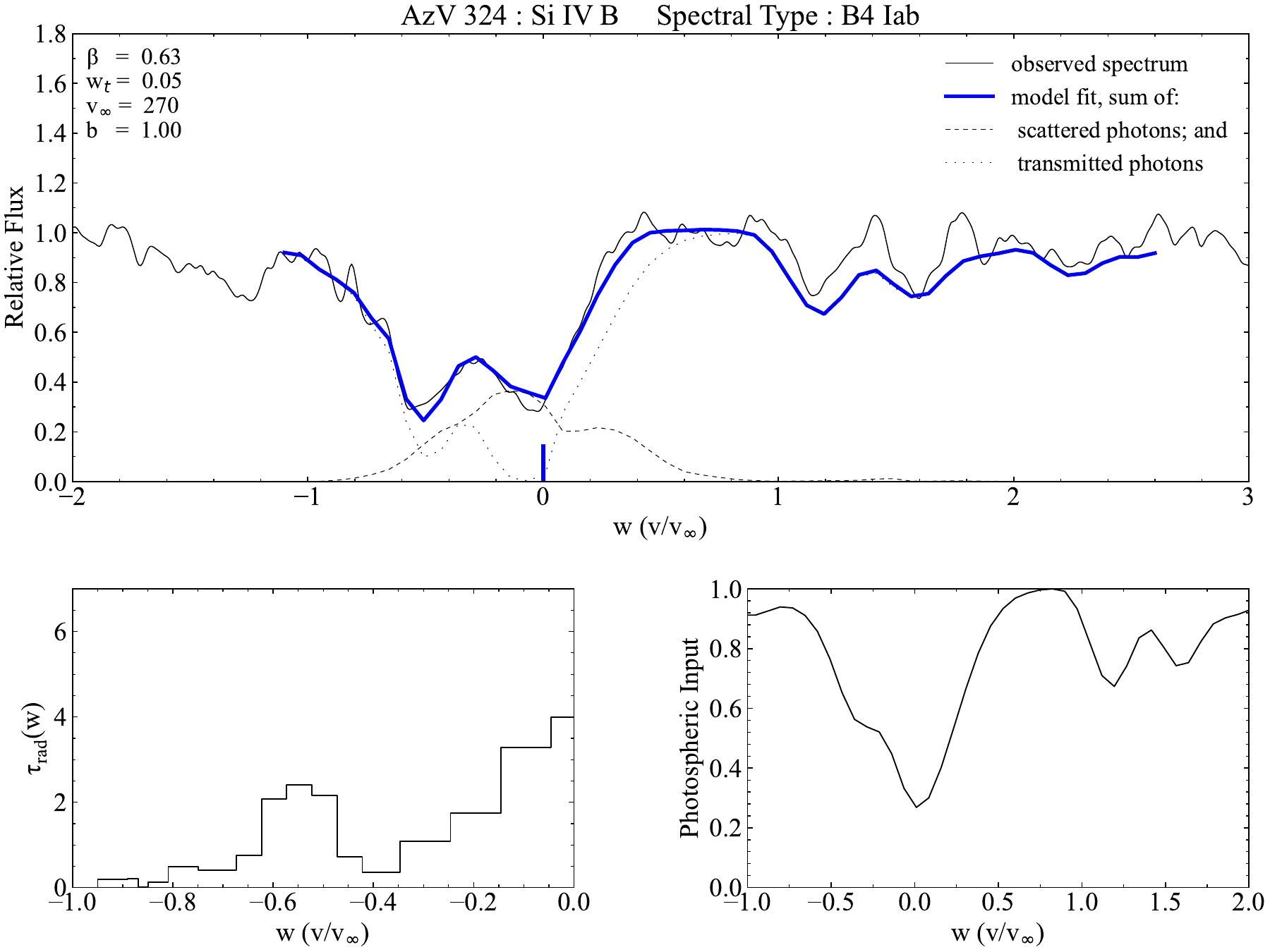} }
 \qquad
 \subfloat[ ]{\includegraphics[width=3.34in]{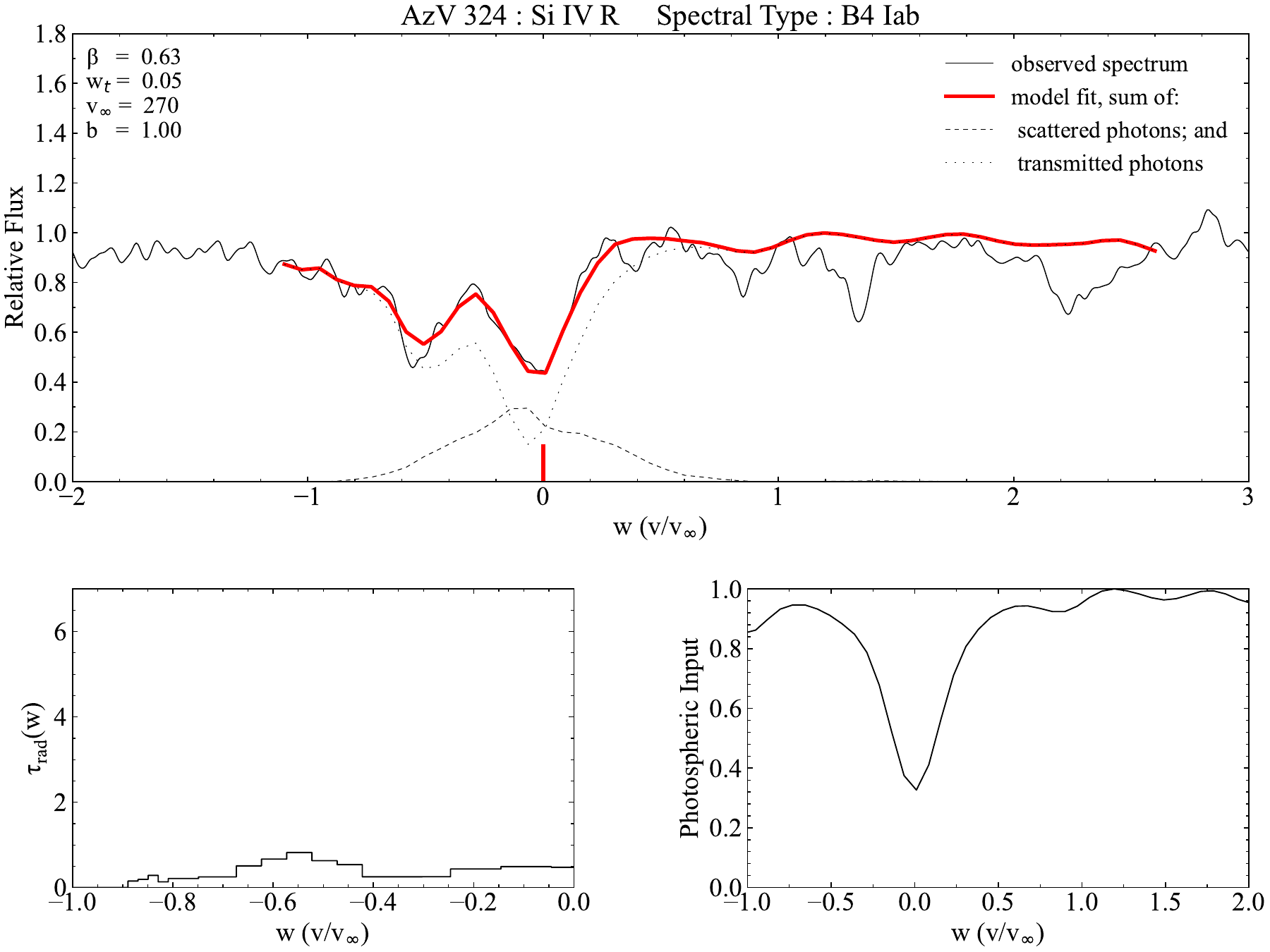} }
 \caption{SEI-derived model fits for (a) blue and (b) red components of the Si \textsc{iv} doublet feature in the UV spectrum of SMC star AzV 324 (spectral type B4 Iab).}
 \label{fig324}
\end{center}
    \end{figure*}

   \begin{figure*}
\begin{center}
 \subfloat[ ]{\includegraphics[width=3.34in]{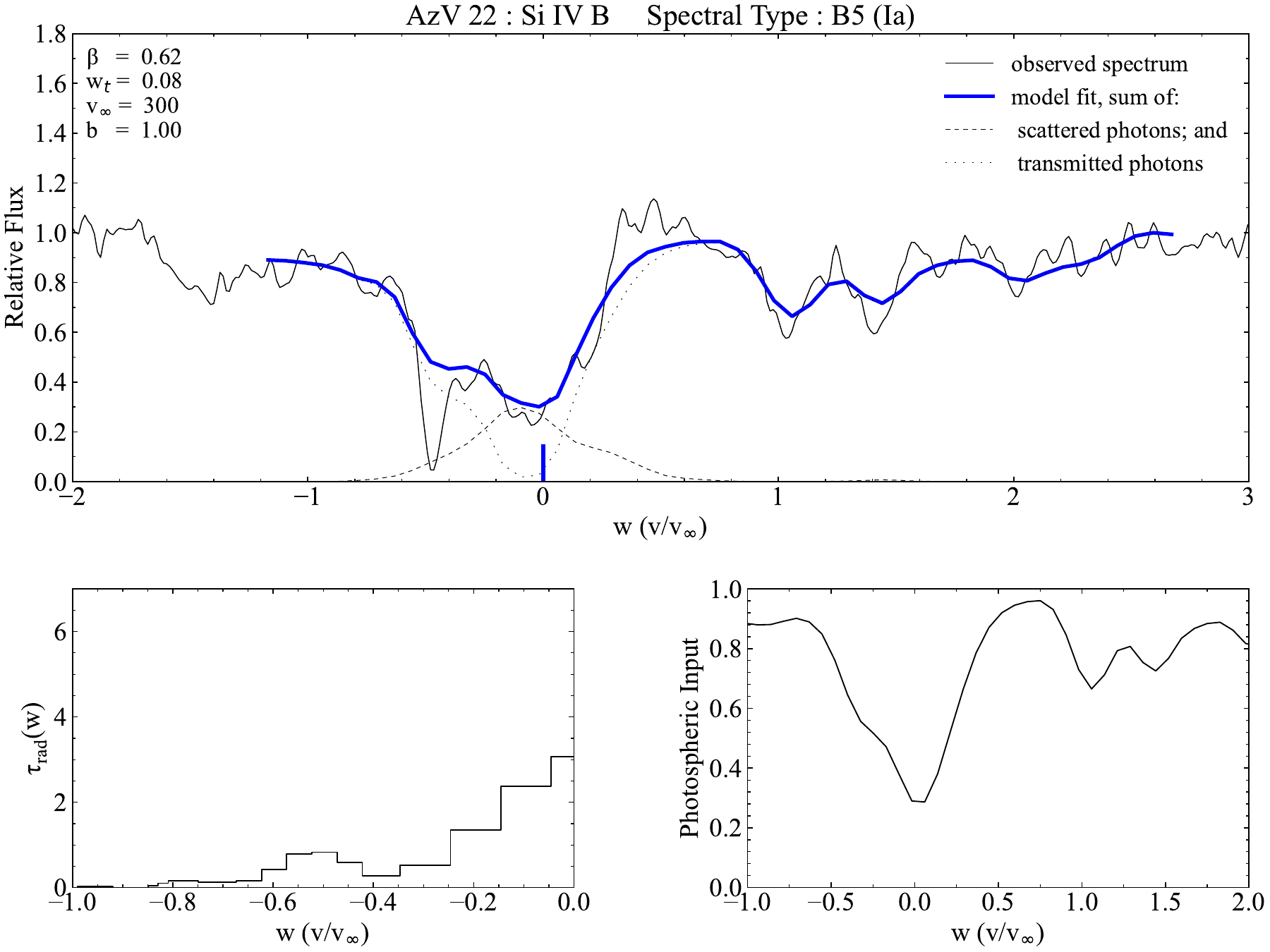} }
 \qquad
 \subfloat[ ]{\includegraphics[width=3.34in]{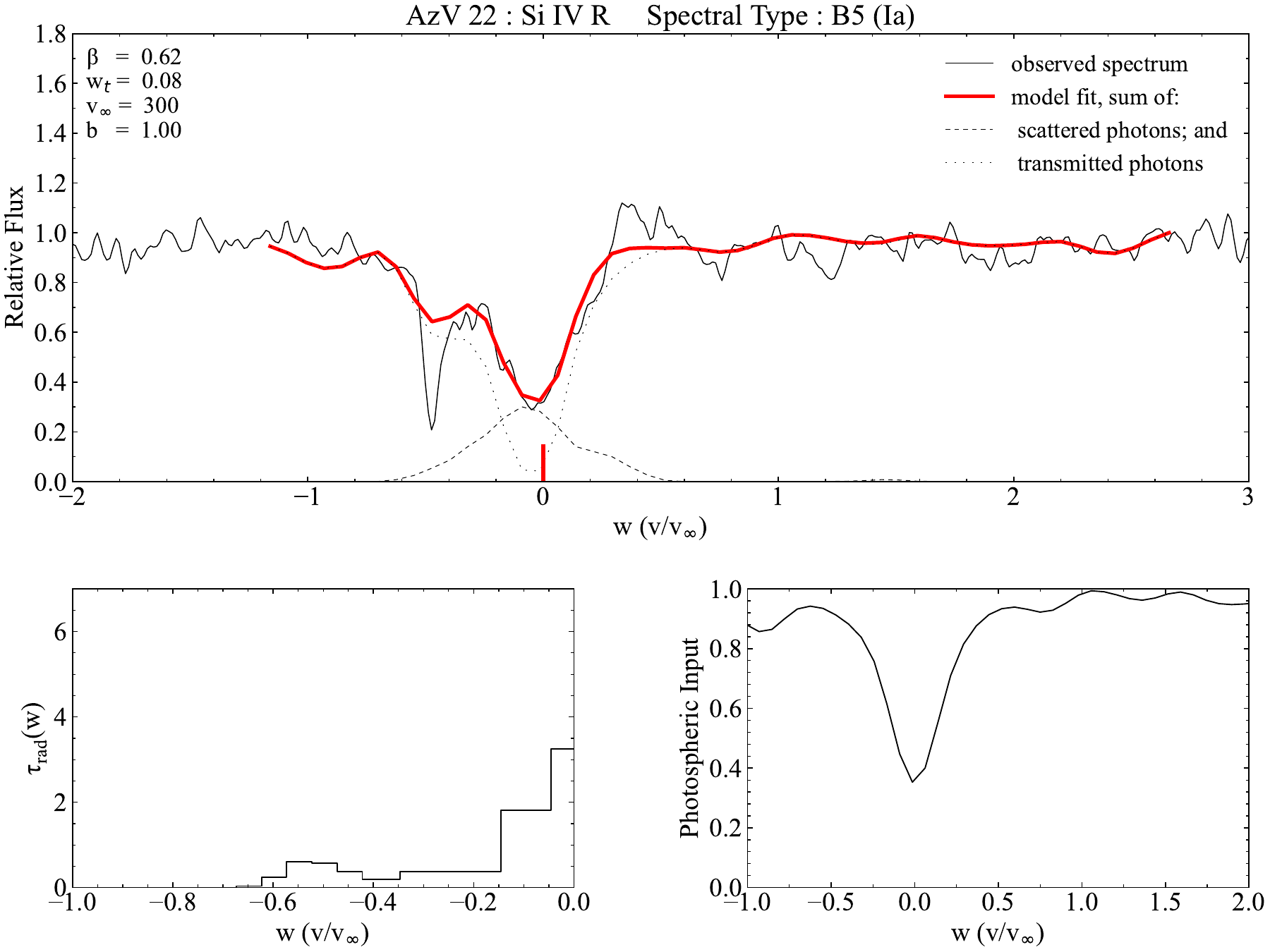} }
 \caption{SEI-derived model fits for (a) blue and (b) red components of the Si \textsc{iv} doublet feature in the UV spectrum of SMC star AzV 22 (spectral type B5 Ia).}
 \label{fig22}
\end{center}
    \end{figure*}

   \begin{figure*}
\begin{center}
 \subfloat[ ]{\includegraphics[width=3.34in]{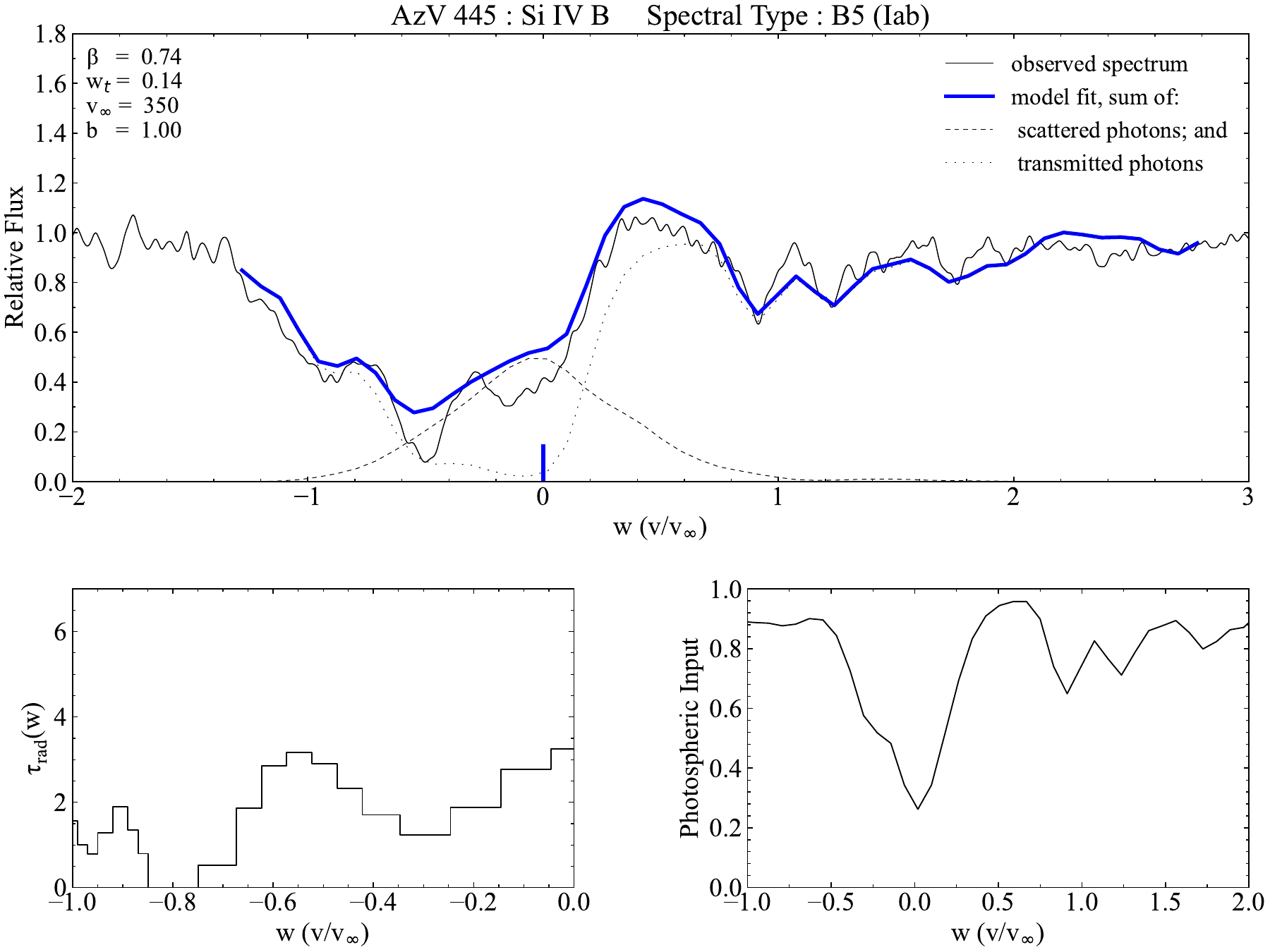} }
 \qquad
 \subfloat[ ]{\includegraphics[width=3.34in]{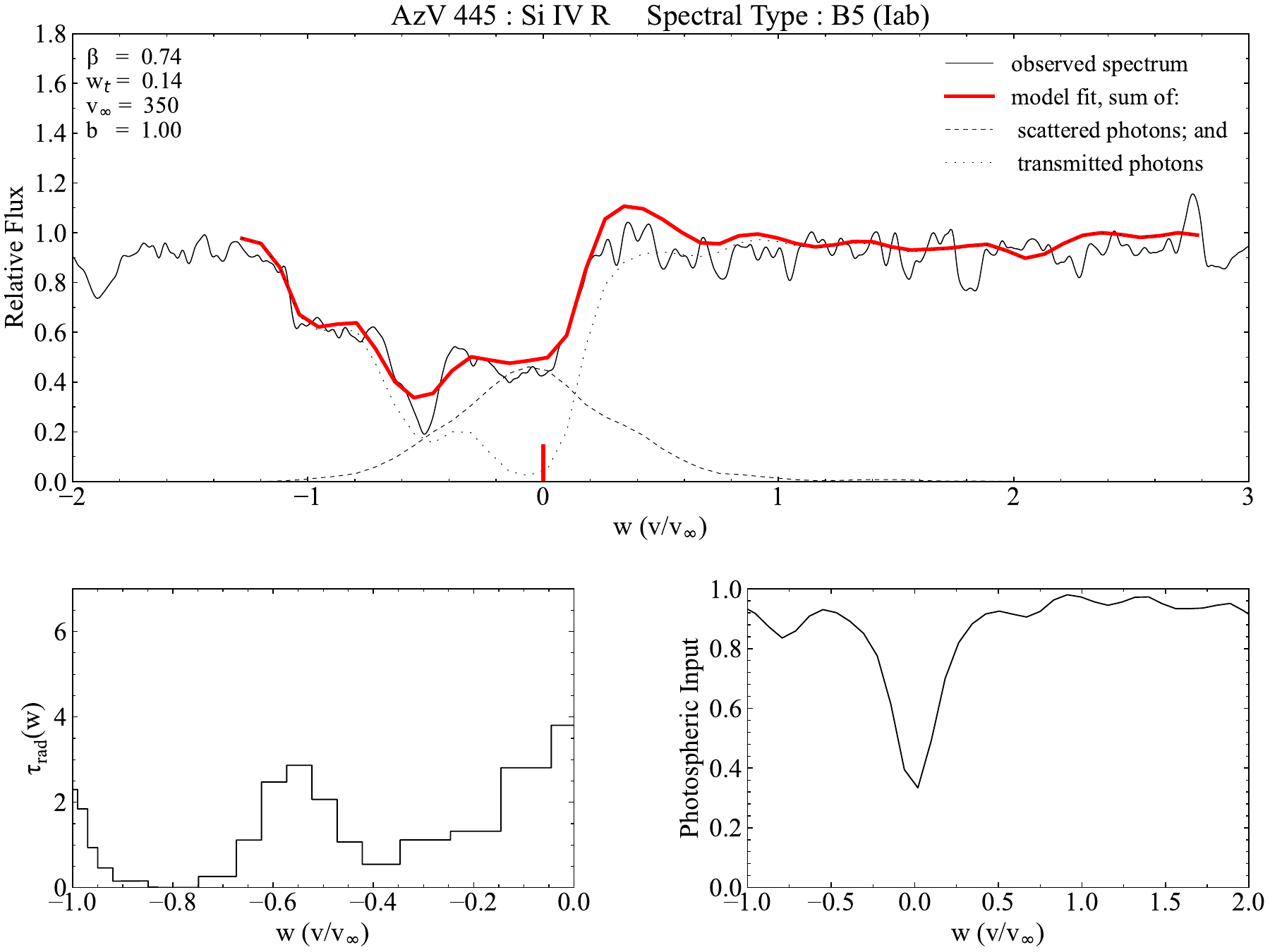} }
 \caption{SEI-derived model fits for (a) blue and (b) red components of the Si \textsc{iv} doublet feature in the UV spectrum of SMC star AzV 445 (spectral type B5 (Iab)).}
 \label{fig445}
\end{center}
    \end{figure*}

\clearpage
\newpage
\section{SEI model line fits for ``low Z'' B supergiants}

   \begin{figure*}
\begin{center}
 \includegraphics[width=5.52in]{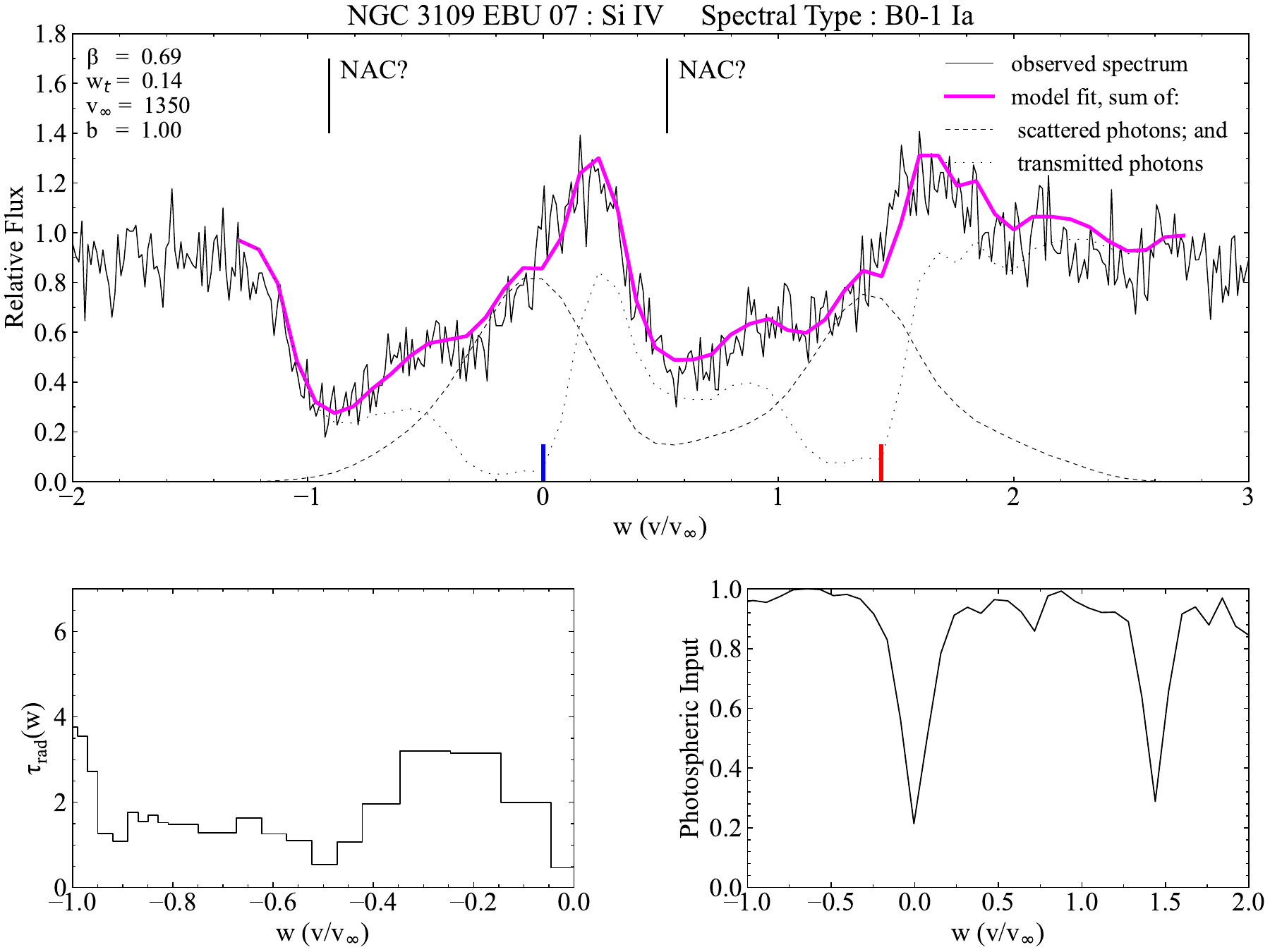}
 \caption{SEI-derived model combined fit for both components of the Si \textsc{iv} doublet feature in the UV spectrum of NGC 3109 star NGC 3109 EBU 07 (spectral type B0-1 Ia). The location of possible narrow absorption components (NAC) in each doublet element are indicated, each at the same distance from the relevant rest wavelength.}
 \label{fig31097}
\end{center}
    \end{figure*}

   \begin{figure*}
\begin{center}
 \subfloat[ ]{\includegraphics[width=3.34in]{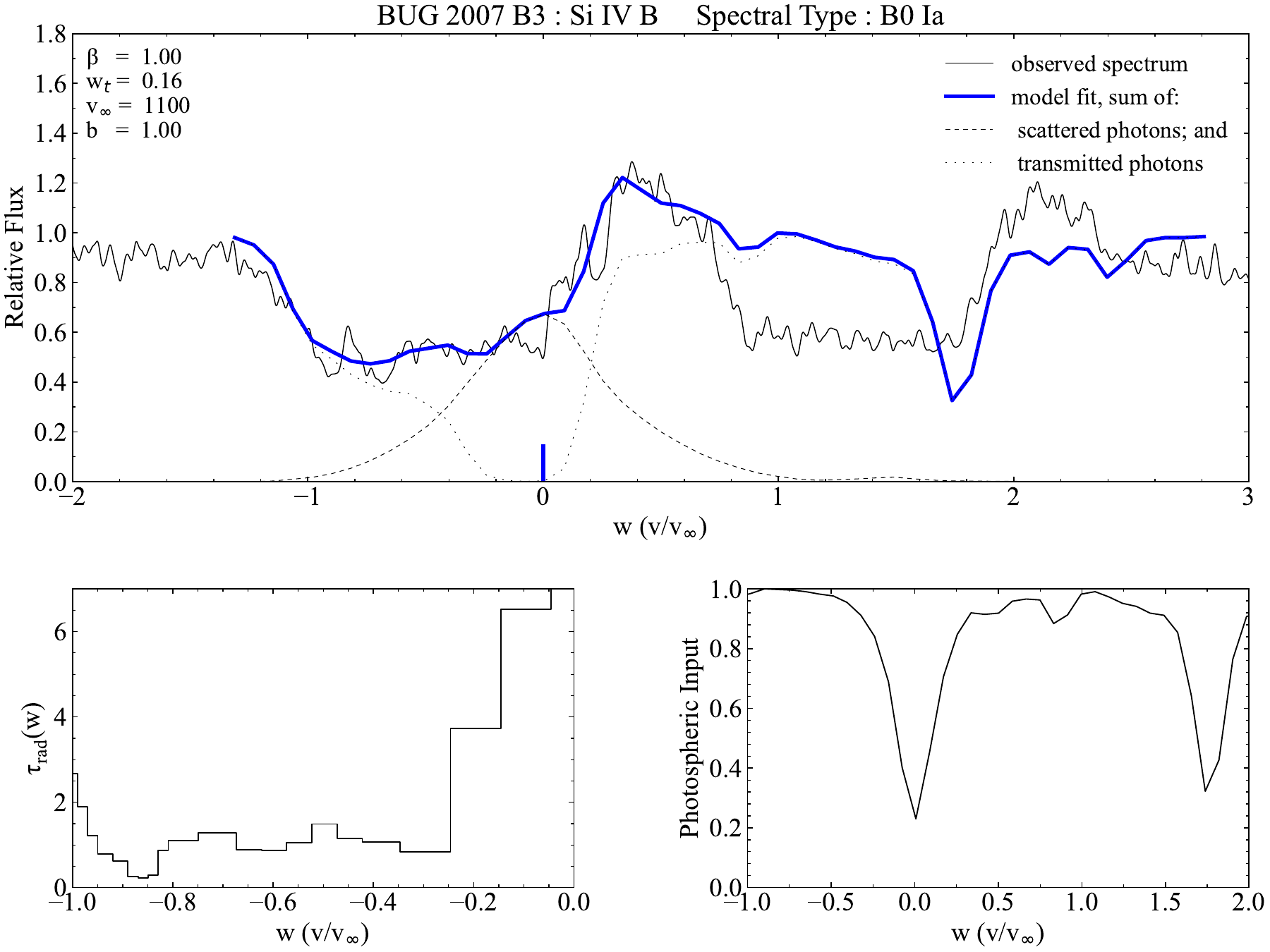} }
 \qquad
 \subfloat[ ]{\includegraphics[width=3.34in]{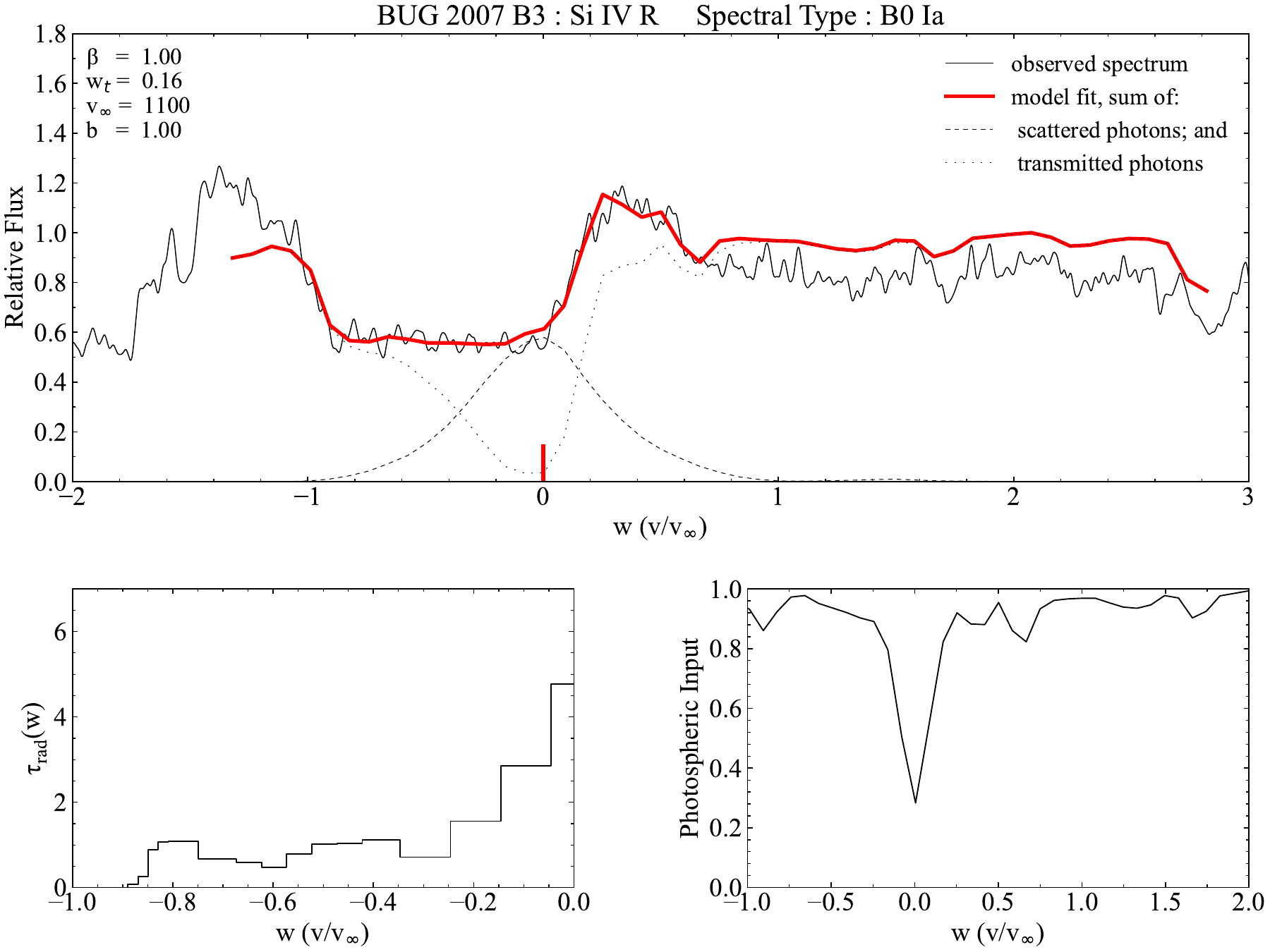} }
 \caption{SEI-derived model fits for (a) blue and (b) red components of the Si \textsc{iv} doublet feature in the UV spectrum of IC 1613 star BUG2007 B3 (spectral type B0 Ia).}
 \label{fig20073}
\end{center}
    \end{figure*}

   \begin{figure*}
\begin{center}
 \subfloat[ ]{\includegraphics[width=3.34in]{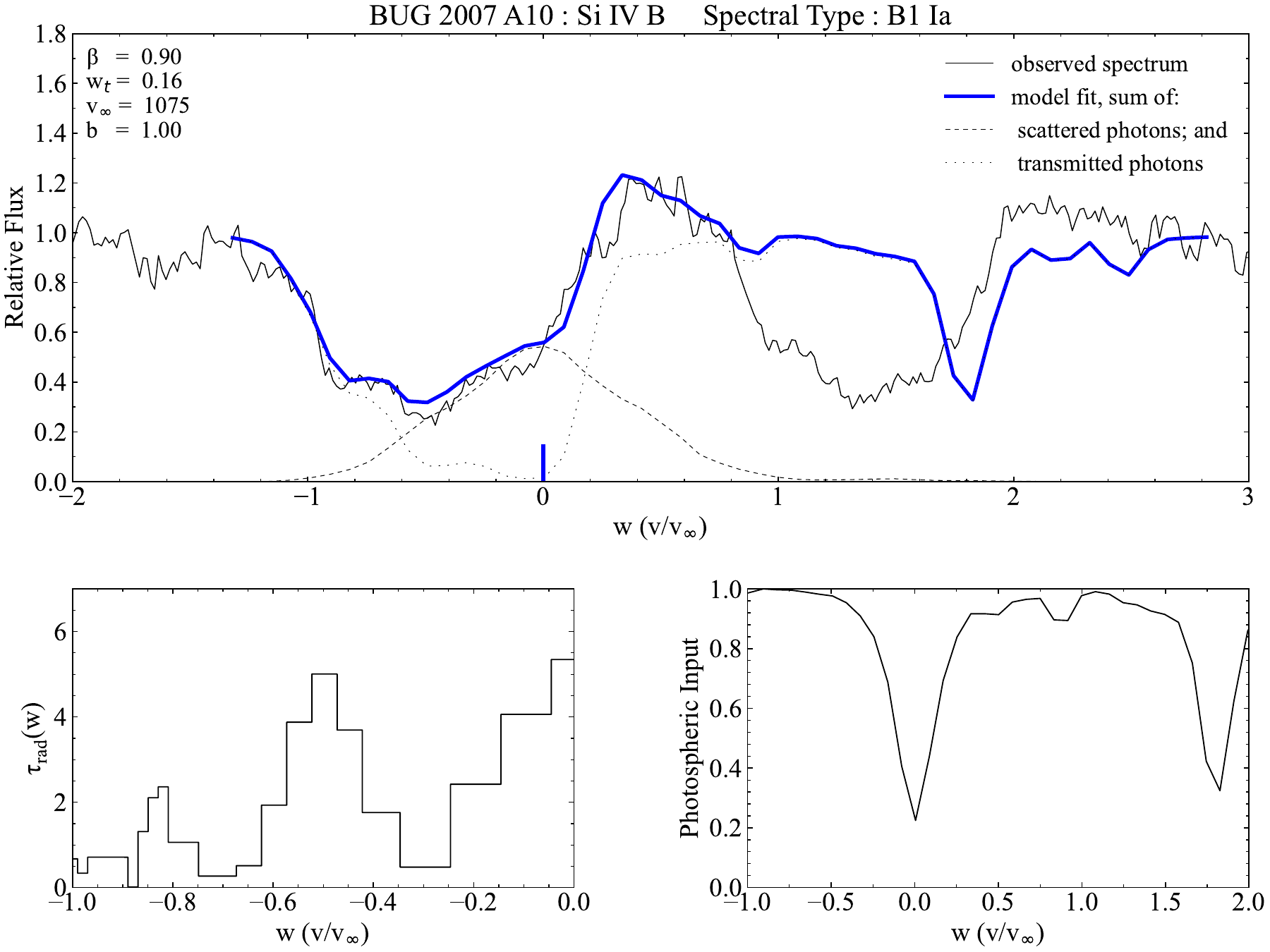} }
 \qquad
 \subfloat[ ]{\includegraphics[width=3.34in]{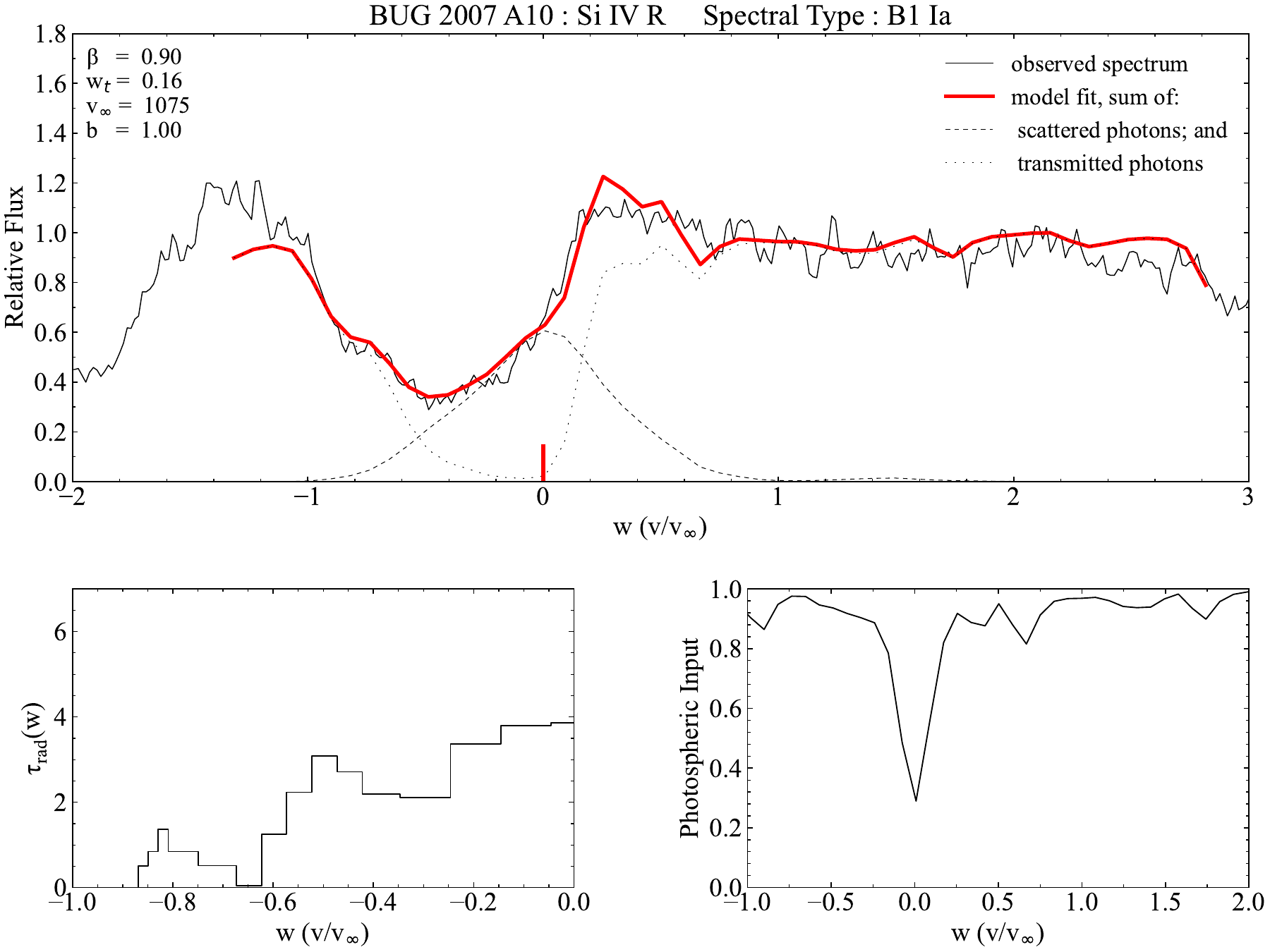} }
 \caption{SEI-derived model fits for (a) blue and (b) red components of the Si \textsc{iv} doublet feature in the UV spectrum of IC 1613 star BUG2007 A10 (spectral type B1 Ia).}
 \label{fig200710}
\end{center}
    \end{figure*}
   
   \begin{figure*}
\begin{center}
 \subfloat[ ]{\includegraphics[width=3.34in]{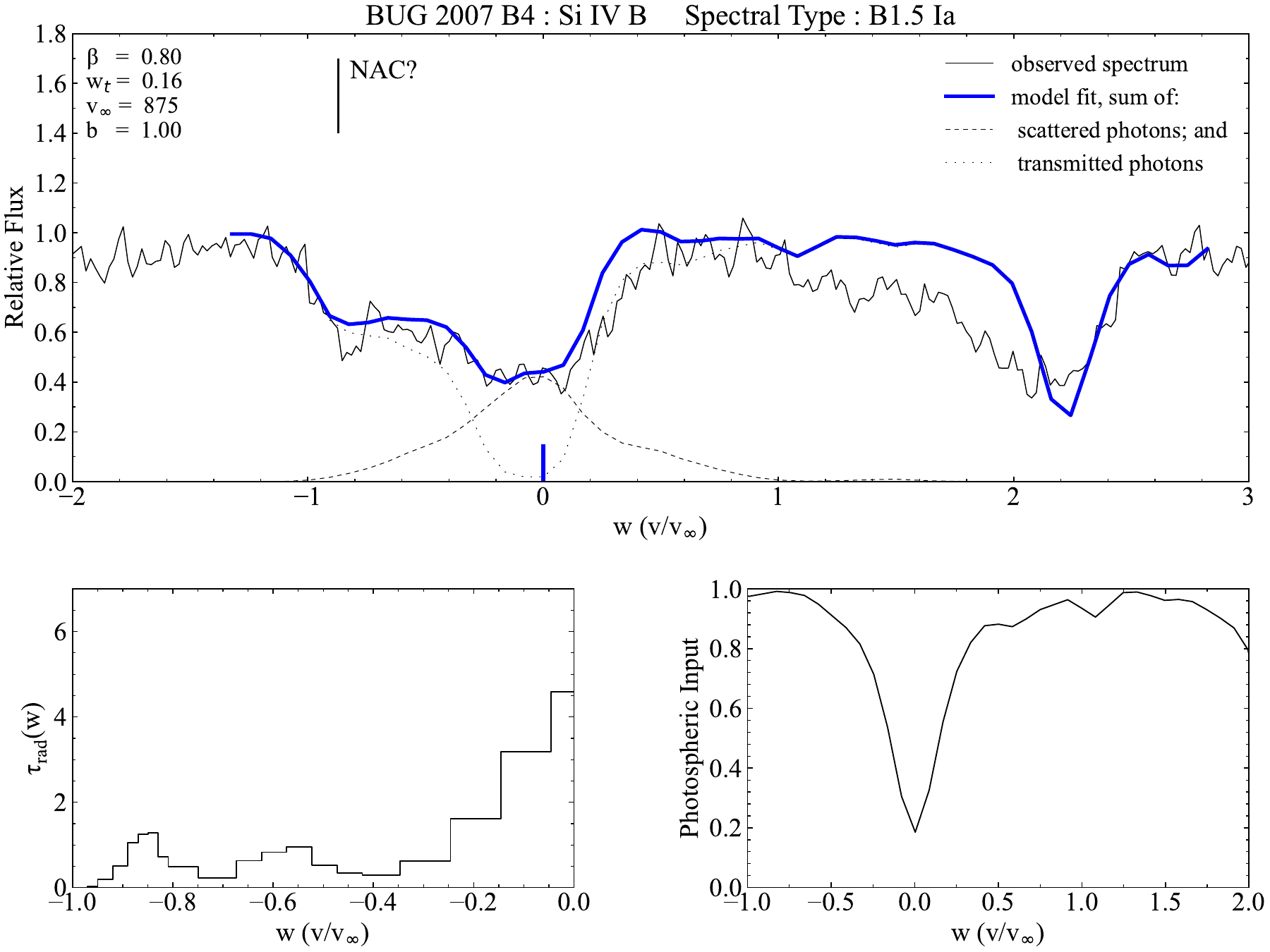} }
 \qquad
 \subfloat[ ]{\includegraphics[width=3.34in]{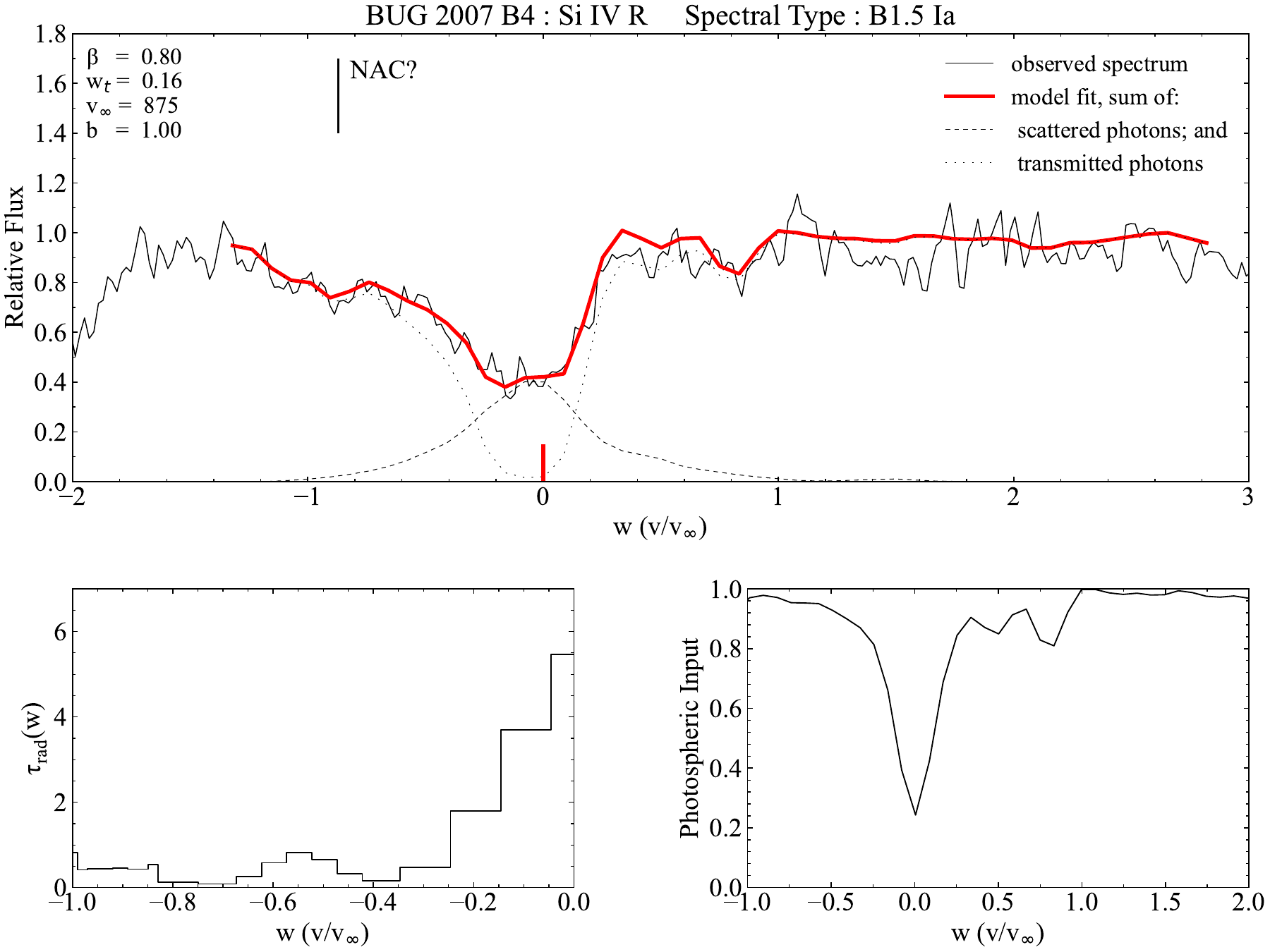} }
 \caption{SEI-derived model fits for (a) blue and (b) red components of the Si \textsc{iv} doublet feature in the UV spectrum of IC 1613 star BUG2007 B4 (spectral type B1.5 Ia). The location of possible narrow absorption components (NAC) in each doublet element are indicated, each at the same distance from the relevant rest wavelength.}
 \label{fig20074}
\end{center}
    \end{figure*}


\bsp	
\label{lastpage}
\end{document}